\documentclass[fleqn,usenatbib,amssymb]{mnras}

\usepackage{newtxtext,newtxmath}
\usepackage{multirow}

\usepackage[T1]{fontenc}

\DeclareRobustCommand{\VAN}[3]{#2}
\let\VANthebibliography\thebibliography
\def\thebibliography{\DeclareRobustCommand{\VAN}[3]{##3}\VANthebibliography}

\usepackage{graphicx}	
\usepackage{amsmath}	

\def\oiiiopt{{\sc{[Oiii]}}$\lambda\lambda$4959,5007\/}
\def\rfe{$R_\mathrm{FeII}$\/}
\def\feii{{Fe\sc{ii}}\/}
\def\hb{{\sc{H}}$\beta$\/}
\def\lledd{$L_{\rm bol}/L_{\rm Edd}$\/}
\def\ledd{$L_{\rm Edd}$\/}
\def\mbh{$M_{\rm BH}$\/}
\def\civ{C{\sc iv}$\lambda$1549\/}
\def\lvc{$L_{\rm 5100}$\/} 
\def\avglvc{$\overline{L}_{\rm 5100}$\/}

\def\kms{km~s$^{-1}$\/}

\def\lbol{$L_{\rm bol}$\/}
\def\feiifull{Fe{\sc ii}$\lambda$4570}
\defcitealias{1987mathewsferland}{MF87}

\title[The spectral energy distribution of extreme population A quasars]{The spectral energy distribution of extreme population A quasars}

\author[Garnica et al.]{
Karla Garnica,$^{1}$\thanks{E-mail: kgarnica@astro.com.mx (KG)}
Deborah Dultzin,$^{1}$
Paola Marziani$^{2}$
and Swayamtrupta Panda$^{3,4,\thanks{Gemini Science Fellow}}$
\\
$^{1}$Universidad Nacional Autónoma de México, Instituto de Astronomía, AP 70-264,  04510, CDMX, México\\
$^{2}$National Institute for Astrophysics (INAF) Astronomical Observatory of Padua, Italy\\
$^{3}$International Gemini Observatory/NSF NOIRLab, Casilla 603, La Serena, Chile\\
$^{4}$Laborat\'orio Nacional de Astrof\'isica - MCTI, R. dos Estados Unidos, 154 - Na\c{c}\~oes, Itajub\'a - MG, 37504-364, Brazil
}

\date{Accepted 2025 May 21. Received 2025 May 20; in original form 2024 August 22}

\pubyear{\the\year{}}

\begin{document}
\label{firstpage}
\pagerange{\pageref{firstpage}--\pageref{lastpage}}
\maketitle

\begin{abstract}
Knowledge of the broad-band active galactic nuclei (AGN) spectral energy distribution (SED) that ionizes the gas-rich broad emission line region is key to understanding the various radiative processes at play and their importance that eventually leads to the emission line formation. We modeled a spectral energy distribution for highly accreting quasars, also known as extreme population A sources, based mainly on observational data available in astronomical databases, and on accretion disk models for the unobservable far-UV domain.
Our selection criterion is the \rfe \ parameter - the ratio of the optical FeII emission between 4434  \AA\ and 4684   \AA\ to the \hb\ $\lambda$4861 \AA\ intensity, \rfe $\ge$ 1. This criterion is satisfied by highly-accreting, possibly super-Eddington, black holes. We analyzed { 155} sources up to a redshift of approximately 1, previously reported in the literature, to construct a median radio-quiet SED spanning from radio to X-ray wavelengths.  We find that the SED of quasars exhibits distinct features compared to lower accreting AGN, including a pronounced { big} blue bump and strong optical/UV emission along with a steep X-ray continuum.  We classify the sources into radio-quiet, radio-intermediate, and radio-loud categories, observing that radio-intermediate and {  a subsample of radio-quiet}  AGN show a significant far-IR excess over the radio-quiet SED and the far-IR excess appears to be related to the prominence of \feii\ emission. There is an overall consistency between the new SED and the one obtained for high Eddington ratio quasars in previous work. We provide the SEDs in digital format for eventual applications.
  
\end{abstract}




\section{Introduction}

The Spectral Energy Distribution (SED) represents the continuum of energy distribution across different wavelengths/frequencies. It serves as a valuable tool in astrophysics for identifying emission processes within celestial sources, thereby offering insights into their nature \citep[e.g.,][]{wilkesetal99}. 

Several authors have put their effort into combining multi-frequency observation to build empirical SED for Active Galactic Nucleus (AGN), and AGN SED modeling, or to create hybrid templates that combine observations and  model inferences. Among the early works a good example is the one of \citet[][hereafter \citetalias{1987mathewsferland}]{1987mathewsferland}. They derived an SED for ``typical quasars'' that more recent works would identify as moderately accreting population A quasars \citep{pandamarziani23}. For low-redshift AGN studies, the most relevant spectral range lies in the unobservable far-ultraviolet (FUV), above the ionization threshold of Hydrogen, where the peak of quasar emission is located \citep{malkansargent82}, the so-called big blue bump (BBB, \citealt{czerny_elvis_1987}). The \citetalias{1987mathewsferland} SED provided a first characterization of the BBB based on the prominence of the He {\sc ii}$\lambda$4686 line. This parameterization of the AGN SED remains a valid reference for typical AGN to date. 

In the 1990s, a more comprehensive observational definition of the AGN SED emerged with the incorporation of X-ray observations. For instance, \citet{1992Puchnarewicz} analyzed a sample of 53 AGN with ultra-soft X-ray excess and observed that they tend to exhibit narrower permitted lines than optically selected samples, a finding supported by \citet{1996boller}. Furthermore, \citet{1993Walter&Fink} investigated a broad wavelength range from ultraviolet (UV) to X-ray, leading to improved constraints on the BBB. Additionally, \citet{1994elvis} compiled a SED atlas of 47 ``normal quasars'', employing a specialized treatment to subtract the starlight of the host galaxy, particularly prominent in the near-infrared (NIR). These efforts covered the SED from the infrared (IR) to the hard X-ray domain, resulting in a mean SED, albeit with considerable dispersion.
Building upon this foundation, new work on AGN SED started to take advantage of increasingly wider optical surveys, such as the Sloan Digital Sky Survey (SDSS). For instance,  \citet{2006richards} constructed a series of mean SEDs as a function of color and luminosity based on a multi-catalog selection of $\sim$300 type 1 quasar.  More recent work continued to take advantage of the increasing availability of multifrequency data \citep{shangetal11,brownetal19,spinolioetal24}. 

A major point in the AGN investigation became the realization that AGN properties do not scatter around the ones of unique prototypical sources \citep{borosongreen92,sulenticetal00a,sulenticetal00c}. Since then, random dispersion in parameters and several weak and stronger trends have been explained as a set of correlations (the so-called Eigenvector 1 of quasars, E1) whose underlying physical parameter is believed to be  Eddington ratio, the ratio between the bolometric luminosity (\lbol) and the  Eddington luminosity (\ledd) \citep{borosongreen92,sulenticetal00a,marzianietal01,sunshen15,pandaetal19,wolfetal20,martinez-aldama_etal_2021}.  Spectral slopes in the UV and X-ray domain are dependent on the AGN accretion status \citep{laoretal97b,1996boller,1993Walter&Fink,2007Vasudeban&Fabian}. For example,  \citet{2007Vasudeban&Fabian} studied a sample of 54 AGN from the FUV up to the X-ray and found a relation between the hard X-ray bolometric correction and   Eddington ratio.  Several works pointed out that quasars accreting at high Eddington ratio show the steepest slopes in the soft and hard X-ray domain \citep{sulenticetal00a,sulenticetal07,grupeetal98,grupeetal10,wangetal13,laurentietal22}. 

In line with the Eigenvector-1 related development, some works modeled the SED for AGN as a function of  Eddington ratio \citep{2020Feland, 2012Jin,2012Done}, based on observational data for low redshift type 1 AGN. Again, they confirmed a steeper slope in the FUV and X-ray domain \citep{2020Feland}. However, all these works constructed SEDs for unobscured type-1 AGN based mainly on color or physical parameters like \lledd\ or luminosity that are subject to several caveats: AGN intrinsic obscuration and variability,  viewing angle effects \citep{runnoeetal13}, as well as large statistical and systematic uncertainties  \citep{vestergaardpeterson06,marzianietal06}. These effects are in turn dependent on the source location along the E1 sequence. For instance,  the bolometric correction is, as a matter of fact, a large factor whose value is dependent on luminosity, Eddington ratio, as well as structural properties of the AGN \citep{2014wang,pandaetal20}.  

Our approach is to construct a semi-observational SED focused on a clearly defined observational criterion. The E1 parameter space is often represented as a sequence in the plane defined by the parameter \rfe\ -- the ratio of the optical FeII emission between 4434  \AA\ and 4684   \AA\ (hereafter \feiifull, \citealt{borosongreen92}) to the  H{\sc i} Balmer line \hb\      intensity and the FWHM of \hb\ itself. This sequence has become known as the ``main sequence" of quasars \citep{sulenticetal00a,marzianietal01,shenho14,sunshen15,wildyetal19,pandaetal19}. The \rfe\  ratio can be measured with a relatively high precision in high signal-to-noise (S/N) spectra, $\delta$\rfe/\rfe $\approx 0.05$ at 1 $\sigma$ confidence level if \rfe$\sim$1 \citep{marzianietal03a}.  Along the quasar main sequence (MS), the \rfe $\ge$ 1  criterion is satisfied by highly accreting quasars or extreme population A sources \citep{2014marziani,duetal16a,panda_etal_2024}. We expect that changes associated with intrinsic optical variability are minimized  \citep{duetal14,duetal15,duetal16,duetal18}. 

The extreme population A of quasars (hereafter, xA) shows strong wind effects from lines of high ionization potential, such as the UV resonance line of C{\sc iv}$\lambda$1549 \citep{marzianietal96,leighlymoore04,sulenticetal07,coatmanetal16,sulenticetal17,vietrietal18,coatmanetal19,vietrietal20}. These winds may enrich of metals the host galaxy or even the intergalactic environment \citep{2015KingPunds}. Recent works derived a metallicity range 10 - 50 Z$_\odot$\ \citep{2021sniegowska,2022garnica}. Such high values were derived from the comparison between measured line intensity ratios and the prediction of photoionization computations. Several workers have pointed out that these values are exceptionally high, in the most extreme cases higher by an order of magnitude to the highest metallicity measures in the host galaxies as well as in the Narrow Line Region (NLR, \citealt{xuetal18,storchi-Bergmannetal98}).  Hence, an important application is to study the metal content of this particular population by modeling the broad line region (BLR) through {\tt  CLOUDY} photoionization computations.  One crucial parameter to carry out the {\tt CLOUDY} modeling is the shape and intensity of the external radiation field striking the medium. Even though there are  AGN SED  carefully constructed for individual quasars available \cite[e.g.,][]{shangetal11,brownetal19}, very few of them satisfy our optical selection criterion (\rfe $\ge$ 1), and no previous work has separated the SED based on the location (spectral type) along the quasar main sequence. Although different SEDs have been employed along the sequence in several works \citep{panda2018,panda_etal_2019_wc,pandaetal19,pandaetal20,marzianietal21a,marzianietal23,florisetal24}, those SEDs were not computed for different spectral types along the sequence, and not even for population A and B specifically.   Therefore,  an important goal of this work is to create a SED for extreme population A sources mainly based on observational data, following a clear criterion based on the quasar main sequence. The aim is to use the  SED as an input to photoionization models of the BLR at low luminosity. While the SED has been derived from low-$z$ type-1 AGN, a SED for AGN of modest black hole masses  \mbh\    may be appropriate also for very high redshift quasars with \mbh\ $\sim 10^7 - 10^8 M_\odot$ that are found by James Webb Space Telescope \citep[e.g., ][]{onoueetal23,larsonetal23}. Photoionization solutions with an appropriate SED would make it possible to measure more precisely the gas metal content, as well as the broad-line region (BLR) wind dynamical parameters \citep{marzianietal17,vietrietal18,vietrietal20,deconto-machadoetal23,deconto-machadoetal24}.

After introducing the sample selection criteria, we describe the origin of the data in the various frequency domains and the sample construction (Section  \ref{sample}). In Section \ref{results} we show the estimated accretion-related parameters and connect them to the SED (Section \ref{parameters}). The obtained SED  is presented in Section \ref{sed building},  first for the radio-quiet sub-population.  We afterward distinguish the SED based on the radio-loudness (Section \ref{RQ-RI-RL}).   The discussion (Section \ref{disc}) is focused on the comparison with several SEDs available in the literature, SED differences for varying  \feii\ prominence, the implication of the new SED for the derivation of the physical condition of the line emitting region and a brief discussion considering the issue of the viewing angle relevance for highly accreting quasars. Last, even if the data sources are diverse and unchecked, we were able to build a tentative Hubble diagram, to confirm that the \hb\ line is emitted in a virialized system, as found previously (\citealt{dultzinetal20,donofrioetal24}, and references therein).

\section{Sample}
\label{sample}

Our selection is based on the flux ratio of the \feiifull\  and \hb, the so-called \rfe\ parameter. This optical criterion identifies highly accreting quasars for \rfe $\ge$ 1   \citep{2014marziani,duetal16a,pandaetal19}.  To create the xA SED we visited multiple catalogs due to the rarity of xA sources. Of over 5000 quasars among the catalogs, we identify nearly $\sim$ 800 xA sources with  0.013 $<  z  <$ 1.521. This number points out that the xA population involves a minority of AGN, $\approx 16\%$, a prevalence consistent with the one found in SDSS-based samples \citep{2013Marziani_composite}.  We set a redshift limit  $z$ $\approx$ 1, due to instrumental limitations: on the one hand, above $z \approx 1$ the \hb\ spectral range is observable in the NIR, accessible only by IR spectrometers; on the other hand, the multifrequency data needed to build the SED become sparse for the general population of AGN, and are rarer for xA sources.  We selected { 155} sources with available multi-frequency photometry and exiting X-ray measurements corrected by HI absorption to model the xA SED.  The catalog search was sub-divided into two branches with the aims: (1)  to identify xA sources, and (2) to retrieve multi-frequency data, which are described below. 

\subsection{Catalogs to identify xA sources}
\begin{figure}
\centering 
\includegraphics[width=9cm]{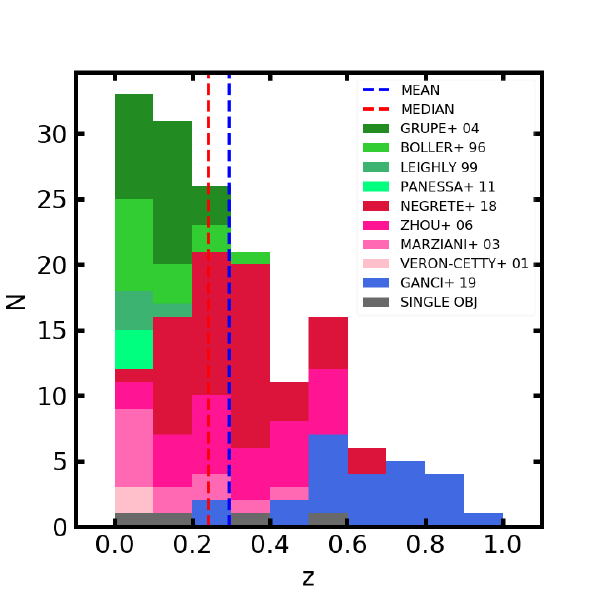} 
    \caption{Stacked redshift distribution of the selected xA low redshift highly accreting quasars in the literature, each color shows the subsamples catalogs to identify xA sources, these catalogs provided us with insight from the x-ray  (green distributions), optical (pink distributions) and radio regions (blue distribution). The blue dashed line shows the general mean of the sample at $\overline{ z_{15{5}}}\approx {0.295}$, red dashed line shows the median of the sample  $\mu({z_{15{5}}}) \approx {0.243}$.\label{fig:z_200}}
\end{figure}  

To identify xA sources we sought catalogs that provided either the \rfe\ parameter measurement or the individual fluxes of \feiifull\ and \hb, in addition to a measurement of the full width at half maximum (FWHM) of \hb\ which is immediately needed to compute the black hole mass (Section \ref{parameters}). All of these parameters are in the optical domain of the spectrum and therefore all catalogs are optically based. 

Figure \ref{fig:z_200} shows the stacked redshift distribution of the AGN retrieved from the catalogs used to identify xA sources. Our first step for creating an xA SED was using the data from the \cite{2003marziani} low-redshift atlas for AGN. This 215-source catalog provides spectroscopic information with S/N  $\approx$ 25 spectra. We select 13 xA sources from this catalog. To extend our sample,  we visited the SDSS Data Release (DR) 7  \citet{2018negrete} catalog for low redshift highly accreting quasars. This catalog contains detailed measurements of the optical region. We selected $\sim$ 44 xA sources. From the  \cite{2006zhou}   SDSS DR3 spectra catalog we selected nearly 30 sources. From the spectrophotometric atlas of narrow-line Seyfert 1  (NLSy1) nuclei of \citet{2001veron-cetty}, we selected 2 xA sources. The previously-described catalogs are shown in Fig. \ref{fig:z_200} by the pink/{ reddish} distributions. Some peculiar xA sources weren't found in catalogs but in particular studies. They are identified by the grey distribution in Fig. \ref{fig:z_200}. That is the case of (1) Mark 231 \citep{2006sulentic}, an    \rfe $\approx$ 1.78  source with a characteristic IR excess, of extraordinarily high luminosity among low-$z$\ type-1 AGN \citep{boksenbergetal77,sandersmirabel96} {,} (2) PHL 1092 \citep{2009miniutti, 2020marinello} an extreme FeII emitter and X-ray weak quasar belonging to the A4 bin (\rfe $\approx$ 1.81) {and (3) PDS 456 \citep{2025Li} a very luminous quasars with extreme X-ray properties \citep{reevesetal00,reevesetal03}}.  The single source SDSS J133602.01+17.2513.1 from \citet{2014marziani} is also included in this histogram.

Although all catalogs provided us with optical parameters to identify them as xA sources, some catalogs were selected because they also provided SED insight in other spectral regions, especially in the radio (Fig. \ref{fig:z_200} blue distribution) and X-ray domains (Fig. \ref{fig:z_200} green distributions). Nearly 30 sources come from the \citet{2019ganci}  radio catalog that classified radio loudness along the Eigenvector 1 optical plane, all of the selected sources show a core-jet morphology. \citet{1996boller} study on soft X-ray properties for narrow-line Seyfert galaxies provided 13 xA sources. The soft X-ray sample of \citet{2004grupe} also included optical spectral and its measurements, and we were able to identify nearly 30 xA sources. A study of narrow-Line Seyfert 1 galaxies observed by ASCA from \cite{1999leighly} reports the X-ray analysis and also some optical properties of the sample, and we select 4 xA sources. From the \citet{2011panessa}  X-ray study on broad lines Seyfert 1 galaxies we selected 3 xA sources. 

\subsection{Catalogs to retrieve multi-frequency data}
\label{sec:sample_retrieve}

Once we selected a low redshift sample, as described in the previous section, we proceeded to download the available photometry.

\subsubsection{Radio to near-UV}

The radio to near-UV photometric data were downloaded from the NASA Extragalactic Database (NED). {All data  is described and provided on Appendix \ref{App:digitalSAMPLE}. A brief description of the observatories/surveys is provided in Table \ref{tab:apertures}.}

{Prior our modeling techniques,} we corrected by Galactic extinction, following the \citet{1989Cardelli} extinction curve, and then applied a redshift correction. The 
 Galactic extinction ($A_\mathrm{V}$) and redshift ($z$) values were also obtained from NED.

\subsubsection{Ultraviolet}
\label{sample:uv}

To enhance the optical-NUV region, we included the composite spectrum from \citet{2013Marziani_composite}. This spectrum is composed of 13 high signal-to-noise population A4 sources within a redshift range of 0.40 $\lesssim z \lesssim$ 0.75  and has a wavelength coverage from 2200 to 6500 \AA. The sample of the 13 spectra is described in \citet{2013Marziani_composite}. Here we use the physical parameters for individual sources that were computed but not reported by those authors.

\subsubsection{X-ray data}
\label{sample:x-ray}

\paragraph{Soft x-rays}

Our soft X-ray database was the \citet{2016Boller} catalog, an improved version of the original ROSAT  quasar catalog in the energy range $\approx$ 0.1 -- 2.4 keV ($\approx$ 2.42$\times 10^{16}$ -- 5.80$\times 10^{17}$\ Hz). We found a match for 250 sources before filtering by error. Our error filter was based on a relative error in the photon index being less than $100$\%, leaving us with over 90 sources from which we removed the outlying steepest and most positive slopes. This left us with 60 sources.

\paragraph{Hard X-rays}

Our hard X-ray database was the \citet{2020Evans}, a Swift X-ray catalog in the energy range 0.3-10 keV (7.254 $\times 10^{16}$- 2.418 $\times 10^{18}$). We found a match for  155 sources. Filtering by a relative error of $100$\%\  leaves us with 70 xA sources. All of \cite{2020Evans} sources show negative slopes in the plane $\nu$ vs. $\nu f_\nu$\ (i.e., spectral index $\alpha > 1$).This results in a systematic decrease in power emitted as frequency increases. However, several sources appear to be intrinsically X-ray weak, beyond what would be expected from this decrease  \citep{saccheoetal23,laurentietal22}.  

\paragraph{Hard X-rays beyond 10 keV}

Our hard X-ray beyond 10 keV database was initially from \citet{2017ricci}, a Swift/BAT X-ray catalog in the energy range 15-150 keV (corresponding to frequencies $\nu \approx 3.6 \times 10^{18}$ Hz -- 3.6 $\times 10^{19}$ Hz). Among the  $\approx$ 900 AGN of this catalog, we only found a match for 7 xA sources, 3 of them known in the literature as radio loud.  We discarded the remaining four for the construction of the median SED in the X-ray domain because they are known $\gamma$-ray emitters. We, therefore, performed a search for NuStar \citep{harrisonetal13} data between 3--80 keV. The results of the search are reported in Table \ref{tab:nustar} and will be further discussed in \S \ref{x-raysed}. In fact, very few radio-quiet xA AGN have been reported as detected in the   X-ray domain beyond 10 keV.

\begin{table*}
    \centering
    \begin{tabular}{llccccccccccc}\hline\hline
       Name & $\Gamma$ & Energy range & Reference \\
       \hline 
      MRK 493 & 2.08$\pm 0.04$ & 1\textemdash20 keV & \citealt{2023Tortosa} \\
      H0707-495 & 3.29$^{+0.02}_{-0.01}$ & 1\textemdash20 keV & \citealt{2018Cao} \\
      IRAS F12397+3333 & 2.28${\pm 0.02}$ & 1 \textemdash20 keV & \citealt{2023Tortosa} \\
      IRAS 04416+1215 & 1.77$^{+0.17}_{-0.09}$ & 1\textemdash20 keV & \citealt{2023Tortosa} \\
    MRK 142 & 2.37${\pm 0.05}$ & 1 \textemdash20 keV & \citealt{2023Tortosa}\\
    \hline
        MEDIAN & 2.28$^{+0.09}_{-0.20}$ \\
       MEAN & 2.36$^{+0.18}_{-0.12}$\\
       $\sigma$ & 0.51 \\
       \hline
    1H0323+342** & 1.85${ \pm 0.03}$ & <25keV & \citealt{2020Panagiotou} \\
      PDS 456 & 2.29$^{+0.04}_{-0.03}$ & <30 keV & \citealt{nardinietal15} \\
       I Zw 1 & 2.15$^{+0.08 }_{-0.06}$ & <40 keV & \citealt{wilkinsetal21} \\
        ARK 564 & 2.48${\pm 0.01}$ & 0.5\textemdash40 keV & \citealt{2022Lewin}\\
       \hline
        MEDIAN & 2.29$^{+0.10}_{-0.07}$ \\
       MEAN & 2.31$^{+0.09}_{-0.07}$\\
       $\sigma$ & 0.14 \\
       \hline
      MRK 766* & 2.24$^{+0.02}_{-0.01}$ & 0.3 \textemdash79 keV & \citealt{2023Mochizuki_mrk766} \\
      MRK 684 & 2.15$^{+0.16}_{-0.19}$ & 3 \textemdash80 keV & \citealt{2021Akylas} \\
      PKS 0558-504* & 2.19$^{+0.07}_{-0.06}$ & 3 \textemdash80 keV & \citealt{2021Akylas}\\
      SWIFT J2127.4+5654& 2.08${\pm 0.01}$ & 3 \textemdash78 keV & \citealt{2018Tortosa}\\ 
      \hline 
       MEDIAN & 2.12$^{+0.02}_{-0.02}$ \\
       MEAN & 2.12$^{+0.06}_{-0.19}$\\
       $\sigma$ & 0.14 \\
         \hline
      SWIFT J2127.4+5654 & 2.50${\pm 0.24}$& 2\textemdash100 keV& \citealt{2011panessa} \\
      IGR J14552\textemdash5133 & 1.90$^{+0.44}_{ -0.46}$ &  2\textemdash100 keV& \citealt{2011panessa} \\
      IGR J16385\textemdash2057 & 3.07$^{+1.00}_{ -0.71}$ &  2\textemdash100 keV& \citealt{2011panessa} \\
      
       \hline
       MEDIAN & 2.50$^{+0.29}_{-0.30}$ \\
       MEAN & 2.50$^{+1.10}_{-0.89}$\\
       $\sigma$ & 0.48 \\
       \hline
    \end{tabular}
    \caption{Photon indices estimated for the energy ranges above 10 keV.    Col. 1: source name; Col. 2: Photon index; Col. 3: Instrumental energy range; Col. 4: source of very hard x-ray data. The last three rows show (1) median of the sample upper and lower semi-interquartile as superior and inferior limits, (2) mean of the sample following Barlow method, and (3) the standard deviation of the sample. Radio-loud (**) and radio-intermediate (*) sources were excluded from the statistical analysis.}
    \label{tab:nustar}
\end{table*}

\begin{figure*}
    \includegraphics[width=7.8cm]{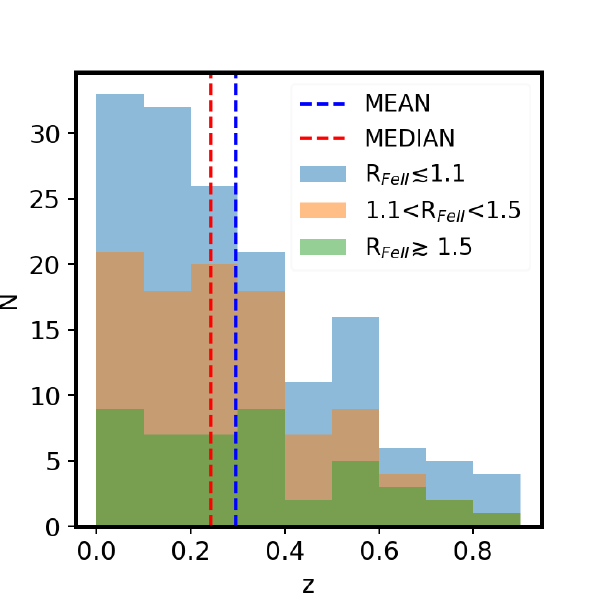}
    \includegraphics[width=7.8cm]{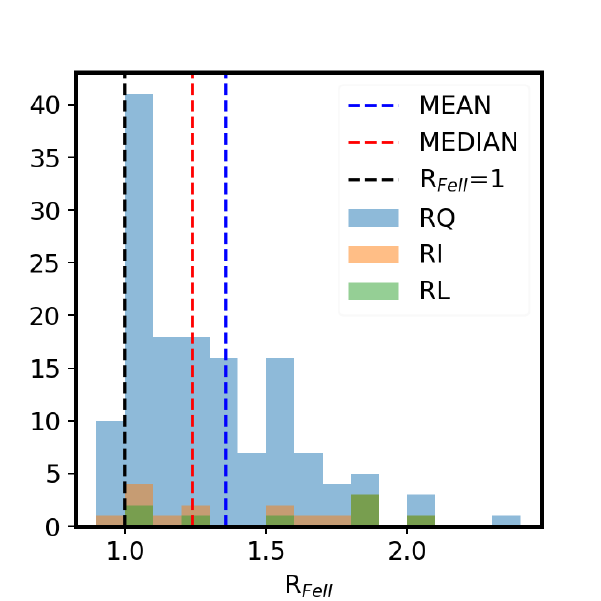}
    \caption{Left: stack distribution of sample redshifts. The green distribution shows the extremely high accretors of the sample (\rfe $\ge$ 1.5 ), which is the less populated subsample (N=45). The orange distribution shows the intermediate high accretors (1..1 < \rfe < 1.5) and the blue distribution shows the high accretors (\rfe $\lesssim$ 1.1) of the sample. This last subsample has the highest prevalence among highly accreting quasars. Right: Stack distribution of sample \rfe\ parameters. The green distribution shows the radio-loud sources of the sample (RL). The orange distribution shows the radio-intermediate (RI) sources and the blue distribution shows the radio-quiet sources (RQ). The last subsample is the most populated one with nearly 90\% of our sample.\label{fig:z+r_fe}} 
\end{figure*}

\subsection{Sample properties}

We restricted our selection to 155 sources with enough information to model their SEDs at least partially,  from the nearly 800 low redshift xA sources.  Our criterion was based on the available multi-frequency photometry and especially on existing X-ray measurements corrected by HI absorption.  Figure \ref{fig:z+r_fe} shows the stack distributions of the redshift (left panel) and the \rfe\ parameter (right panel) for the 155 xA sample. 

The redshift shows a skewed distribution with median $\mu(z)\approx{0.243}$, and average $\overline{z} \approx${0.295}).  {Sources with \rfe\ $\lesssim 1.1$ are found up to redshift $z\approx 0.9$, while sources with \rfe$\gtrsim 1.1$ are detected up to redshift $z\approx 0.7$,  in part because they are much rarer}.  The redshift distributions of the samples separated through the \rfe\ values appear very similar.    

The right side of Fig. \ref{fig:z+r_fe} shows the \rfe\ distribution.  Most of our sample shows  \rfe$\lesssim$1.5  which classifies them as belonging to the A3 spectral type with $1.0 <$ \rfe\ $\le$ 1.5 \citep{sulenticetal02}.  The \rfe\ parameter shows a skewed  distribution  ($\overline{R_\mathrm{FeII}} \approx $ 1.36, $\mu({R_\mathrm{FeII}}) \approx $ 1.24) with a mode at \rfe $\approx$ 1.05. The lowest bin (0.9 $<$ \rfe$\le$1.0) includes borderline objects that can be considered bona-fide xA within the measurement uncertainties. 

  {We subdivided our 155 sources sample by identifying }radio-quiet, radio-intermediate, and radio-loud sources.  We measured the radio-loudness of our sample according to the radio to optical specific flux ratio $R_\mathrm{K} = f_{1.4GHz}/f_{5100\AA}$\ from \citet{2019ganci} (see Sect. \ref{RQ-RI-RL}; this criterion is a modified version of the classical Kellermann's ratio \cite{kellermannetal89} involving 6cm radio and B band specific fluxes). We adopt the following limits from \cite{zamfiretal08}: radio-quiet (RQ) $ \log R_\mathrm{K} < 1.0$; radio-intermediate (RI) $ 1 \le \log R_\mathrm{K} < 1.8$; radio-loud (RL, jetted, \citealt{padovani17,padovani17a}): $ \log R_\mathrm{K} \ge 1.8$.  
Of the 155 sources, $\approx$90\%  are radio-quiet while the remaining 10\%\  are radio-intermediate and radio-loud in almost equal percentages. 

We only found 1.4 GHz data for 20\% of the sample (28 sources). Of the 28 sources, 8 were classified as radio-loud and 8  were classified as radio-intermediate. { Additional information was collected on the RLs from literature, confirming that most of them are jetted. } The remaining 12 sources with  $\log R_\mathrm{K}   < 1.0$ or the sources lacking radio data were considered radio-quiet. This corresponds to the 90\% of the sample ($\sim$140 sources). 

{ The three samples, separated by radio-loudness, do not show significant differences in the distribution of \rfe\ } {($p_\mathrm{RI-RL}$=0.28, $p_\mathrm{RQ-RI}$=0.89), indicating that the \rfe\ parameter is not strongly connected to the radio-loudness in the domain of xA sources. }
  \begin{figure*}%
    \includegraphics[width=7.8cm]{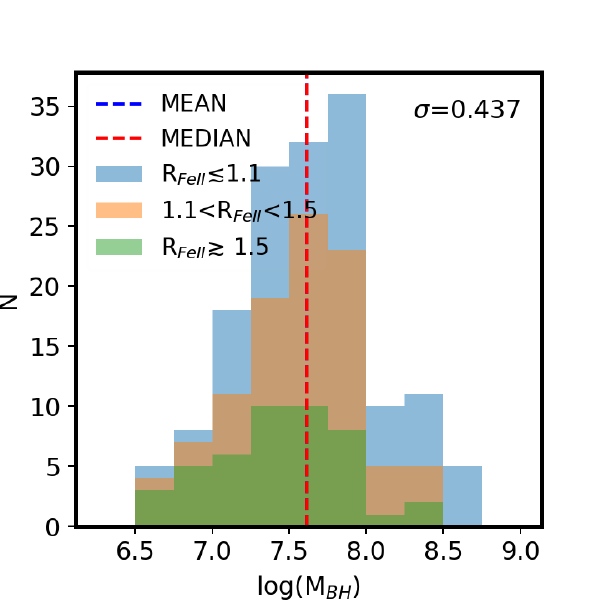}
    \includegraphics[width=7.8cm]{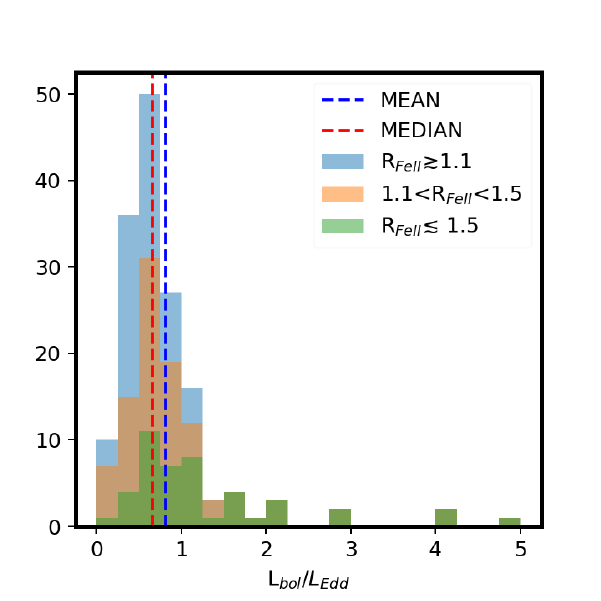}
    \caption{Stack distributions of the logarithmic black hole mass (left)  and the \lledd\ (right) on a linear scale.  The green distribution shows the extremely high accretors of the sample (\rfe $\gtrsim$ 1.5 ), this is the less populated subsample.  The orange distribution shows the intermediate high accretors (1.1< \rfe<1.5) and the blue distribution shows the high accretors (\rfe$\lesssim $ 1.1) of the sample. This last subsample corresponds to the most common of highly accreting quasars.\label{fig:Mass+L/Ledd}} 
\end{figure*}

\begin{figure*}
\includegraphics[trim={0cm 0cm 0cm 0cm},clip,width=\textwidth]{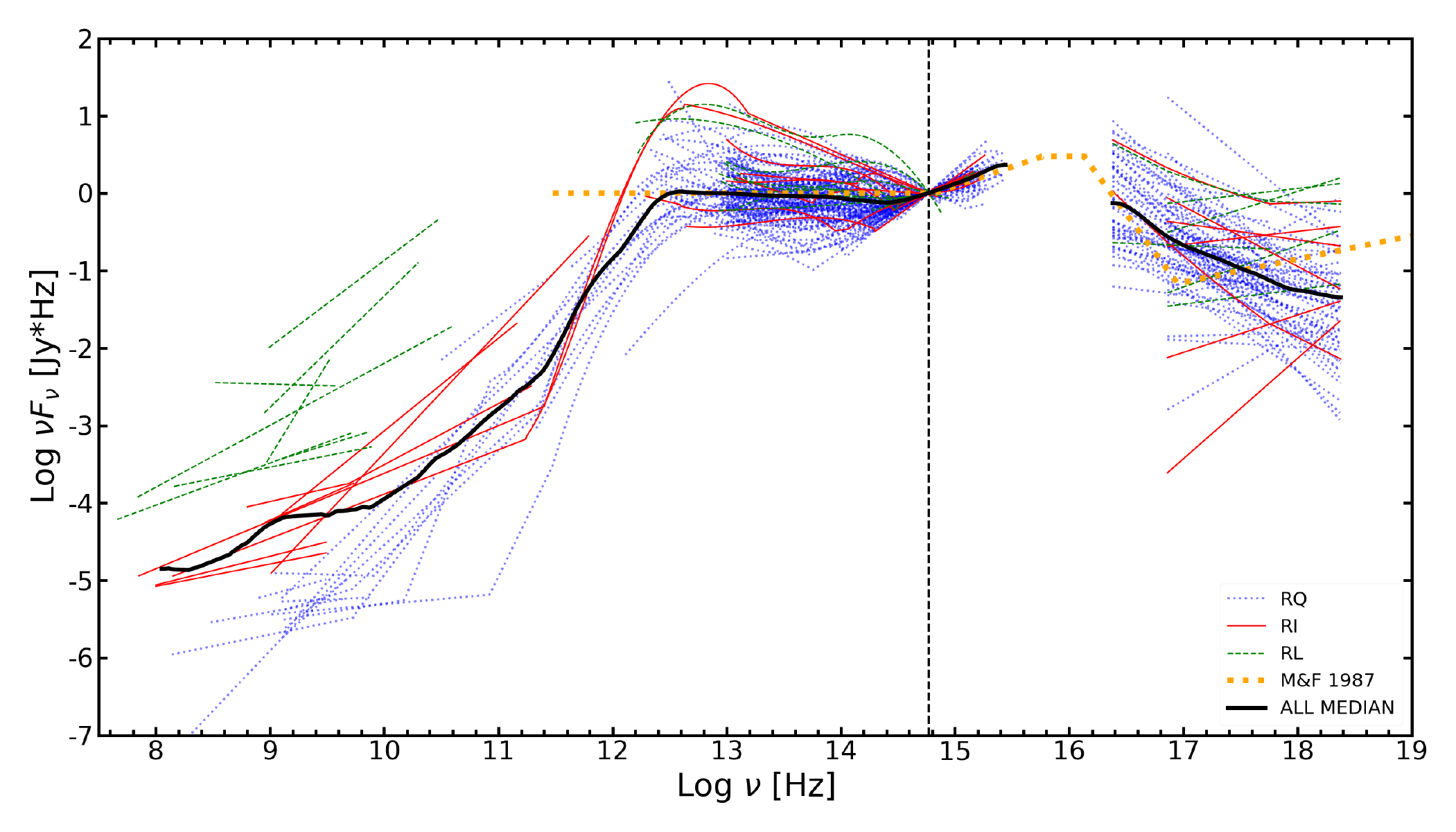} 
\caption{All data modeled. The dotted lines show the individual SEDs modeled for RQ (blue), RI ({red solid line}) and RL (green {dashed line}) sources to construct a characteristic SED for xA sources.{ The black solid line shows the median of all 155 sources.} The gold thick dotted line shows the \citetalias{1987mathewsferland} SED for classical quasars. Abscissa corresponds to the logarithm of the frequency in units of hertz and ordinate the logarithmic of specific flux times the frequency. \label{fig:alldata}}
\end{figure*}  

\section{Results}
\label{results}

\subsection{Estimation of accretion parameters}
\label{parameters}

We determined some basic physical parameters using the continuum 5100 \AA\ measurements from our SED model such as the bolometric luminosity ($L_\mathrm{bol}$), the central supermassive black hole mass ($M_\mathrm{BH}$), and the Eddington ratio (\lledd). Figure \ref{fig:Mass+L/Ledd} shows the stack distributions of \mbh\ and \lledd. The estimated parameter values are listed in Table \ref{tab:150_basic}, and details of the computations are reported below. 

\par{\textit{Bolometric luminosity:}} We used the conventional way to derive the luminosity, \mbox{$L = 4 \pi d_\mathrm{c}^{2}f$}  where $d_\mathrm{c}$ is the comoving distance  according to \citet[][ $\mathbf{\Omega_\mathrm{M} =0.3, \Omega_\Lambda =0.7}$]{2006sulentic}:

 \begin{equation}
 \centering
 d_\mathrm{c} \approx \frac{c}{H_0}\left[ 1.500(1-e^{-\frac{z}{6.309}}) + 0.996(1-e^{-\frac{z}{1.266}})\right],
\label{dcaprox}
\end{equation}

\noindent and $c$ is the speed of light, $H_0$ the Hubble constant assumed as 70 \kms\ {Mpc$^{-1}$} and $z$\ the  source redshift. The flux $f$ = $\lambda f_\lambda$\ is in the {quasar rest frame}.  {We apply  three values of bolometric correction computed as described in Sect. \ref{sed building}}.

\par\textit{Black hole mass and Eddington ratio:} To obtain an accurate black hole estimation we combined the \hb-based mass scaling relation proposed by \citet{2018mejia-restrepo} and  \cite[][{c.f. \citealt{donofrioetal24}}]{2019Du&Wang}: 
\begin{equation}
\begin{split}
    \log  {M_\mathrm{BH}}(\mathrm{H}\beta)  & \approx  5.220 + 0.45\log L_{5100,44} \\ &  +  0.83\log\mathrm{FWHM}(\mathrm{H}{\beta})  - 0.35 R_\mathrm{FeII}. 
\end{split}
\label{MassHB}
\end{equation} 

\noindent This relation requires $L_{5100,44}$  the continuum luminosity at 5100 \AA \ normalized by $10^{44}$ {erg s$^{-1}$}, the FWHM of \hb\ in units of km s$^{-1}$  and a correction using the \rfe\ parameter, an accretion rate indicator. 

Using the \mbh\ obtained from the previous step, we estimated the Eddington luminosity following:

\begin{equation}
L_\mathrm{Edd}=1.3 \mathsf{x} 10^{38} \left(\frac{M}{M_{\odot}}\right) \mathrm{erg\, s}^{-1}. \label{Le}
\end{equation}

  Given the Eddington  luminosity, we computed the Eddington ratio, \lledd, which is an indicator of the BH accretion rate at any given \mbh. 

Figure \ref{fig:Mass+L/Ledd} shows the stacked distributions of the previous parameters, the colors represent three ranges of \rfe\ parameter. 
 
The \mbh (left side) shows a symmetrical distribution with $\sigma=0.437$.  The \lledd\ distributions (right) are asymmetrical showing with a mode close to 1 for all values of \rfe\ parameters and the highest end of the distribution dominated by the { extremely high accretors which  have the smallest masses. } 

\subsection{SED building}
\label{sed building}

Our goal with a large sample is to create a median SED which characterizes the population of high accreting quasars. A risk to take when modeling broad-band SED based on observational data are that the composite result may not correspond to any individual source and its features may not describe a real source. However, the super-Eddington candidates of \citet{2022garnica} show spectra that are almost a carbon copy of each other.  Given the spectral similarity, we chose to create a semi-empirical SED based mostly on databases of observational data with the expectation that, in a photoionization scenario, the SED should too be rather similar with a small scatter. 

We address the modeling of the xA SED in three parts: (1) the radio to NUV SED; (2) the UV accretion disk SED and (3) the X-ray SED which will be described in the sections below.  Figure \ref{fig:alldata} shows all data normalized to 5100 \AA\ and compared to the \citetalias{1987mathewsferland} SED.

\subsubsection{The radio to NUV SED}
\label{sec:radioopt_sed}

The radio to NUV photometry was downloaded from NED. As mentioned, we corrected for Galactic extinction using $A_\mathrm{V}$\ retrieved from NED, and the \citet{1989Cardelli} extinction curve. 
We afterward applied a redshift correction to obtain the SED in each quasar rest frame, rescaling the frequencies by a factor $(1+z)$ and the $\nu f_\nu$ by a factor $(1+z)^2$. 

The modeling was carried out using simple linear regression and cubic splines. If the data distribution allowed us to, we interpolated a continuum from radio to NUV, but for most cases mostly in the submillimeter and FIR, we left gaps in modeling due to lack of data. Extrapolation was completely avoided.  To compare our data, we normalized each model at 5100 \AA, the continuum near \hb. The normalization constant was taken from our models per each source and is reported in Table \ref{tab:150_basic}.

We also include the composite optical-NUV spectrum from \citet{2013Marziani_composite}, previously described in Sec. \ref{sample:uv}, with a weight of 13. Since individual measurements are not available, and the A4 definition is based on a restricted range of \rfe\ between 1.5 and 2.0, we assume \rfe $\approx 1.75 \pm 0.25$\ and the average \hb\ FWHM as reported in \citet{2013Marziani_composite}.

As shown in Figure \ref{fig:alldata} the models tend to concentrate between the MIR and optical regimens which are the bands covered by most telescopes, leaving a gap from the submillimiter to FIR making us unable to fully connect with the radio regime which tends to show a turning point for RI sources. 

\subsubsection{The UV accretion disk SED: SEDs from sub- and super-Eddington accretion disk models}
\label{superedddisks}

The \citet{kubotadone18} model incorporates three main components:  (1) a hot, optically thin corona, located closest to the black hole,  where higher-energy photons are produced through Compton scattering of softer photons from the inner regions; (2) an inner, warm Comptonizing intermediate region that produces the soft X-ray excess often observed in AGN spectra. It acts as a transition zone where thermal emission from the outer disk is Compton upscattered by a warm, optically thick plasma.
(3) An outer standard disk: the outer regions of the accretion disk are modeled traditionally, where the disk emission is assumed to thermalize locally to a blackbody spectrum. These components are energetically tied together using the Novikov–Thorne emissivity profile to determine a size scale for the hard X-ray corona equal to the radius where the remaining accretion energy down to the black hole can power the observed X-ray emission. 

In other words,  this connection establishes a physical basis for the size and luminosity of the hard X-ray corona, posited to match the radius where remaining accretion energy can generate the observed X-ray emission. The model adjusts the hard X-ray spectral index based on the ratio of the hard X-ray luminosity to the intercepted seed photon luminosity from the disk.  Therefore, the model successfully reproduces the observed correlation between a steeper spectral index and increasing Eddington ratio     {\citep{wangetal13}, or at least a $\Gamma_\mathrm{soft} > 2$\  at high \lledd\ \citep[][Garc\'\i a-Soto, in preparation, although see \citealt{laurentietal22,2023Tortosa}]{shemmeretal08,risalitietal09,brightmanetal13,fanalietal13,trakhtenbrotetal17,huangetal20,liuetal21}. The model successfully reproduces the optical and UV \feii\ emission for a wide range of continuum and BLR properties in Type-1 AGNs \citep{panda_etal_2019_wc}. Interestingly, it predicts a decreasing amount of optical variability with an increasing Eddington ratio, which aligns with observations, although additional processes might be needed to explain the observed optical variability fully. 

A second model by \citet{kubotadone19} adjusts the standard disk emissivity to account for the super-Eddington flows characterized by slim disks. These slim disks manage to maintain surface luminosity at the local Eddington limit through radial advection, which alters the expected luminosity distribution, particularly under conditions of high accretion rates where $\epsilon(r) \propto r^{-2}$ rather than $ \epsilon(r) \propto r^{-3}$ as seen in standard sub-Eddington disks. Likewise to the first model, in \cite{kubotadone19}  the disk is subdivided into three regions: an outer standard disk where the emission follows the traditional disk law with $ \propto r^{-3}$, and two innermost warm and hot Comptonizing regions. The transition between the Comptonizing regions and the outer disk is determined by the local disk flux reaching the Eddington limit, where the emissivity becomes  $ \propto r^{-2}$. The treatment is meant to include the effect of advection in the super-critical regime albeit with a simplified disk geometry, and neglecting the presence of winds that are especially powerful in super-Eddington quasars \citep[][]{sulenticetal07,marzianietal16a,sulenticetal17,bischettietal17,vietrietal18,2020marinello,rodriguez-ardila_etal_2024}.  Powerful winds can significantly affect the mass accretion rate onto the black hole. 

The UV  {domain right} above the Lyman limit is mostly hidden by Galactic absorption.  Figure \ref{fig:alldata} shows this deficiency in the observational SED by the gap of data from NUV to the soft X-ray regime. We supplemented the observations with SEDs computed by models of accretion disk emission. Specifically, we adjust the model by \citet{kubotadone18}  to fit the observed SEDs, for \mbh\ $\approx 6.5 \cdot 10^7 M_\odot$\ { in the case of the median composites}. Figure \ref{fig:seddisks} shows the SEDs corresponding to the median and the upper-lower semi-interquartile ranges {for the composites obtained normalizing the individual SEDs at 5100 \AA}. The parameters of the \citet{kubotadone18} models are somewhat different (see Table \ref{tab:my_label}) but align well with the median black hole mass and the range of masses and Eddington ratios of the sample.{   The bolometric corrections from 5100 \AA\ are 8.84, 14.34, 14.72  for the $q1, q2, q3$, respectively.}  In the case of the third quartile SED, we also considered a more proper super-Eddington disk model following \citet{kubotadone19}. 
Regarding the effect on the SED, the right panel of 
 Fig. \ref{fig:seddisks} shows an enlargement centered on the UV spectral ranges for comparison between the sub-Eddington and super-Eddington. In this case, a model is shown with $\dot{m} =10$ following \citet{kubotadone19}, for the same values of the \mbh\ associated with the three quartiles (dashed lines). For the median and $q3$, there is a slight softening of the SED peak, from $\log h\nu_\mathrm{Peak} \approx $ 1.43 to 1.35 ($q2$) and $\log h\nu_\mathrm{Peak} \approx $ 1.62 to 1.55 ($q3$). The implications on the ionizing continuum are minor, and the photoionization predictions for the main broad-line diagnostic intensity ratios customarily employed for estimating metallicity should not be affected.

\begin{table}
    \centering
    \begin{tabular}{cccccccc}\hline\hline
               & \multicolumn{3}{c}{$f$5100 normalized} && \multicolumn{3}{c}{luminosity scaled}\\ \cline{2-4}\cline{6-8}
      Quartile & $M_\mathrm{BH}$ & \multicolumn{2}{c}{$\log\dot{m}$}  &&    $M_\mathrm{BH}$ & \multicolumn{1}{c}{$\log\dot{m}$}    \\ \cline{3-4}    & [$M_\odot$] & &  &   & [$M_\odot$] & & \\  \hline 
       q2   &       $6.31 \cdot 10^7$          & 0.0     & 1.0   && $6.31 \cdot 10^7$ & 0.0 &  \\ 
       q1   &       $10^8$         & -0.5 &     \ldots   && $10^8$  & -0.5\\
       q3    &       $10^7$   &      0.0    & 1.0     && $6.31 \cdot 10^8$ & 0.0      \\
       \hline
    \end{tabular}
    \caption{UV disk model parameters from \citet{kubotadone18} used to supplement the observed SED in the unobservable FUV. }
    \label{tab:my_label}
\end{table}

\begin{figure*}
\centering   
\includegraphics[scale=0.5, trim={0 6.5cm 0 0},clip]{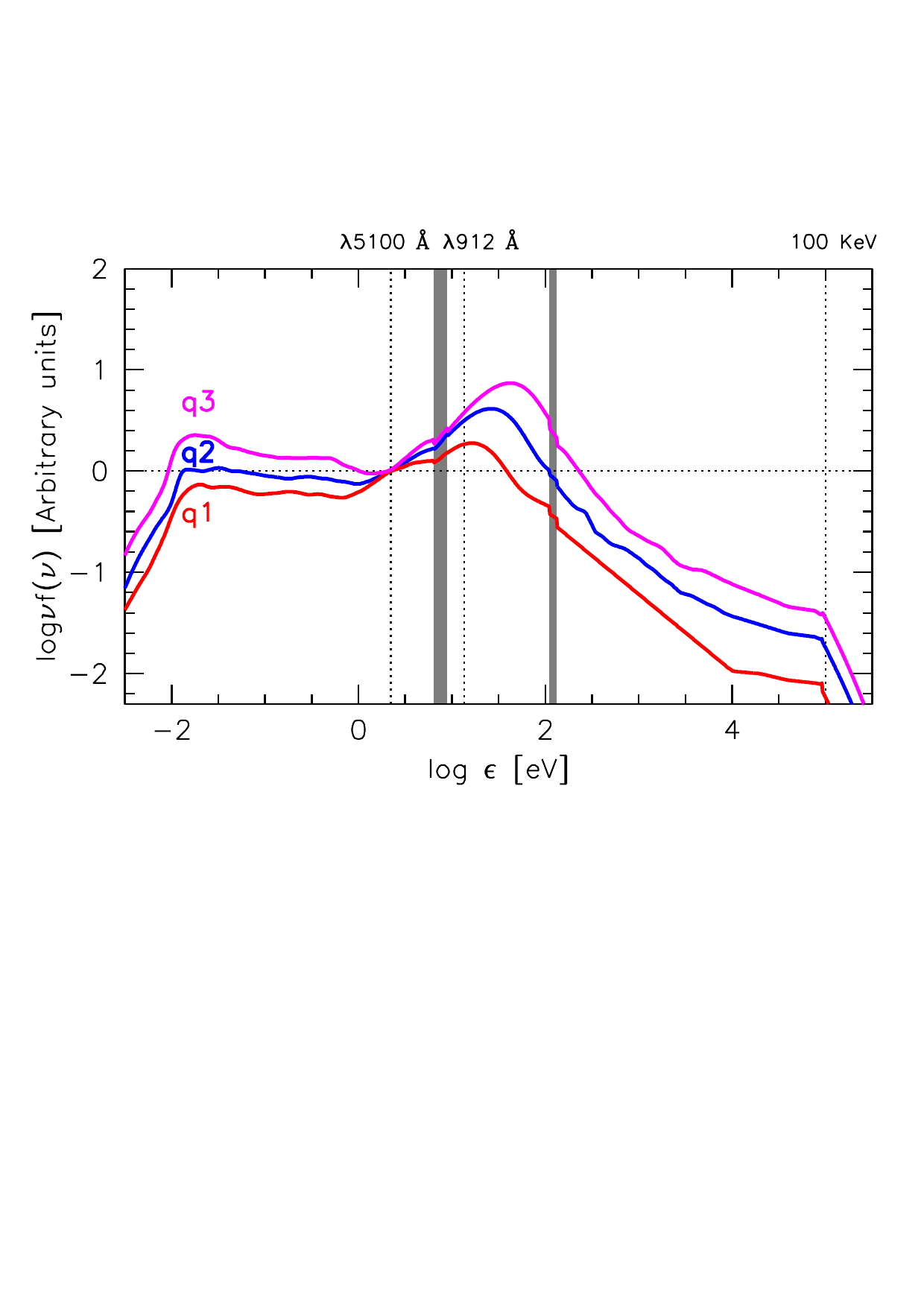}  
\hspace{0.0cm}
\vspace{-0cm}
\includegraphics[scale=0.35]{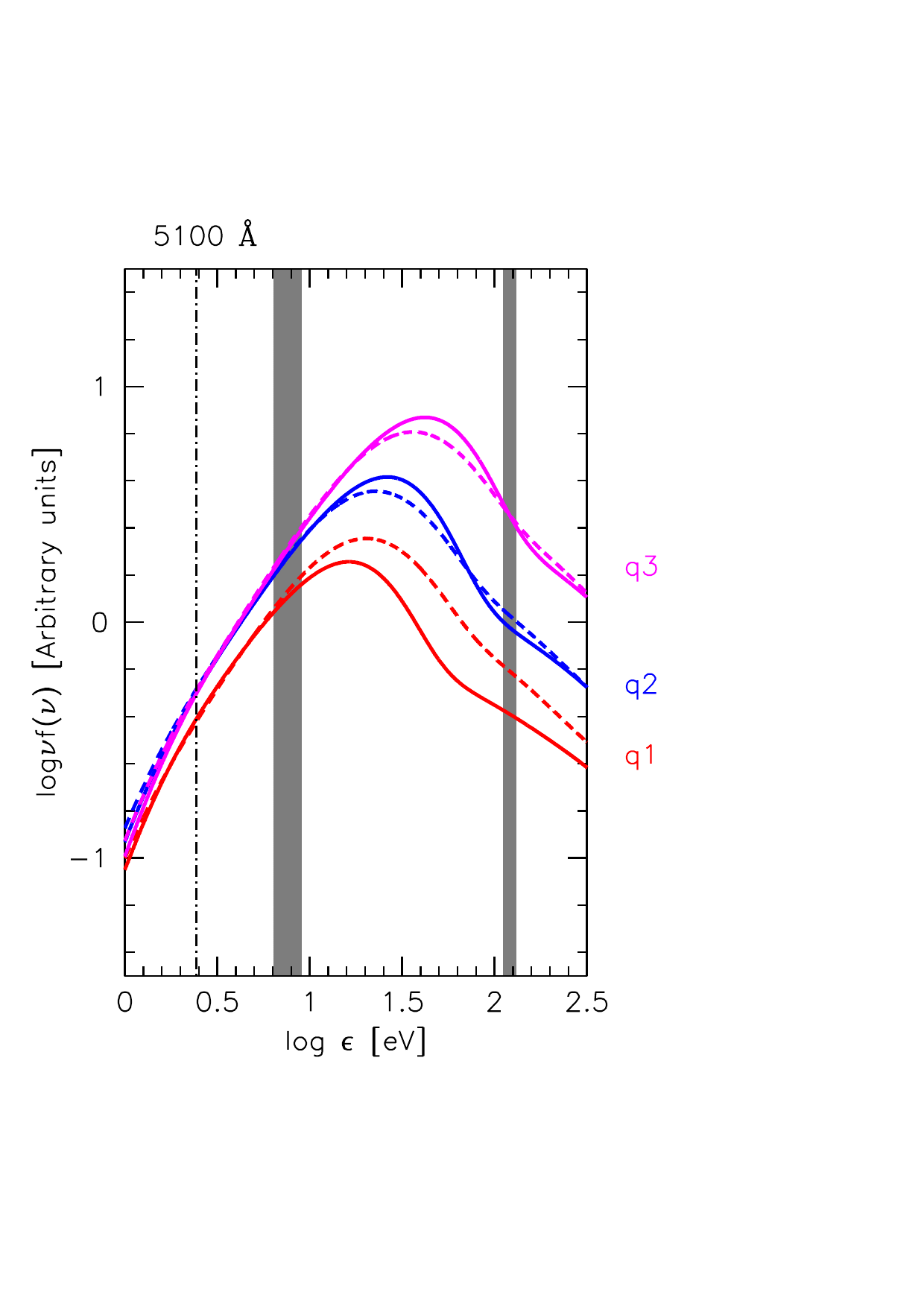} 
\\
\caption{Left: Spectral energy distributions corresponding to the median SED (second quartile, q2, thick black line), and the first and third quartile (q1 and q3, red and magenta lines). Abscissa is the logarithm of energy in units eV; ordinate is flux $\nu f(\nu)$ in arbitrary units normalized at 5100 \AA.  Disk model SEDs were added in the unobservable FUV following \citet{kubotadone18}. Right: the panel shows a zoomed-in view of the unobservable ultraviolet (UV) region for a comparison with a super-Eddington model. The meaning of color is the same and black hole masses are the same as in the left panel. The dashed line represents a solution with an accretion rate of $\log \dot{m} = 1$ following \citet{kubotadone19}.   \label{fig:seddisks} }

\end{figure*}

\subsubsection{The X-ray SED}
\label{x-raysed}

The various X-ray spectral ranges are usually represented by power-law spectra,  $f(\nu) = f_0 ( \frac{\nu}{\nu_0} )^{-\alpha}$, where $\alpha$ is the slope of the power law spectra and is negative by convention. This slope is also related to the photon index by $\Gamma = \alpha +1$. The following expression yields the specific flux   in units of [erg s$^{-1}$ cm$^{-2}$ Hz$^{-1}$] for each source at the lower frequency end of the band ($\nu_0$), given the total flux $F_\mathrm{T}$, the photon index $\Gamma$ and the frequency range $\nu_0 - \nu_1$:

\begin{equation}
 {  F_{\nu_0}=  \frac{F_T} {[ \left( \frac{\nu_1}{\nu_0}\right)^{-\Gamma +2} - 1 ]} \frac{-\Gamma +2}{ \nu_0} } . 
\end{equation}

To avoid the H I absorptions in the X-ray region, we selected our X-ray data from catalogs reporting the results of fits to the X-ray spectrum that had previously performed the corrections for each source (see Sec. \ref{sample:x-ray}).  The median $\Gamma_\mathrm{soft}$ resulting from the reanalyis of the ROSAT survey \citep{2016Boller}  is $ \approx  2.87^{+0.34}_{-0.45}$. This value of $\Gamma_\mathrm{soft}$ is consistent with the values found in early studies that yielded the formulation of the so-called 4D Eigenvector 1 space \citep{1996boller,sulenticetal00a,sulenticetal00c}. The steep SED in the soft X-ray domain has been associated with a prominent soft X-ray excess, in turn, associated with high accretion rate \citep[e.g.,][]{1996boller,wangnetzer03,grupe04,benschetal15}.

In the hard X-ray domain, we ended up with X-ray data from 2 keV to 100 keV from the  {\it Swift}/BAT, \textit{Integral}, and \textit{NuStar} observatories.  In the spectral range between 2 and 10 keV, the {\it Swift} data \citep{2020Evans} yield $\Gamma_\mathrm{hard} \approx $ 
2.53$^{+0.34}_{-0.39}$}.  XMM spectra confirm that super-Eddington candidates do show $\Gamma > 2$ in the range 2-10 keV \citep{laurentietal22,2023Tortosa}, and $\Gamma$\ might be correlated with Eddington ratio \citep{2023Tortosa}. \citet{laurentietal22} found that a significant fraction of an xA sample \citep{2014marziani}  show weak X-ray emission, with a large optical to X-ray ratio.        Table \ref{tab:nustar} reports individual,  median, and average $\Gamma$ for several ranges that are used to estimate $\Gamma$ above 10 keV.  The median photon index above 10 keV and up to 20 keV has been estimated to be around $\Gamma \approx 2.3$, with values reaching $\approx$2.5 for the most extreme super-Eddington sources, in agreement with \citet{wangetal13}. 
{Even greater care should be taken in considering the extension beyond 20 keV. We carried out a search on the High Energy Astrophysics Science Archive Research Center (HEASARC) archive for data in the range 3--80 keV obtained with \textit{NuStar} \citep{harrisonetal13}, and references to published fits. 
 
 Results for different ranges are reported in Table \ref{tab:nustar}. Sources marked with asterisks are RL or $\gamma$-ray emitters and were excluded from averages.   $\gamma$-ray emission from a beamed core may affect the high-energy end of the SED, as found in $\gamma$-ray emitting NLSy1s, producing a net flattening and hence enhancing the probability of the detection \citep{paliya19,dammando19,dammando20,foschinietal21}. 
 The data in Table \ref{tab:nustar} suggest a slight flattening between 40 and 80 keV.  The flattening could be associated with a weak reflection component \citep{haardtmaraschi91,haardtmaraschi93,nandrapounds94}, whose prominence could be related to the $\Gamma$. However, the origin of the reflection is not clear, as the accretion geometry is expected to be basically different for super-Eddington and sub-Eddington AGN \citep{karaetal16,zhangetal24,danekhar24}. In the super-Eddington case, the X-ray reflection might be due to a Compton thick outflow encircling the funnel of an optically thick, geometrically thick accretion disk. At any rate,  a steep or faint X-ray SED implies the absence of the Compton reflection hump in the form assumed in the \citetalias{1987mathewsferland} SED.

\begin{figure*}
\centering 
\includegraphics[scale=0.5, trim={0 6.5cm 0 0},clip]{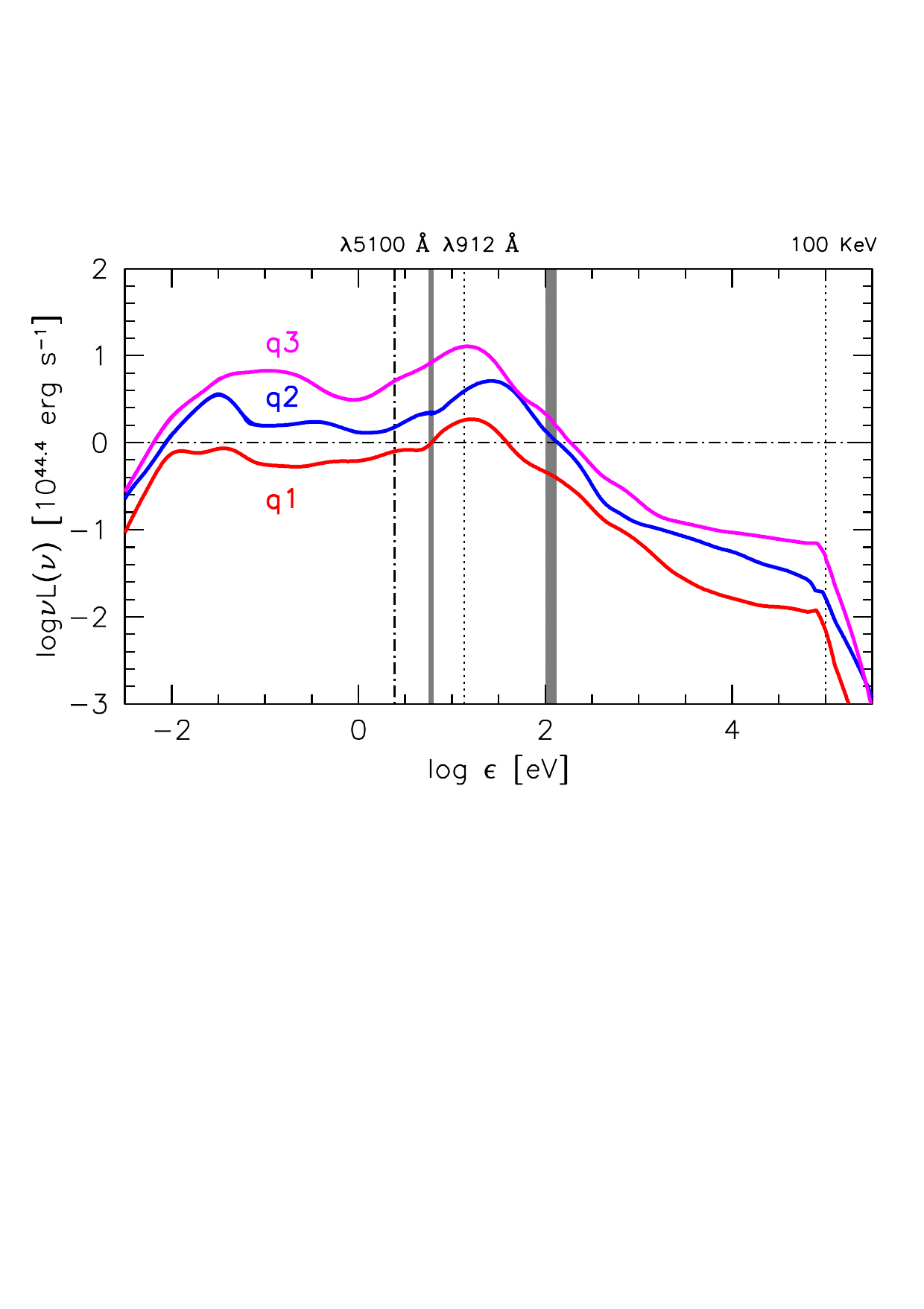}  
\hspace{0.0cm}
\vspace{-0cm}
\caption{ Luminosity-scaled SEDs (second quartile, q2, thick blue line), and the first and third quartile (q1 and q3, red and magenta lines). Abscissa is the logarithm of energy in units eV; ordinate is optical luminosity at 5100 \AA\ normalized by 10$^{44.4}$ erg s$^{-1}$.  Disk model SEDs were added in the unobservable FUV following \citet{kubotadone18}.  \label{fig:sedlum} }

\end{figure*}

 Observations of NLSy1s with {\em Integral} include 3 sources that are xA with \rfe$\gtrsim$1: SWIFT J2127.4+5654 ($\Gamma \approx 2.50^{+0.24}_{-0.24}$), IGR J16385\textemdash2057 ($\Gamma \approx  3.07^{+1.00}_{ -0.71}$),  IGR J14552\textemdash5133  ($\Gamma \approx 1.90^{+0.44}_{ -0.46}$, between 2 and 100 keV \citep{2011panessa}, a slope that is consistent with the one of \citet{2017ricci} up to $\approx 200$ \ keV.  These results are confirmed by recent observations with NuStar that yield $\Gamma$\ values consistently $\gtrsim 2$ for I Zw 1 ($\Gamma \approx  2.15^{+0.08 }_{-0.06}$, up to 40 keV, \citealt{wilkinsetal21}),  PDS 456 ($\Gamma \approx  2.29^{+0.04}_{-0.03}$, up to 30 keV, \citealt{nardinietal15}).  We therefore assigned $\Gamma \approx 2.5$\ for the spectral slope between 80 and 100 keV. Beyond 100 keV, the SED decreases exponentially, with a turnover at 100 keV \citep{1987mathewsferland,tortosaetal22,2021Akylas}.

\subsection{Luminosity-scaled SED}  

{The SED obtained from the normalization at 5100 \AA\ is helpful to analyze   features toward the IR and toward the UV with respect to the optical emission, partially suppressing luminosity effects.  In order to build a SED preserving information on  luminosity, we computed the median and the quartiles by   first normalizing the SED by the flux at 5100 \AA\ and multiplying by  optical luminosity at 5100 \AA.\ Since there are few  individual SEDs (14) that  were complete from radio to UV ,  we used a bootstrapping method to include the SEDs for which the information was less complete, averaging over several hundreds bootstrap replications. This approach was adopted  to ensure that the resultant SED was smooth enough to be representative of continuum emission. }

{Fig. \ref{fig:sedlum} shows the SED obtained giving a weight proportional to the optical luminosity of each individual object. The interpolation between the optical and X-ray range has been modeled as done for the SED normalized at 5100 \AA, namely choosing a disk model in agreement within the energy ranges shown by the grey strips in Fig. \ref{fig:sedlum}. The accretion parameters are reported in Table \ref{tab:my_label}. The $\log \dot{m} = 0$\ and \mbh $= 10^8$ M$_\odot$ for the median are consistent with the SED normalized at 5100 \AA. The first quartile also involves the same parameters, with sub-Eddington accretion, \mbh $= 10^8$ M$_\odot$, $\log \dot{m} = -0.5$.    Larger luminosity sources, represented by the third quartile and  necessarily associated with larger \mbh, since \lledd\ shows small scatter) are well fit by a disk model assuming $ \approx5\cdot 10^8$ M$_\odot$, still within the \mbh\ covered by our sample, albeit close to the largest masses in the distribution of Fig. \ref{fig:Mass+L/Ledd}.  The larger \mbh\ yields a softening of the spectrum, with a displacement of the peak emission toward the Lyman limit (magenta line of Fig. \ref{fig:sedlum}). The bolometric corrections are $\approx 11.2, 15.7, 11.0$\ for the $q1, q2, q3$\ in this order. A second important aspect is that the  IR excess seen in the SED normalized at 5100 \AA\  is confirmed. Intriguingly, the median composite shows an excess that is restricted toward the FIR, while the $q3$\ SED includes a broader excess encompassing emission at around $\approx 20 \mu$m. These features are observed in luminous AGN \citep{netzeretal16}, and might be connected to powerful emission from the molecular torus \citep{esparza-arrendondoetal21} and enhanced star formation.  } {The same excess is visible also in the 3rd quartile of the SED normalized at 5100 \AA, and to some extent also in the median. Nonetheless, the excess is definitely stronger in the luminosity-scaled SED, implying that it is associated with the most luminous sources in the sample. }


\section{The radio variance among xA sources}
\label{RQ-RI-RL}

When we first compared all the individual SEDs we modeled for this work (see Figure \ref{fig:alldata}), we observed an unexpected variance in the radio frequencies. To generate a more representative SED for xA sources, we built SEDs for the for the RQ, RI and RL classes. The measurements were taken from the  individual SED models described in Sec. \ref{sec:radioopt_sed}.

\subsection{The SED of radio-quiet xAs}
\begin{figure}
\centering 
\includegraphics[trim={0cm 0cm 0cm 0cm},clip,width=0.46\textwidth]{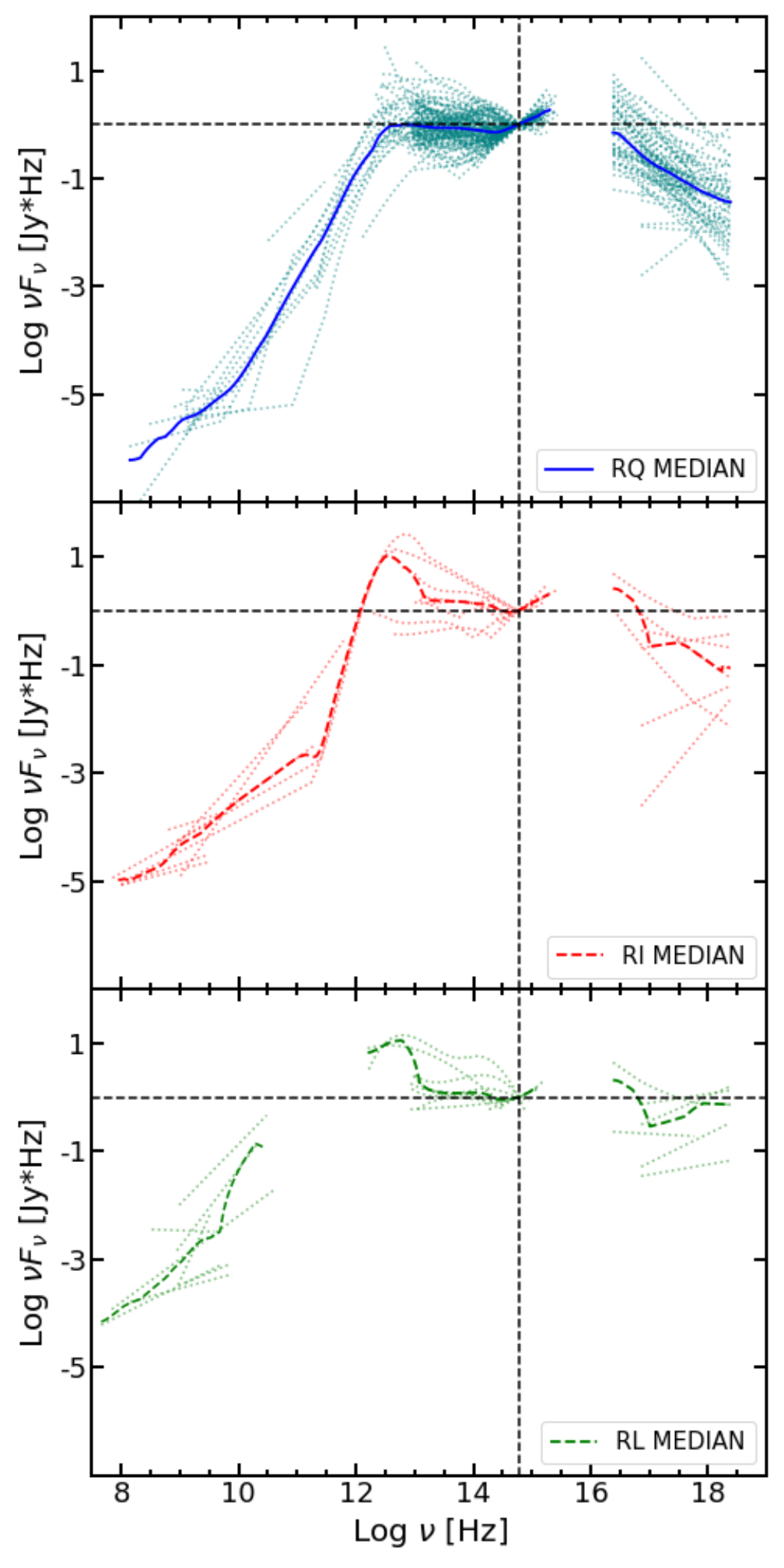} \caption{{Comparison between the median RQ, RI and RL SEDs. Top: blue solid line shows the median from 139 radio-quiet xA sources, blue dotted lines show the 139 individual SEDs (90\% of the sample). Center and bottom: red and green dashed lines shows the median for radio-intermediate and radio-loud sources, respectively. The RQ and RI medians were constructed with only \%5 of the sample (8 sources) and shown as dashed lines in each panel. Given the small number of sources involved, the RI and RL SEDs are shown only for illustrating the difference with respect to the RQ template in the FIR and radio domains.  Dashed vertical lines in each panel identify  the normalization frequency (5100 \AA, abscissa) and $\nu F_\nu$=1 (ordinate). Abscissa corresponds to the logarithm of the frequency in units of hertz and ordinates the logarithmic of specific flux times the frequency.\label{fig:RQ}}}

\end{figure}  

Since 90\% of our sample is radio-quiet, our most representative xA SED is the radio-quiet one. The blue solid line in Figure \ref{fig:RQ} shows the median from all radio-quiet models ($N = $139), the grey shadow encompasses the first and third quartile of the sample and is shown as a representation of the data diversity. The SED has been normalized to the optical flux $\lambda f_\lambda (5100)$ \AA.  From radio to FIR, the SED shows an upward slope. In the infrared region, the SED shows a flat distribution with no trace of any FIR excess possibly associated with star formation. The optical to NUV region shows us an upward steep slope which leads to higher fluxes to the ones of the \citetalias{1987mathewsferland} SED.

\subsection{The SED of radio-intermediate xAs}

Only 5\% of the sample was identified as radio-intermediate. From these 8 sources, only MRK 766 and PKS 0558-504 had data from the  X-ray catalogs above 10 keV but unfortunately, the data wasn't consistent enough for taking a median. Figure \ref{fig:RQ} (top panel) shows the individual SEDs of all radio-intermediate xA sources, {and a composite.} 
The tentative SED for the RI class reveals an FIR excess that is suppressed in the median of RQ (Fig. \ref{fig:RQ} mid panel). This FIR excess is not entirely unexpected if the enhancement in radio emission is mostly due to star formation \citep{condonetal13,caccianigaetal15,2019ganci}. { We therefore associate the RI with the RQ quasars of our sample, keeping them separated from the RL ones that meet a very restrictive criterion on $R_\mathrm{K}$. } 

\subsection{The SED of radio-loud xAs}

Approximately five percent of the sample was identified as radio-loud ($N = 8$).  We couldn't derive a { reliable} median SED from this subsample due to the lack of data { and the diversity of the few SEDs available (Figure \ref{fig:RQ} bottom panel)}.

\begin{table*}[h!]
    \centering

    \caption{Observational and derived parameters for the 150 xA sources used to construct the composite SEDs}
    \label{tab:150_basic}
    \begin{tabular}{c|c|c|c|c|c|c|c|c|c}
\hline\tabcolsep=0pt
No. 	&	NAME 	&	 $z$ 	&	 \rfe 	&	 FWHM \hb\ 	&	 $\lambda f_\lambda$ 5100 \AA\ 	&	$\log$ \mbh\ 	&	 \lledd\ 	&	Radio 	&	 DATABASE \\
&  & & & [\kms] & [Jy Hz]\ & [$M_\odot$] \\
(1) & (2) &(3) & (4) & (5) & (6) & (7) & (8) & (9) &(10) \\
\hline
1	&	I Zw 1	&	0.061	&	1.3	&	1092	&	12.57	&	7.5	&	0.994	&	RQ	&	\cite{2003marziani}	\\
2	&	IRAS 07598+6508 	&	0.148	&	1.21	&	4845	&	12.64	&	8.44	&	0.447	&	RQ	&	\cite{2003marziani}	\\
3	&	Ton 28	&	0.329	&	0.92	&	1759	&	12.17	&	8.26	&	1.023	&	RQ	&	\cite{2003marziani}	\\
4	&	Mark 766 	&	0.013	&	1.14	&	1981	&	12.18	&	7	&	0.099	&	RI	&	\cite{2003marziani}	\\
5	&	PG 1244+026 	&	0.048	&	1.06	&	833	&	11.87	&	7.09	&	0.547	&	RQ	&	\cite{2003marziani}	\\
6	&	PG 1259+593 	&	0.476	&	1.07	&	3388	&	12.44	&	8.69	&	1.347	&	RQ	&	\cite{2003marziani}	\\
7	&	Mark 662 	&	0.055	&	1.18	&	6209	&	12.35	&	8.04	&	0.086	&	RQ	&	\cite{2003marziani}	\\
8	&	PG 1404+226 	&	0.098	&	0.93	&	907	&	11.94	&	7.47	&	0.651	&	RQ	&	\cite{2003marziani}	\\
9	&	PG 1415+451 	&	0.114	&	1.03	&	2555	&	11.94	&	7.86	&	0.344	&	RQ	&	\cite{2003marziani}	\\
10	&	Mark 478 	&	0.079	&	0.93	&	1703	&	12.46	&	7.85	&	0.588	&	RQ	&	\cite{2003marziani}	\\
11	&	PG 1444+407 	&	0.268	&	1.2	&	2944	&	12.2	&	8.28	&	0.703	&	RQ	&	\cite{2003marziani}	\\
12	&	PG 1552+085 	&	0.119	&	1.4	&	1716	&	12.1	&	7.68	&	0.830	&	RQ	&	\cite{2003marziani}	\\
13	&	PG 1700+518 	&	0.289	&	0.98	&	2148	&	12.5	&	8.41	&	1.207	&	RQ	&	\cite{2003marziani}	\\
14	&	PHL 1092 	&	0.396	&	1.81	&	1850	&	11.77	&	7.85	&	2.448	&	RQ	&	\cite{2009miniutti} \cite{2020marinello}	\\
15	&	SDSS J071933.35+403253.0 	&	0.515	&	1	&	3552	&	11.24	&	8.22	&	0.289	&	RQ	&	\cite{2019ganci}	\\
16	&	SDSS J080037.62+461257.9 	&	0.239	&	1	&	2018	&	11.57	&	7.9	&	0.561	&	RQ	&	\cite{2019ganci}	\\
17	&	SDSS J081929.48+522345.2 	&	0.624	&	2	&	2382	&	10.86	&	7.62	&	1.130	&	RQ	&	\cite{2019ganci}	\\
18	&	SDSS J083558.42+261444.4 	&	0.789	&	1.5	&	2532	&	10.7	&	7.81	&	0.717	&	RQ	&	\cite{2019ganci}	\\
19	&	SDSS J094248.08+112934.3 	&	0.566	&	1	&	1492	&	10.62	&	7.66	&	0.500	&	RQ	&	\cite{2019ganci}	\\
20	&	SDSS J095150.49-025545.5 	&	0.903	&	1	&	2582	&	11.2	&	8.26	&	0.587	&	RQ	&	\cite{2019ganci}	\\
21	&	SDSS J095633.93+562216.0 	&	0.895	&	1.5	&	3150	&	11.51	&	8.3	&	1.097	&	RQ	&	\cite{2019ganci}	\\
22	&	SDSS J101952.59+073050.8 	&	0.524	&	1.5	&	1595	&	10.8	&	7.56	&	0.827	&	RQ	&	\cite{2019ganci}	\\
23	&	SDSS J102818.15+535113.6 	&	0.513	&	1.5	&	1718	&	10.97	&	7.66	&	0.944	&	RQ	&	\cite{2019ganci}	\\
24	&	SDSS J103346.39+233220.0 	&	0.47	&	1	&	1674	&	11.16	&	7.88	&	0.760	&	RQ	&	\cite{2019ganci}	\\
25	&	SDSS J104011.18+452125.9 	&	0.548	&	1	&	2855	&	11.54	&	8.29	&	0.532	&	RQ	&	\cite{2019ganci}	\\
26	&	SDSS J105525.26+154433.8 	&	0.819	&	1	&	2629	&	10.54	&	7.94	&	0.390	&	RQ	&	\cite{2019ganci}	\\
27	&	SDSS J105957.21+274150.7 	&	0.243	&	1	&	1290	&	10.92	&	7.45	&	0.363	&	RQ	&	\cite{2019ganci}	\\
28	&	SDSS J114339.53+205921.1 	&	0.752	&	1.5	&	1903	&	11.08	&	7.87	&	1.406	&	RQ	&	\cite{2019ganci}	\\
29	&	SDSS J114915.30+393325.4 	&	0.629	&	1	&	1693	&	10.77	&	7.8	&	0.599	&	RQ	&	\cite{2019ganci}	\\
30	&	SDSS J120910.61+561109.2 	&	0.454	&	1.4	&	1644	&	10.68	&	7.51	&	0.560	&	RQ	&	\cite{2019ganci}	\\
31	&	SDSS J123640.35+563021.4 	&	0.698	&	2	&	2467	&	10.84	&	7.65	&	1.177	&	RQ	&	\cite{2019ganci}	\\
32	&	SDSS J124511.25+335610.1 	&	0.711	&	1	&	3447	&	11.99	&	8.65	&	1.029	&	RQ	&	\cite{2019ganci}	\\
33	&	SDSS J130631.63+435100.4 	&	0.755	&	1	&	2381	&	11.83	&	8.46	&	1.192	&	RQ	&	\cite{2019ganci}	\\
34	&	SDSS J132146.53+265150.1 	&	0.846	&	1	&	2770	&	11.84	&	8.56	&	1.181	&	RQ	&	\cite{2019ganci}	\\
35	&	SDSS J142549.19+394655.0 	&	0.505	&	1	&	2175	&	10.7	&	7.79	&	0.362	&	RL	&	\cite{2019ganci}	\\
36	&	SDSS J163345.22+512748.4 	&	0.629	&	1	&	3632	&	11.36	&	8.35	&	0.397	&	RQ	&	\cite{2019ganci}	\\
37	&	SDSS J170300.48+410835.8 	&	0.894	&	1	&	3138	&	10.94	&	8.21	&	0.355	&	RQ	&	\cite{2019ganci}	\\
38	&	SDSS J171749.62+253908.7 	&	0.797	&	1	&	2028	&	10.9	&	8	&	0.440	&	RQ	&	\cite{2019ganci}	\\
39	&	SDSS J000410.80-104527.1 	&	0.24	&	1.54	&	854	&	11.28	&	7.26	&	1.228	&	RQ	&	\cite{2018negrete} 	\\
40	&	SDSS J000834.71+003156.1 	&	0.264	&	1.2	&	1330	&	11.51	&	7.68	&	0.955	&	RQ	&	\cite{2018negrete} 	\\
41	&	SDSS J002141.01+003841.7 	&	0.311	&	1.74	&	4055	&	10.79	&	7.63	&	0.275	&	RQ	&	\cite{2018negrete} 	\\
42	&	SDSS J004052.14+000057.2 	&	0.406	&	1.46	&	1002	&	11.41	&	7.6	&	2.004	&	RQ	&	\cite{2018negrete} 	\\
43	&	SDSS J011110.04-101631.8 	&	0.179	&	1.53	&	3208	&	11.68	&	7.82	&	0.499	&	RQ	&	\cite{2018negrete} 	\\
44	&	SDSS J020028.37-093859.0 	&	0.321	&	1.26	&	1210	&	10.89	&	7.42	&	0.600	&	RQ	&	\cite{2018negrete} 	\\
45	&	SDSS J021707.87-084743.5 	&	0.291	&	1.31	&	2151	&	11.52	&	7.86	&	0.780	&	RQ	&	\cite{2018negrete} 	\\
46	&	SDSS J021859.86+002855.8 	&	0.352	&	1.29	&	3030	&	11.26	&	7.94	&	0.499	&	RI	&	\cite{2018negrete} 	\\
47	&	SDSS J025627.76-080134.9 	&	0.476	&	1.66	&	1262	&	11.48	&	7.69	&	2.456	&	RQ	&	\cite{2018negrete} 	\\
48	&	SDSS J030000.00-080356.9 	&	0.564	&	2.44	&	1762	&	12	&	7.83	&	7.968	&	RQ	&	\cite{2018negrete} 	\\
49	&	SDSS J032255.49+001859.8 	&	0.384	&	1.67	&	1419	&	11.38	&	7.62	&	1.627	&	RQ	&	\cite{2018negrete} 	\\
50	&	SDSS J033901.67-055139.9 	&	0.224	&	1.6	&	1475	&	11.54	&	7.53	&	1.058	&	RQ	&	\cite{2018negrete} 	\\
51	&	SDSS J044428.77+122111.7 	&	0.09	&	1.47	&	1260	&	12.31	&	7.53	&	1.111	&	RQ	&	\cite{2018negrete} 	\\
52	&	SDSS J073955.14+331236.8 	&	0.321	&	1.24	&	1029	&	10.75	&	7.31	&	0.567	&	RQ	&	\cite{2018negrete} 	\\
53	&	SDSS J074151.15+423443.6 	&	0.325	&	1.41	&	2117	&	10.8	&	7.53	&	0.383	&	RQ	&	\cite{2018negrete} 	\\
54	&	SDSS J074644.79+294059.0 	&	0.292	&	1.36	&	1305	&	11.4	&	7.61	&	1.052	&	RQ	&	\cite{2018negrete} 	\\
55	&	SDSS J075005.28+292944.3 	&	0.328	&	3.07	&	1366	&	11.06	&	6.91	&	4.912	&	RQ	&	\cite{2018negrete} 	\\
56	&	SDSS J075141.56+353914.8 	&	0.306	&	1.67	&	2800	&	10.65	&	7.45	&	0.292	&	RQ	&	\cite{2018negrete} 	\\
57	&	SDSS J080131.58+354436.4 	&	0.179	&	1.33	&	727	&	11.31	&	7.19	&	1.516	&	RQ	&	\cite{2018negrete} 	\\
58	&	SDSS J081636.18+294135.7 	&	0.26	&	1.56	&	1900	&	11.1	&	7.5	&	0.556	&	RQ	&	\cite{2018negrete} 	\\
59	&	SDSS J082140.74+371518.7 	&	0.315	&	1.52	&	5042	&	10.77	&	7.78	&	0.190	&	RQ	&	\cite{2018negrete} 	\\
60	&	SDSS J082811.55+123359.3 	&	0.362	&	1.49	&	2475	&	11.07	&	7.72	&	0.561	&	RQ	&	\cite{2018negrete} 	\\

\hline
    \end{tabular}
    \end{table*}

\setcounter{table}{2}
\begin{table*}[h!]
    \centering
    \caption{Cont.}
    \begin{tabular}{c|c|c|c|c|c|c|c|c|c}
\hline\tabcolsep=0pt
No. 	&	NAME 	&	 $z$ 	&	 \rfe 	&	 FWHM \hb\ 	&	 $\lambda f_\lambda$ 5100 \AA\ 	&	$\log$ \mbh\ 	&	 \lledd\ 	&	Radio 	&	 DATABASE \\
&  & & & [\kms] & [Jy Hz]\ & [$M_\odot$] \\
(1) & (2) &(3) & (4) & (5) & (6) & (7) & (8) & (9) &(10) \\
\hline
61	&	SDSS J083325.69+391204.7 	&	0.322	&	2.5	&	1202	&	11.1	&	7.08	&	3.574	&	RQ	&	\cite{2018negrete} 	\\
62	&	SDSS J083337.99+342122.0 	&	0.597	&	1.43	&	2112	&	11.32	&	7.96	&	1.345	&	RQ	&	\cite{2018negrete} 	\\
63	&	SDSS J083453.39+384708.5 	&	0.184	&	1.8	&	3011	&	11.07	&	7.44	&	0.312	&	RQ	&	\cite{2018negrete} 	\\
64	&	SDSS J083550.97+142344.3 	&	0.353	&	1.34	&	2141	&	11.4	&	7.86	&	0.83	&	RQ	&	\cite{2018negrete} 	\\
65	&	SDSS J083753.88+193900.2 	&	0.242	&	1.22	&	3941	&	10.94	&	7.78	&	0.175	&	RQ	&	\cite{2018negrete} 	\\
66	&	SDSS J092247.03+512038.0 	&	0.16	&	1.75	&	820	&	11.43	&	7.1	&	2.000	&	RQ	&	\cite{2018negrete} 	\\
67	&	SDSS J093302.68+385228.0 	&	0.178	&	1.34	&	1231	&	11.37	&	7.4	&	0.635	&	RQ	&	\cite{2018negrete} 	\\
68	&	SDSS J100541.86+433240.4 	&	0.179	&	1.32	&	2034	&	11.81	&	7.78	&	0.723	&	RQ	&	\cite{2018negrete} 	\\
69	&	SDSS J101000.68+300321.5 	&	0.256	&	1.27	&	1858	&	11.66	&	7.83	&	0.894	&	RQ	&	\cite{2018negrete} 	\\
70	&	SDSS J101325.43+221229.4 	&	0.273	&	1.39	&	2400	&	11.52	&	7.84	&	0.707	&	RQ	&	\cite{2018negrete} 	\\
71	&	SDSS J102237.44+393150.1 	&	0.603	&	1.84	&	7269	&	11.67	&	8.43	&	0.624	&	RL	&	\cite{2018negrete} 	\\
72	&	SDSS J103457.27+235638.1 	&	0.42	&	1.25	&	1306	&	11.47	&	7.8	&	1.515	&	RQ	&	\cite{2018negrete} 	\\
73	&	SDSS J104009.33+560343.2 	&	0.392	&	1.23	&	2030	&	10.97	&	7.72	&	0.513	&	RQ	&	\cite{2018negrete} 	\\
74	&	SDSS J104613.72+525554.2 	&	0.503	&	1.55	&	2004	&	11.62	&	7.98	&	1.926	&	RQ	&	\cite{2018negrete} 	\\
75	&	SDSS J104816.57+222238.9 	&	0.33	&	1.6	&	822	&	11.2	&	7.31	&	1.652	&	RQ	&	\cite{2018negrete} 	\\
76	&	SDSS J105205.57+364039.6 	&	0.61	&	1.33	&	2356	&	11.97	&	8.34	&	1.558	&	RQ	&	\cite{2018negrete} 	\\
77	&	SDSS J120226.75-012915.2 	&	0.151	&	3.13	&	751	&	11.49	&	6.59	&	6.653	&	RQ	&	\cite{2018negrete} 	\\
78	&	SDSS J120548.14+584814.4 	&	0.276	&	1.56	&	1310	&	11.01	&	7.35	&	0.715	&	RQ	&	\cite{2018negrete} 	\\
79	&	SDSS J124938.40+464724.9 	&	0.15	&	1.24	&	1496	&	11.52	&	7.51	&	0.505	&	RQ	&	\cite{2018negrete} 	\\
80	&	SDSS J130416.99+020537.0 	&	0.229	&	1.53	&	1120	&	11.53	&	7.46	&	1.271	&	RQ	&	\cite{2018negrete} 	\\
81	&	SDSS J152939.29+203906.8 	&	0.152	&	1.37	&	5803	&	11.64	&	8.01	&	0.126	&	RQ	&	\cite{2018negrete} 	\\
82	&	SDSS J155119.59+255047.0 	&	0.593	&	1.51	&	2261	&	11.52	&	8.05	&	1.027	&	RQ	&	\cite{2018negrete} 	\\
83	&	SDSS J133602.01+172513.1 	&	0.552	&	1	&	2440	&	12.11	&	8.5	&	1.263	&	RI	&	\cite{2014marziani} 	\\
84	&	1H0323+342 	&	0.063	&	2	&	1650	&	11.78	&	7.07	&	0.788	&	RL	&	\cite{2011panessa}	\\
85	&	IGR J16385-2057 	&	0.026	&	1.2	&	1700	&	12.24	&	7.23	&	0.28	&	RQ	&	\cite{2011panessa}	\\
86	&	SWIFT J2127.4+5654 	&	0.014	&	1.3	&	2000	&	13.2	&	7.44	&	0.269	&	RQ	&	\cite{2011panessa}	\\
87	&	5C3.100 	&	0.071	&	2.7	&	685	&	12.69	&	6.96	&	10.368	&	RQ	&	\cite{1996boller} 	\\
88	&	PKS 0129-066 	&	0.22	&	1.8	&	1310	&	11.11	&	7.22	&	1.309	&	RL	&	\cite{1996boller} 	\\
89	&	E0132-411 	&	0.27	&	1.2	&	1930	&	11.27	&	7.72	&	0.527	&	RQ	&	\cite{1996boller} 	\\
90	&	E0944+464 	&	0.351	&	1.3	&	1320	&	10.95	&	7.5	&	0.680	&	RQ	&	\cite{1996boller} 	\\
91	&	Mrk 42	&	0.024	&	1	&	670	&	12.38	&	7.00	&	0.566	&	RQ	&	\cite{1996boller} 	\\
92	&	WISEA J122926.42+132020.5 	&	0.152	&	1.1	&	1120	&	11.21	&	7.32	&	0.392	&	RQ	&	\cite{1996boller} 	\\
93	&	E1228+123 	&	0.116	&	1.15	&	1680	&	11.22	&	7.35	&	0.223	&	RQ	&	\cite{1996boller} 	\\
94	&	IRAS 13224-3809 	&	0.066	&	2.4	&	650	&	11.9	&	6.66	&	2.863	&	RQ	&	\cite{1996boller} 	\\
95	&	Mrk 291 	&	0.036	&	1.1	&	700	&	12.46	&	7.16	&	0.998	&	RQ	&	\cite{1996boller} 	\\
96	&	Mrk 493 	&	0.031	&	1.31	&	410	&	12.35	&	6.79	&	1.388	&	RQ	&	\cite{1996boller} 	\\
97	&	Mrk 507 	&	0.055	&	2.7	&	960	&	12.55	&	6.92	&	4.992	&	RQ	&	\cite{1996boller} 	\\
98	&	E0337-267 	&	0.11	&	1.1	&	1340	&	11.05	&	7.19	&	0.326	&	RQ	&	\cite{1996boller} 	\\
99	&	0707-495 	&	0.041	&	2.77	&	1000	&	12.47	&	6.76	&	3.287	&	RQ	&	\cite{1996boller} 	\\
100	&	Mrk 231 	&	0.042	&	1.78	&	2100	&	12.83	&	7.55	&	0.793	&	RI	&	\cite{2006sulentic}	\\
101	&	RX J0100.4-5113 	&	0.063	&	0.97	&	3190	&	12.34	&	7.91	&	0.24	&	RQ	&	\cite{2004grupe}	\\
102	&	RX J0349.1-4711 	&	0.299	&	1.16	&	1700	&	11.63	&	7.88	&	0.985	&	RQ	&	\cite{2004grupe}	\\
103	&	RX J0439.6-5311 	&	0.243	&	2.66	&	700	&	10.73	&	6.56	&	3.013	&	RQ	&	\cite{2004grupe}	\\
104	&	ES 0614-584 	&	0.055	&	1	&	1080	&	12.4	&	7.49	&	0.566	&	RQ	&	\cite{2004grupe}	\\
105	&	RX J1034.6+3938 	&	0.043	&	1.37	&	700	&	11.52	&	6.71	&	0.46	&	RQ	&	\cite{2004grupe}	\\
106	&	RX J1117.1+6522 	&	0.147	&	0.99	&	1650	&	11.76	&	7.73	&	0.505	&	RQ	&	\cite{2004grupe}	\\
107	&	PG 1115+407 	&	0.155	&	0.98	&	1740	&	12.08	&	7.92	&	0.753	&	RQ	&	\cite{2004grupe}	\\
108	&	Z 186-18 	&	0.032	&	1.02	&	1450	&	11.56	&	7.01	&	0.147	&	RQ	&	\cite{2004grupe}	\\
109	&	RX J1209.8+3217 	&	0.145	&	1.09	&	1320	&	11.5	&	7.49	&	0.465	&	RQ	&	\cite{2004grupe}	\\
110	&	MCG+08-23-067 	&	0.03	&	1.08	&	730	&	11.45	&	6.66	&	0.219	&	RI	&	\cite{2004grupe}	\\
111	&	IRAS F12397+3333 	&	0.044	&	1.79	&	1640	&	12.42	&	7.28	&	0.605	&	RQ	&	\cite{2004grupe}	\\
112	&	IRAS 13349+2438 	&	0.108	&	1.25	&	2800	&	12.72	&	8.15	&	0.573	&	RQ	&	\cite{2004grupe}	\\
113	&	RX J1355.2+5612 	&	0.122	&	1.62	&	1100	&	11.64	&	7.24	&	1.377	&	RI	&	\cite{2004grupe}	\\
114	&	PG 1402+261 	&	0.164	&	1.1	&	1623	&	12.25	&	7.95	&	1.17	&	RQ	&	\cite{2004grupe}	\\
115	&	RX J1413.6+7029 	&	0.107	&	0.97	&	4400	&	12.09	&	8.12	&	0.14	&	RQ	&	\cite{2004grupe}	\\
116	&	Mrk 684 	&	0.045	&	1.5	&	1260	&	12.42	&	7.3	&	0.616	&	RQ	&	\cite{2004grupe}	\\
117	&	RX J1618.1+3619 	&	0.034	&	1.14	&	950	&	11.43	&	6.77	&	0.204	&	RQ	&	\cite{2004grupe}	\\
118	&	RX J1702.5+3247 	&	0.163	&	0.98	&	1680	&	12.45	&	8.09	&	0.778	&	RQ	&	\cite{2004grupe}	\\
119	&	RX J2216.8-4451 	&	0.135	&	1.13	&	1630	&	12.04	&	7.77	&	0.743	&	RQ	&	\cite{2004grupe}	\\
120	&	RX J2217.9-5941 	&	0.159	&	0.96	&	1430	&	11.52	&	7.61	&	0.445	&	RI	&	\cite{2004grupe}	\\

\hline
    \end{tabular}
\end{table*}

\setcounter{table}{2}
\begin{table*}{}[h!]
    \centering
        \caption{Cont.} 
    \begin{tabular}{c|c|c|c|c|c|c|c|c|c}
\hline\tabcolsep=0pt
No. 	&	NAME 	&	 $z$ 	&	 \rfe 	&	 FWHM \hb\ 	&	 $\lambda f_\lambda$ 5100 \AA\ 	&	$\log$ \mbh\ 	&	 \lledd\ 	&	Radio 	&	 DATABASE \\
&  & & & [\kms] & [Jy Hz]\ & [$M_\odot$] \\
(1) & (2) &(3) & (4) & (5) & (6) & (7) & (8) & (9) &(10) \\
\hline
121	&	RX J2242.6-3845 	&	0.22	&	1.01	&	1900	&	11.41	&	7.77	&	0.446	&	RQ	&	\cite{2004grupe}	\\
122	&	RX J2317.8-4422 	&	0.134	&	1.09	&	1010	&	11.65	&	7.44	&	0.649	&	RQ	&	\cite{2004grupe}	\\
123	&	SDSS J085900.49+383211.7 	&	0.345	&	2.47	&	1873	&	11.34	&	7.38	&	2.095	&	RQ	&	\cite{2006zhou} 	\\
124	&	SDSS J150521.92+014149.7 	&	0.158	&	1.63	&	1125	&	11.14	&	7.12	&	0.943	&	RQ	&	\cite{2006zhou} 	\\
125	&	SDSS J150506.47+032630.8 	&	0.408	&	1.53	&	1082	&	10.98	&	7.41	&	1.156	&	RL	&	\cite{2006zhou} 	\\
126	&	WISEA J100131.15+593939.5 	&	0.305	&	1.35	&	1043	&	10.84	&	7.29	&	0.651	&	RQ	&	\cite{2006zhou} 	\\
127	&	SDSS J144111.54-021235.1 	&	0.082	&	1.28	&	1209	&	12.15	&	7.47	&	0.729	&	RQ	&	\cite{2006zhou} 	\\
128	&	SDSS J133623.76+031059.6 	&	0.49	&	1.25	&	970	&	10.78	&	7.44	&	0.93	&	RQ	&	\cite{2006zhou} 	\\
129	&	SDSS J090654.48+391455.3 	&	0.241	&	1.24	&	1140	&	11.35	&	7.51	&	0.829	&	RQ	&	\cite{2006zhou} 	\\
130	&	SDSS J094857.31+002225.5 	&	0.584	&	1.22	&	1432	&	11.29	&	7.88	&	1.482	&	RL	&	\cite{2006zhou} 	\\
131	&	SDSS J144902.16+472138.1 	&	0.513	&	1.19	&	1717	&	11.22	&	7.88	&	1.012	&	RQ	&	\cite{2006zhou} 	\\
132	&	SDSS J125100.44+660326.8 	&	0.282	&	1.18	&	1342	&	11.85	&	7.87	&	1.527	&	RQ	&	\cite{2006zhou} 	\\
133	&	SDSS J160028.72+011026.6 	&	0.429	&	1.15	&	1931	&	11.54	&	8.02	&	0.663	&	RQ	&	\cite{2006zhou} 	\\
134	&	SDSS J141419.84+533815.3 	&	0.164	&	1.13	&	1873	&	11.3	&	7.56	&	0.318	&	RQ	&	\cite{2006zhou} 	\\
135	&	SDSS J083553.46+055317.1 	&	0.204	&	1.13	&	1259	&	11.62	&	7.65	&	0.837	&	RQ	&	\cite{2006zhou} 	\\
136	&	SDSS J083437.09+532818.1 	&	0.586	&	1.13	&	1656	&	11.25	&	7.94	&	1.169	&	RQ	&	\cite{2006zhou} 	\\
137	&	SDSS J221918.53+120753.1 	&	0.081	&	1.11	&	982	&	11.47	&	7.15	&	0.526	&	RQ	&	\cite{2006zhou} 	\\
138	&	SDSS J143940.27+030528.6 	&	0.268	&	1.11	&	828	&	11.32	&	7.46	&	1.043	&	RQ	&	\cite{2006zhou} 	\\
139	&	SDSS J144519.03+583945.2 	&	0.566	&	1.08	&	1008	&	10.57	&	7.46	&	0.693	&	RQ	&	\cite{2006zhou} 	\\
140	&	SDSS J074511.27+191942.9 	&	0.384	&	1.08	&	1303	&	11.29	&	7.75	&	0.961	&	RQ	&	\cite{2006zhou} 	\\
141	&	WISEA J150245.37+405437.3	&	0.233	&	1.06	&	680	&	11.06	&	7.24	&	1.235	&	RQ	&	\cite{2006zhou} 	\\
142	&	SDSS J085038.63+520145.7 	&	0.491	&	1.06	&	1811	&	10.98	&	7.82	&	0.615	&	RQ	&	\cite{2006zhou} 	\\
143	&	SDSS J151302.60+542227.8 	&	0.508	&	1.05	&	1513	&	10.82	&	7.7	&	0.595	&	RQ	&	\cite{2006zhou} 	\\
144	&	SDSS J094233.25+093837.7 	&	0.205	&	1.04	&	927	&	10.58	&	7.1	&	0.449	&	RQ	&	\cite{2006zhou} 	\\
145	&	WISEA J093749.84+585522.7 	&	0.132	&	1.04	&	978	&	10.61	&	6.97	&	0.28	&	RQ	&	\cite{2006zhou} 	\\
146	&	WISEA J130955.96+530636.7 	&	0.321	&	1.03	&	973	&	11.01	&	7.47	&	0.69	&	RQ	&	\cite{2006zhou} 	\\
147	&	SDSS J163323.58+471858.9 	&	0.116	&	1.02	&	909	&	11.45	&	7.28	&	0.447	&	RL	&	\cite{2006zhou} 	\\
148	&	SDSS J110542.72+020250.8 	&	0.454	&	1.01	&	1256	&	11.08	&	7.73	&	0.849	&	RQ	&	\cite{2006zhou} 	\\
149	&	Ton S180 	&	0.062	&	1.03	&	1085	&	12.54	&	7.59	&	0.79	&	RQ	&	\cite{2001veron-cetty} 	\\
150	&	IRAS 09571+8435 	&	0.092	&	1.05	&	1185	&	11.91	&	7.48	&	0.514	&	RQ	&	\cite{2001veron-cetty} 	\\
151	&	PKS 0558-504 	&	0.137	&	1.56	&	1250	&	12.52	&	7.75	&	2.45	&	RI	&	\cite{1999leighly}	\\
152	&	Mrk 142 	&	0.045	&	1.36	&	1470	&	11.94	&	7.19	&	0.435	&	RQ	&	\cite{1999leighly}	\\
153	&	IRAS 17020+4544 	&	0.06	&	1.86	&	1040	&	11.66	&	6.88	&	0.844	&	RL	&	\cite{1999leighly}	\\
154	&	Ark 564 	&	0.025	&	1	&	950	&	12.75	&	7.29	&	0.412	&	RQ	&	\cite{1999leighly}	\\
155	&	PDS 456	&	0.185	&	1	&	2883	&	13.03	&	8.59	&	1.194	&	RQ	&	\cite{2025Li}	\\

\hline
    \end{tabular}   
Col 1: progressive number; Col. 2: common name; Col. 3: redshift; Col. 4: \feii\ prominence parameter \rfe; Col. 5: FWHM \hb\ broad emission in \kms; Col. 6: flux in Jy Hz; Col. 7: decimal logarithm of black hole mass in solar masses; Col. 8: Eddington ratio; Col. 9: radio class; Col. 10: source of optical data for xA classification. 
    
\end{table*}

\section{Discussion}
\label{disc}

\subsection{Comparison between the xA SED and SEDs for the general quasar population. }

\subsubsection{\citeauthor{1987mathewsferland} SED}

The  SED  of \citetalias{1987mathewsferland}   has been frequently used in photoionization computations to represent the typical AGN continuum and can serve as a useful benchmark.  It appears to be appropriate for Population  A sources, radiating as an Eddington ratio $\gtrsim 0.1 - 0.2$\ \citep{marzianietal03b,pandaetal19,2020Feland}, because of its prominent BBB.  The {\citetalias{1987mathewsferland}  SED} is { mainly } an observationally-derived SED, inferred from direct constraints on the continuum and on emission lines. The BBB  continuum \citep{malkansargent82,malkan83} is peaking in the (un-observable) extreme ultraviolet but its prominence was derived for ``classical quasars'' \, considering the equivalent width of a recombination line whose ionic species has an ionization threshold a few Rydbergs. An estimate of the flux at 4 Ryd was made assuming that the continuum short ward of  228 \AA\ is represented as a power law, $\nu f_\nu = const.$\ between 1 and 4 Ryd.  The flux at 4 Ryd can be derived from the equivalent width of the He{\sc ii}$\lambda$ 1640 line, assuming case B conditions \citep{macalpineetal85,2020Feland}. However, recent disk models suggest that a flat behavior between 1 and 4 Ryd might be appropriate only for the super-Eddington case \citep{kubotadone18,kubotadone19}.

 The shapes of the \citetalias{1987mathewsferland} SED and the RQ SED derived in this work are fairly consistent in the range from the mid-IR and near UV, with just a slightly steeper slope in the NIR and NUV for the xA sources (Fig. \ref{fig:RQ}). A major difference occurs in the soft/hard X-ray domain: at $\log \nu \gtrsim 16.3$, the shape is shallower and decreasing ($\Gamma > 2$) with increasing frequency up to $\log \nu \approx 19$ [Hz]. The reflection hump of the \citetalias{1987mathewsferland} SED is not present. The absence of the hump is consistent with previous measurements of the X-ray domain for super-Eddington candidates \citep{wangetal13,laurentietal22}, and with a general trend of the steepening slope with Eddington ratio \citep{risalitietal09,brightmanetal13,trakhtenbrotetal17}.  However, given the weakness of the hard X-ray emission, a Compton hump excess above 10 keV leading to a flattening of the SED cannot be excluded \citep{lanzuisietal16}. 

\subsubsection{Eddington ratio-dependent SEDs}

A more up-to-date approach is to consider the SEDs categorized by the Eddington ratio, which helps explore correlations with emission-line properties \citep[][and references therein]{2020Feland}. The SEDs are based on a sequence of AGN spectra stacked according to their Eddington ratios. As the Eddington ratio increases, the SEDs show more far-ultraviolet brightness. \citet{2020Feland} considered four cases: low \lledd\  SEDs that show a standard accretion disk emission profile with fairly pronounced high-energy X-ray tails; intermediate \lledd\  SEDs, where  there is a noticeable increase in the FUV brightness relative to lower Eddington ratios. The SEDs begin to show changes in the soft X-ray region, indicating a transition in the energy distribution that affects the ionization states more significantly. High \lledd\ SEDs appear to be more FUV-bright, reflecting higher levels of energy and more significant emission in the ultraviolet spectrum, and the presence of an X-ray excess. The SEDs here have a more pronounced soft X-ray excess. The super-Eddington (highest) SEDs 
display the most extreme features, with dominant disc emission and very steep X-ray tails. These SEDs are characterized by the highest FUV brightness and significant ionizing potential, affecting the emission properties of helium and hydrogen lines observed. Fig. \ref{fig:sedlledd} compares the high and highest SEDs with the one derived for the xA sample. 
The median, the 3rd, and 1st quartile are compared with the \lledd-derived SEDs.

SEDs built according to the Eddington ratio \citep{2020Feland} are valuable as long as they reflect the trend of the E1 sequence, although the estimation of the Eddington ratio is, in practice, affected by statistical and systematic uncertainties. 
At any rate, the comparison between the xA SED and the one for high values of Eddington ratio (Fig. \ref{fig:sedlledd}) reveals an overall consistency, with the 3rd quartile SED following closely the same trend in the soft and hard X-ray domain up to $\sim 10 - 20$ keV. The ``highest" case in Fig. \ref{fig:sedlledd} (due to only one object, PHL 1092) corresponds to the extreme SEDs in our sample. The ``mid" case is instead ruled out by the increase toward the NUV of the xA SED. 

\begin{figure}

\includegraphics[scale=0.44, trim={0cm 4cm 0cm 0cm},clip]{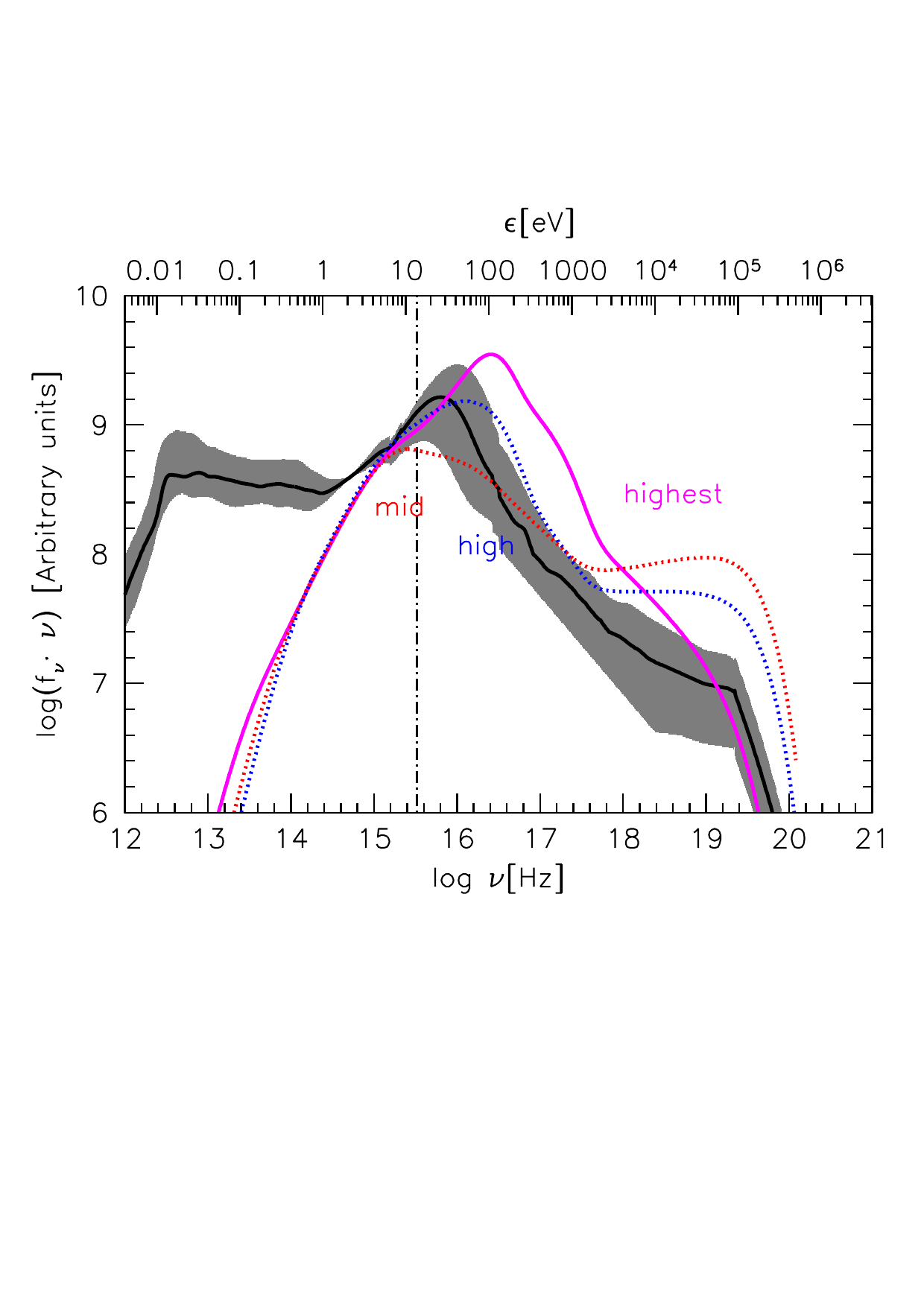} \\
\caption{Spectral energy distributions corresponding to the median SED (second quartile, q2, thick black line), with the shaded area between the first and third quartile, and the mid, high and highest \lledd\ SED derived by \citet[][red, blue and magenta lines, respectively]{2020Feland}. Abscissa is the logarithm of the frequency in Hz (bottom) and energy in eV (top); ordinate is flux $\nu f(\nu)$ in arbitrary units.   \label{fig:sedlledd} }
\end{figure}

\subsection{ Origin of radio and FIR emission}

 { The full sample is made only of xA or borderline xA quasars, with \rfe$\gtrsim 0.9$, and it is not intended to be representative of the general population of AGN. Previous work \citep[][see also   \citealt{marzianietal21,marzianietal25}]{sanietal10,bonzinietal15,caccianigaetal15,2019ganci} showed that the FIR emission in quasars accreting at relatively high rates can be associated with star formation.  The RQ+RI median composite shows no FIR excess with respect to the RQ (both SEDs are almost flat from the optical to the submillimeter), but the RI SED shows a clear FIR excess (Fig. \ref{fig:RQ}). It is therefore reasonable to test whether FIR and radio emission are consistently indicating star formation, and whether the strongest \feii\ emitter could be associated with the strongest FIR and radio emission.  } 
 
 We measured {  the radio power and the FIR luminosity by scaling the emission at 70 $\mu$m   in the FIR domain and at 1.4 GHz in the radio domain} relative to the median  optical luminosity of  composites. }   The power values in the FIR and radio are reported in Table \ref{tab:sed_excess}. 
{The FIR and radio spectral regions shown in Figure \ref{fig:SED_Rfe} can be converted into luminosity and star formation rate (SFR)  scaling the luminosity and the radio power, { knowing} the optical luminosity of the sample. The parameter $q$, defined as the specific power ratio between the FIR and the radio \citep{yunetal01,bressanetal02} reveals that the emission can be accounted for { by star formation in the case of the median RQ composite, as its value is consistent with } $\mathbf{q \approx 2,} $\ as observed in star-forming galaxies (first row of Table \ref{tab:sed_excess}). 

\subsection{The most extreme among the extreme: \rfe\  $\gtrsim$ 1.5}
\label{sec:rfe_SEDs}

\begin{figure*}
    \includegraphics[width=8.8cm]{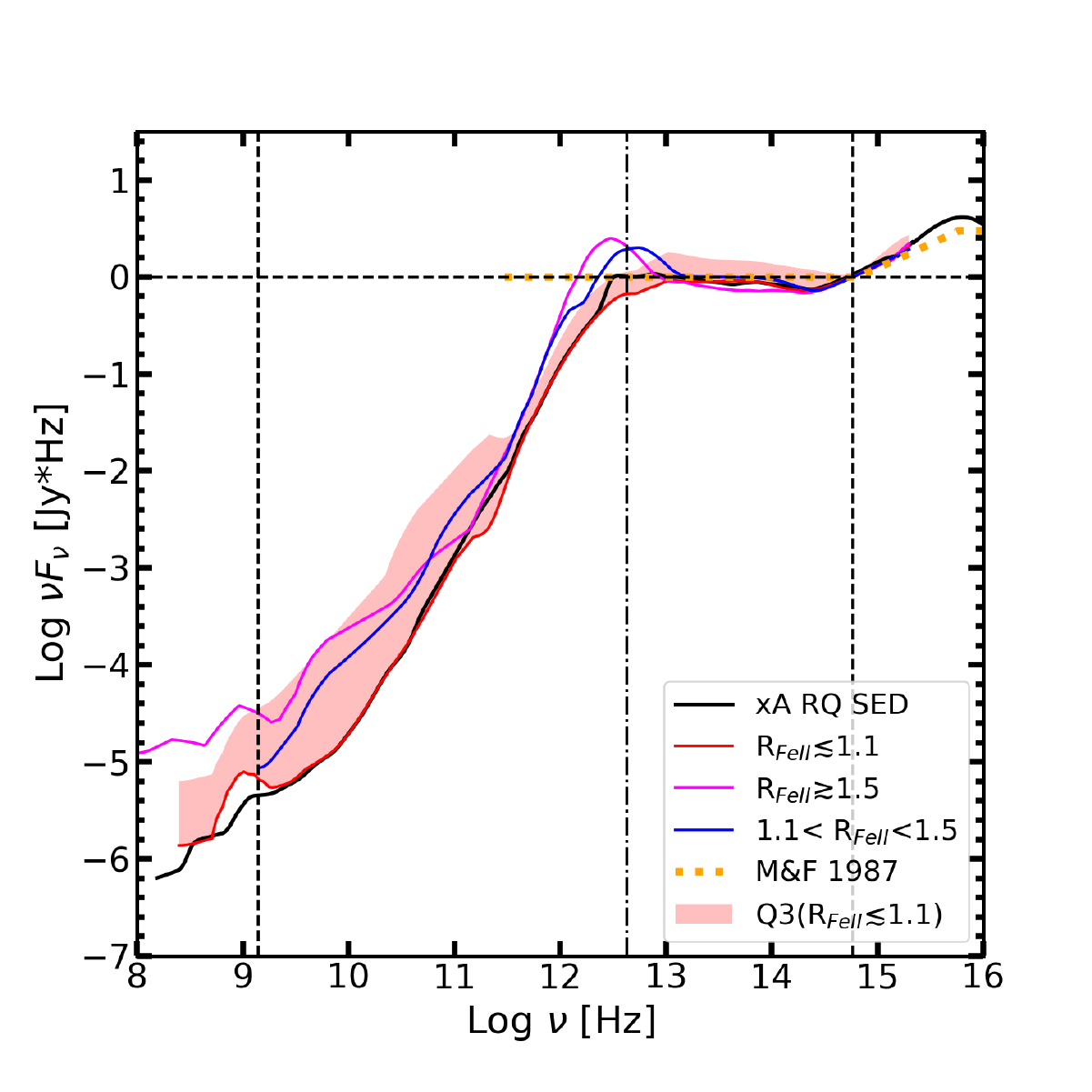}
    \includegraphics[width=8.8cm]{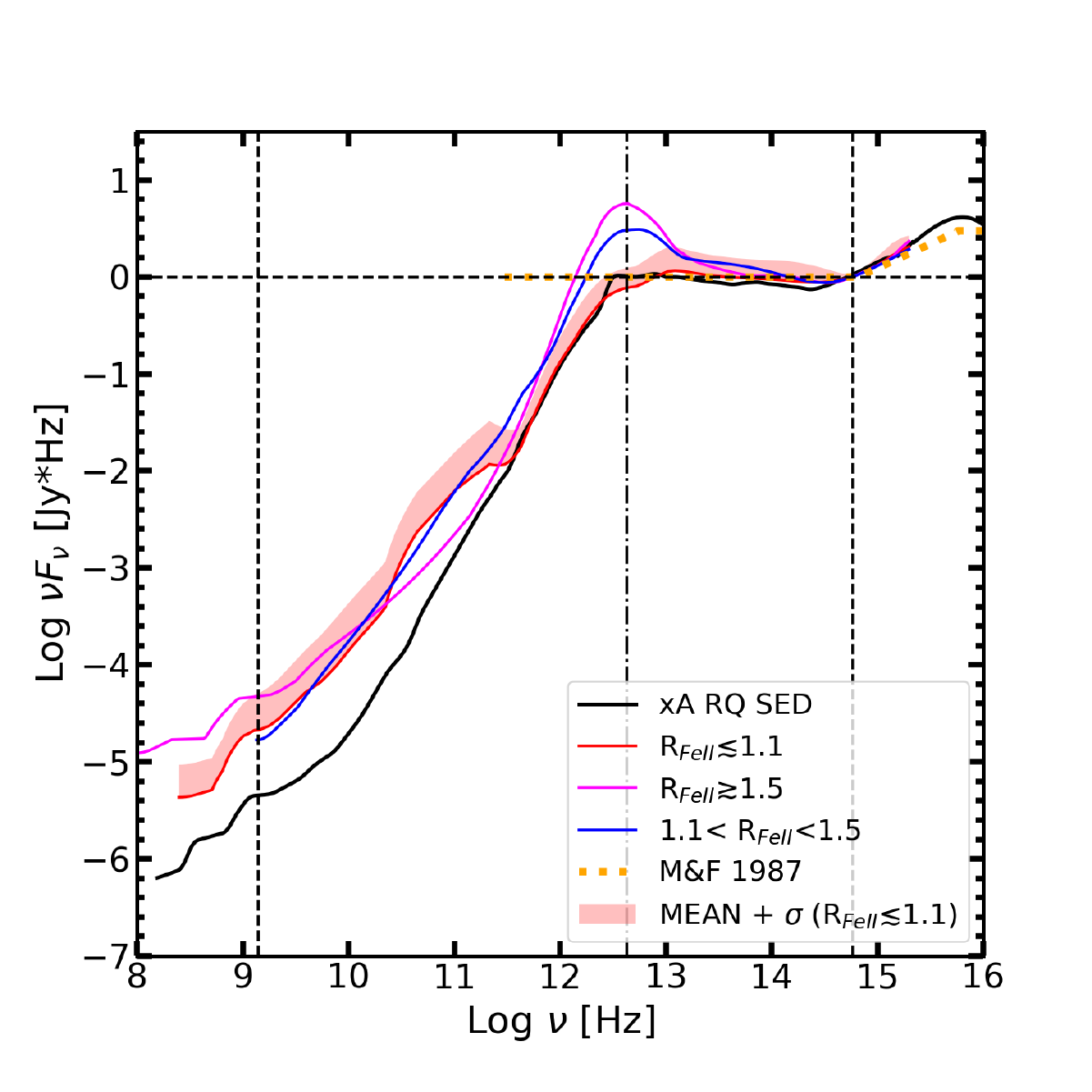}

\caption{ { RQ+RI SEDs  grouped by} \rfe\ parameter. { Left: Median SEDs. The magenta } line shows the SED for the sources with the highest \rfe \ parameter (\rfe $\ge$1.5), the blue line shows the SED for an intermediate range of \rfe \ parameter (1.1$<$ \rfe$<$1.5), and the red line shows the SED for the lowest \rfe\ parameter  (\rfe $\le$1.1). { The halftone band traces the 3rd quartile for the SED with the lowest  \rfe\ parameter. Right: same, for average composites. In this case the halftone band traces the average composite plus 1 standard deviation. In both cases, the halftone bands are meant to show the statistical significance of the IR excess for the SEDs restricted to the intermediate and high \rfe\ parameter.} In both panels, black solid line shows our RQ xA median SED, the gold thick dotted line shows the \citet{1987mathewsferland} SED for classical quasars. Abscissa corresponds to the logarithm of the frequency in units of hertz and ordinate the logarithmic of specific flux times the frequency. \label{fig:SED_Rfe}}
\end{figure*}

\begin{table*}
    \centering
    \begin{tabular}{lccccccccccc}\hline\hline
       Composite SED & $N$ & \multicolumn{2}{c}{FIR} && \multicolumn{3}{c}{Radio} & $q$ \\ \cline{3-4}\cline{6-8}
      
       & &     $  L$(70$\mu$m) & $\log$SFR$_\mathrm{70\mu m}$ &&      $  P_\mathrm{\nu,1.4GHz}$ & $\log$ SFR$_\mathrm{1.4GHz}$  &$\log R_\mathrm{K}$ &  \\ 
            &    & [erg s$^{-1}$]  & [$M_\odot$ yr$^{-1}$]    &    & [erg s$^{-1}$Hz$^{-1}$]   & [$M_\odot$ yr$^{-1}$] \\    
\multicolumn{1}{c}{(1)} & (2) & (3) & (4) && (5) & (6) & (7) & (8)     \\
\hline
RQ Median             & 139 &  2.9E+44  &  1.44   &&   7.9E+29$^\mathrm{a}$ &  1.66 & 0.25 & 1.99 \\ 
RQ+RI Median             & 147 &  2.7E+44  &  1.40   &&   2.0E+30$^\mathrm{a}$ &  2.06 & 0.618 & 1.56 \\ 

       \hline
        \rfe $\ge$1.5        & 40&     5.7E+44 & 1.73     &     & 4.3E+30 &  2.40   & 0.96&    1.55  \\ 
       1.1 $<$\rfe$\le$ 1.5  & 54&   5.2E+44 & 1.60      &      & 1.3E+30 & 1.88   &0.44 &   2.03    \\ 
       \rfe $\le$ 1.1        & 53&   2.0E+44  &1.27    &&   1.0E+30 & 1.76&    0.32    & 1.72     \\ 
\hline \hline
      
    \end{tabular}
    \caption{{ Col. 1: \rfe \ range. First row refers to the full \rfe\ range for the RQ plus RI sub-samples; Col. 2: number of sources in each sample. Cols. 3--4: FIR luminosity at 70$\mu m$, SFR associated with FIR emission, computed following \citet[][c.f. \citealt{2019ganci}]{lietal10}; Cols. 5--7: radio parameters, in this order:   excess power at 1.4 GHz, ``pseudo" SFR \citep{yunetal01}, and $\log R_\mathrm{K}$. Col. 8: $q$\ parameter i.e., the ratio between FIR and radio specific fluxes \citep{yunetal01}. FIR and radio  of the SED as a function of the \rfe\ parameter are normalized  to the xA SED . The full RQ+RI sample shows a median  $\mu$(\lvc) $\approx  2.8 \cdot 10^{44}$ erg s$^{-1}$\ at 5100 \AA, quite similar to the median luminosity of the three subsamples,  (3.0, 2.7, 2.8) $\cdot 10^{44}$\ erg s$^{-1}$\   for the low, intermediate and high \rfe\ range respectively. }}
    \label{tab:sed_excess}
\end{table*}

To inquire into the role of the \rfe\ parameter in the SED, we subdivided our sample into three categories: (1) \rfe $\lesssim$1.1 which identifies borderline xA sources with $0.9 \lesssim$ \rfe $\lesssim 1.1$;  (2) 1.1$<$ \rfe$<$1.5, an intermediate range, corresponding to the spectral type A3 following \citet{sulenticetal02}, and (3) \rfe $\gtrsim$1.5, the most extreme xA sources (A4 and very rare sources with \rfe $\gtrsim 2$). The three \rfe\ ranges each encompass roughly a third of the sample (Tab. \ref{tab:sed_excess}).  Figure \ref{fig:SED_Rfe} shows the median (left panel) and mean (right panel) SEDs for each range of the \rfe \ parameter compared to the \citetalias{1987mathewsferland} SED (gold) and our xA RQ SED (black solid line). The \citetalias{1987mathewsferland} SED and the xA RQ SED are in very good agreement from the sub-mm break to the optical range (5100 \AA). The low \rfe\ ($\le 1.1$) shows a slight deficit emission  with respect to the RQ composite, beginning at $\log \nu ~ 10^{13}$ Hz, and  a downward sloping $\nu f_\nu$ behavior from the optical down to the sub-mm break at $\log \nu \approx 12.4$\ [Hz].

 { Sources with \rfe$\gtrsim$ 1.1, however, do show a clear excess over the composite RQ  (Fig. \ref{fig:SED_Rfe}), a factor $\approx 2$\ in the FIR domain.}  

{ If we compare the excess at 70 $\mu$m, between the median SED of the subsample satisfying the condition  \rfe$\le 1.1$ and of the ones \rfe$> 1.1$, the excess reaches a factor $\approx 3$. At 100 $\mu$m, the excess is a factor $\approx$ 2.6     (1.1 $<$ \rfe $<1.5$) and $\approx 5$ (\rfe $\ge 1.5$).}  The conclusions are reinforced if the {\rm average}  SED in each \rfe\ range is computed. In this case, the excess is larger, reaching  a factor $\approx 3 $ and $\approx 5$\ for the intermediate and high \rfe\ cases, respectively. { With a dispersion $\sigma \approx 0.2$ dex for the low \rfe\ SED,  the significance is at a $\approx 4 \sigma$\ level if the 70 $\mu$m\ luminosities of the average SED for \rfe $\le 1.1$ and \rfe$\ge 1.5$ are compared, and $\approx 2 \sigma$\ if the comparison is carried out between the intermediate 1.1 $<$\rfe $<$1.5 and the low \rfe $\le 1.1$ SED.  }

The sample has a \lvc\ median $\mu$(\lvc) $\approx$ 10$^{44.45}$ erg s$^{-1}$ and an average \avglvc  $\approx$ 10$^{44.47}$ erg s$^{-1}$.  The highest \rfe\ SED  has an average MIR-FIR luminosity $L$(MIR-FIR) $\approx$ 10$^{45.4}$ erg s$^{-1}$.
{ This power}  would position the source within the range of luminous infrared galaxies. The discrepancy between the median reported in Table \ref{tab:sed_excess} and average FIR luminosity can be tentatively explained by the assumption that, alongside moderately FIR luminous quasars, there exists a significant minority of quasars (even within the xA spectral type) with luminosity well above the sample median.

This analysis, albeit very tentative, suggests that strong, singly-ionized iron emission could be related to star formation in the host galaxy, revealed by a consistent excess of FIR and radio emission. The connection between a strong \feii\ emission and the IR was early discussed by  \citet{1993Lipari}. They suggested a possible scenario for  ``young QSOs" in the last stages of a strong Starburst, similar to the evolutionary scenario for AGN and quasars described by \citet{2018Donofrio}, and in line with earlier suggestion of a connection between enhanced star formation and the feeding of nuclear activity (for more recent perspectives see \citealt{rafanellietal11,kauffmannetal24}).  \citet{1993Lipari} and \citet{sandersetal88}   also suggested a possible origin for broad absorption line (BAL) sources through a violent star formation episode \citep{perrydyson92,liparietal94,liparietal03}. An analysis of a large SDSS sample with FIRST detections revealed that xA sources show a high prevalence of RI quasars along the quasar MS \citep{2019ganci}, a result consistent with an enhancement due to star formation and with the results presented in this paper.  Recent works \citep{2022garnica,2021sniegowska} have found a high metal content in luminous xA sources that might be well associated with peculiar forms of nuclear star formation \citep{Wang2023}.

\subsection{Photoionization analysis: intensity ratios predicted by the xA SED}
\label{photoion}

The SED derived in this work is somewhat different from the SEDs provided by {\tt CLOUDY}. We therefore performed a tentative evaluation of the implication of the new SEDs for the estimate of the BLR physical conditions. The heatmaps of Fig. \ref{fig:photoion} show the predictions on \rfe, equivalent width of \civ\ and intensity ratio \civ/\hb, obtained from {\tt CLOUDY 23.01}\ simulations {\citep{2023Chatzikos, gunasekeraetal23}}, as a function of metallicity, ionization parameter $U$\ and Hydrogen density $n_\mathrm{H}$. The basic scenario of extreme BLR properties developed by previous works \citep{negreteetal12,martinez-aldamaetal15,martinez-aldamaetal18} is confirmed. The \rfe\ measured on the  virialized component 
 is well accounted for by $\log n_\mathrm{H} \sim 12-13$ cm$^{-3}$, with   ionization parameter $\log U \sim -1.5 - 2$. The $W$ \civ \ and the \civ/\hb\ ratios are known to be low in the virialized component of xAs \citep{marzianietal10,florisetal24}, with values $\ll 10$. This rules out the solutions at high $U$ which predict \civ/\hb\ $\sim 60$. The combination of low \civ\ equivalent width, low \civ/\hb, and high \rfe\ (in the range 1 -- 1.5) are best matched by $\log U \sim -2.5$, $\log n_\mathrm{H} \sim 12 - 13$ cm$^{-3}$, and $Z \sim 10 - 20 Z_\odot$. Higher $Z$\ would make it possible to achieve \rfe$\gtrsim $2. The highest \rfe\  value  ($\sim 3.5$) is obtained for $\log U = -1.5$,  $Z = 50 Z_\odot$,    $\log n_\mathrm{H} = 13$ cm$^{-3}$. This solution is ruled out for the real xA because the predicted \civ/\hb\ ($\sim 60$)   is way higher than observed. It is also important to consider that sources with \rfe$\gtrsim 2$\ are a tiny minority of the general population of type-1 AGN. If the xA sources with \rfe$\gtrsim 1$\ are 10 \%\ of optically selected samples at $z \lesssim 1$ \citep{marzianietal13a}, sources with \rfe$\gtrsim 1.5$\ might be just 2 -- 3 \%, and with \rfe$\gtrsim 2$\ less than 1 \%. Most likely values of metallicity appear to be within 10 -- 20 $Z_\odot$. These values are consistent with previous studies of luminous quasars \citep{hamannferland93} or of xA samples \citep{2021sniegowska,2022garnica}.

{ This scenario is confirmed if the luminosity scaled-SED is used (bottom panels of Fig. \ref{fig:photoion}). In this case, the optical luminosity at 5100 \AA\ from the median luminosity-scaled composite is used, and the radii of the line emitting regions scaled to obtain the exact value of the ionization parameters as assumed for the SED normalized at 5100\AA. The trends induced by the two SEDs as a function of density and ionization parameters  are consistent, and the low \civ/\hb\ ratio observed in the virialized component requires $\log n_\mathrm{H} = 11 - 12$ cm$^{-3}$, ionization parameter $\log U \sim -2.5$. In no case with $Z \lesssim 10$, \rfe\ reaches value above 1, also for the high ionization solution $\log U \sim -1.5$. However, the high-ionization solution is again ruled out by the very high \civ/\hb\ $\gtrsim$ 70. }

\begin{figure*}
\centering 
\includegraphics[scale=0.25]{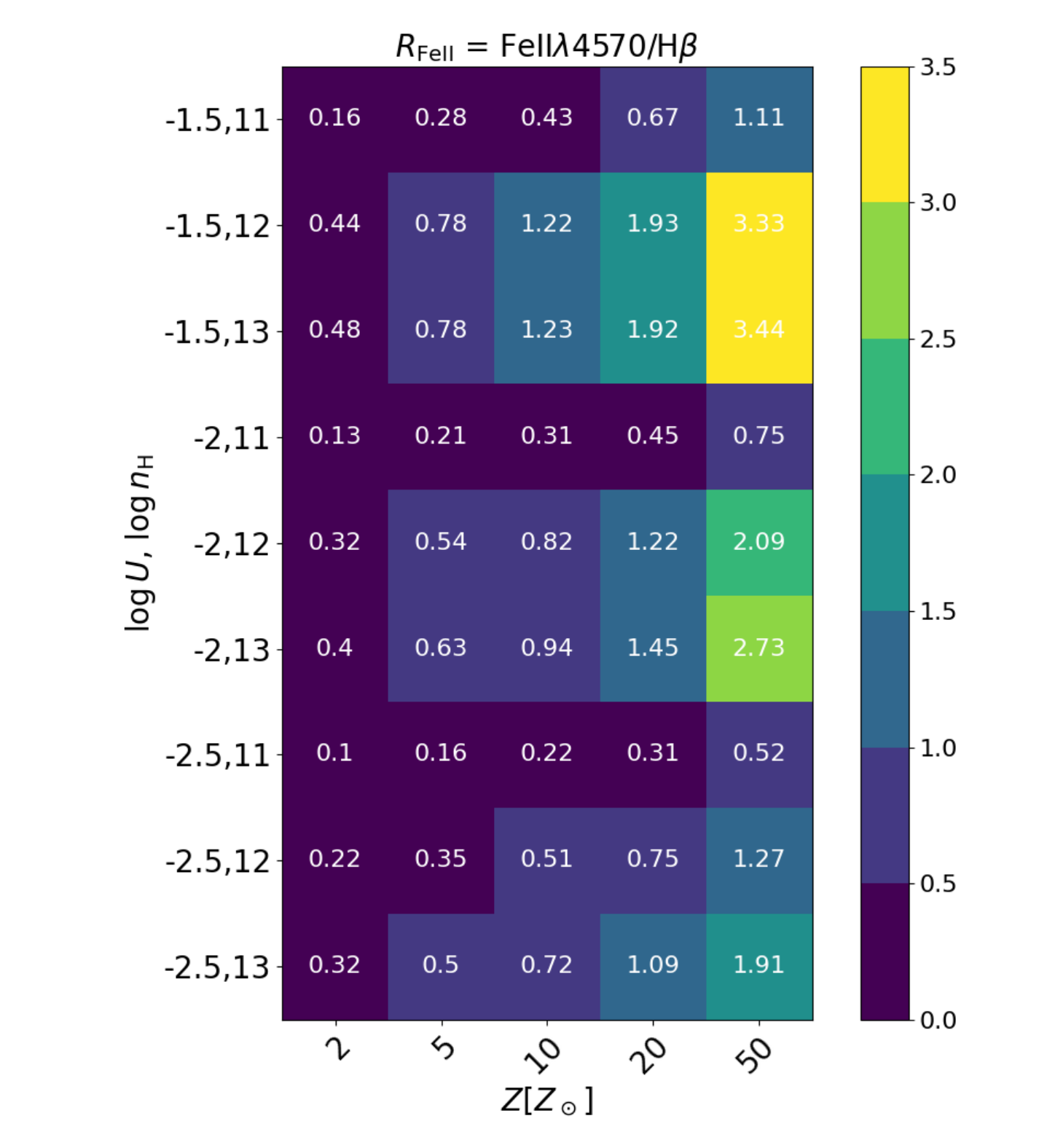}\hspace{-0.45cm}
\includegraphics[scale=0.25]{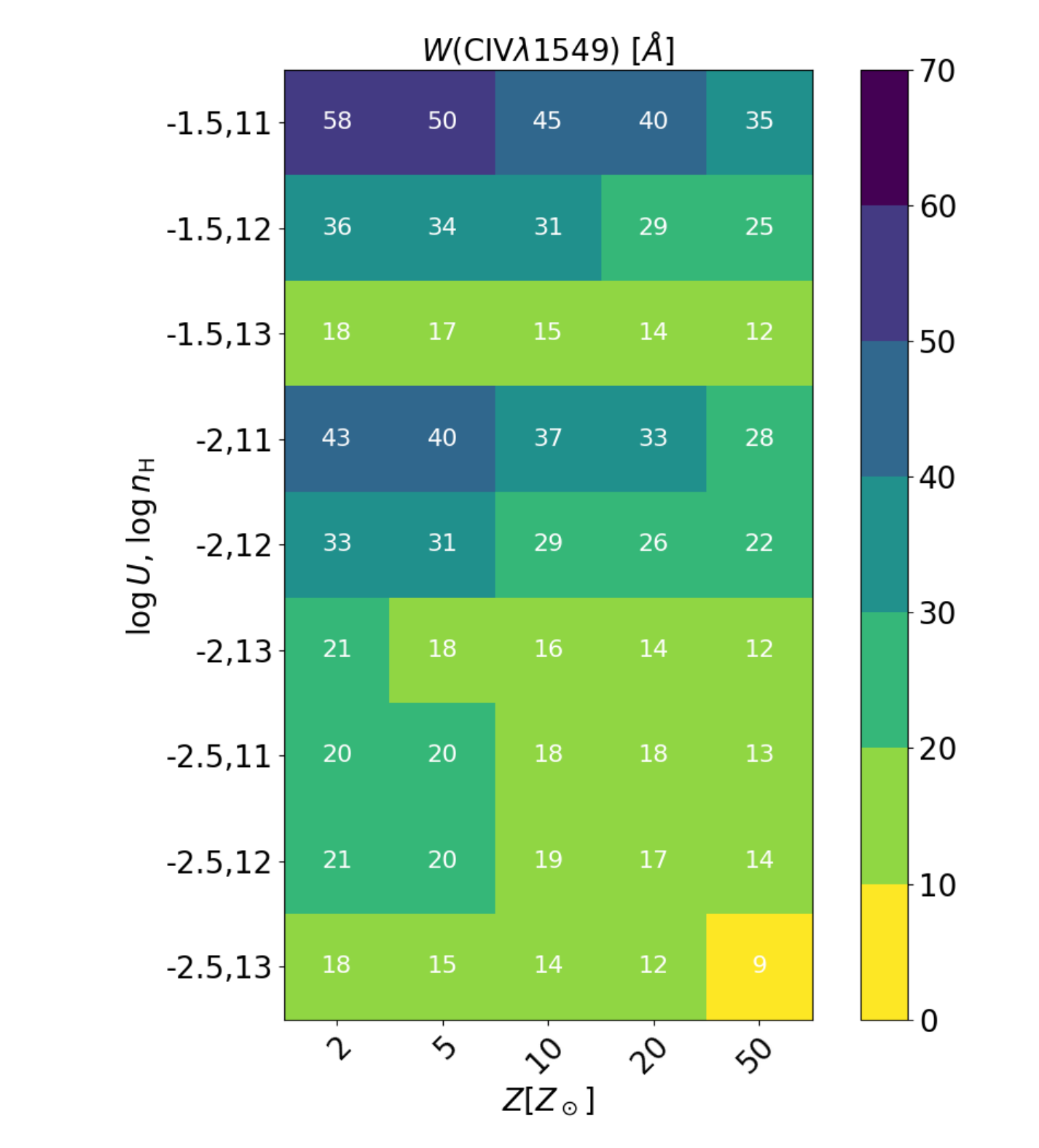}\hspace{-0.45cm}
\includegraphics[scale=0.25]{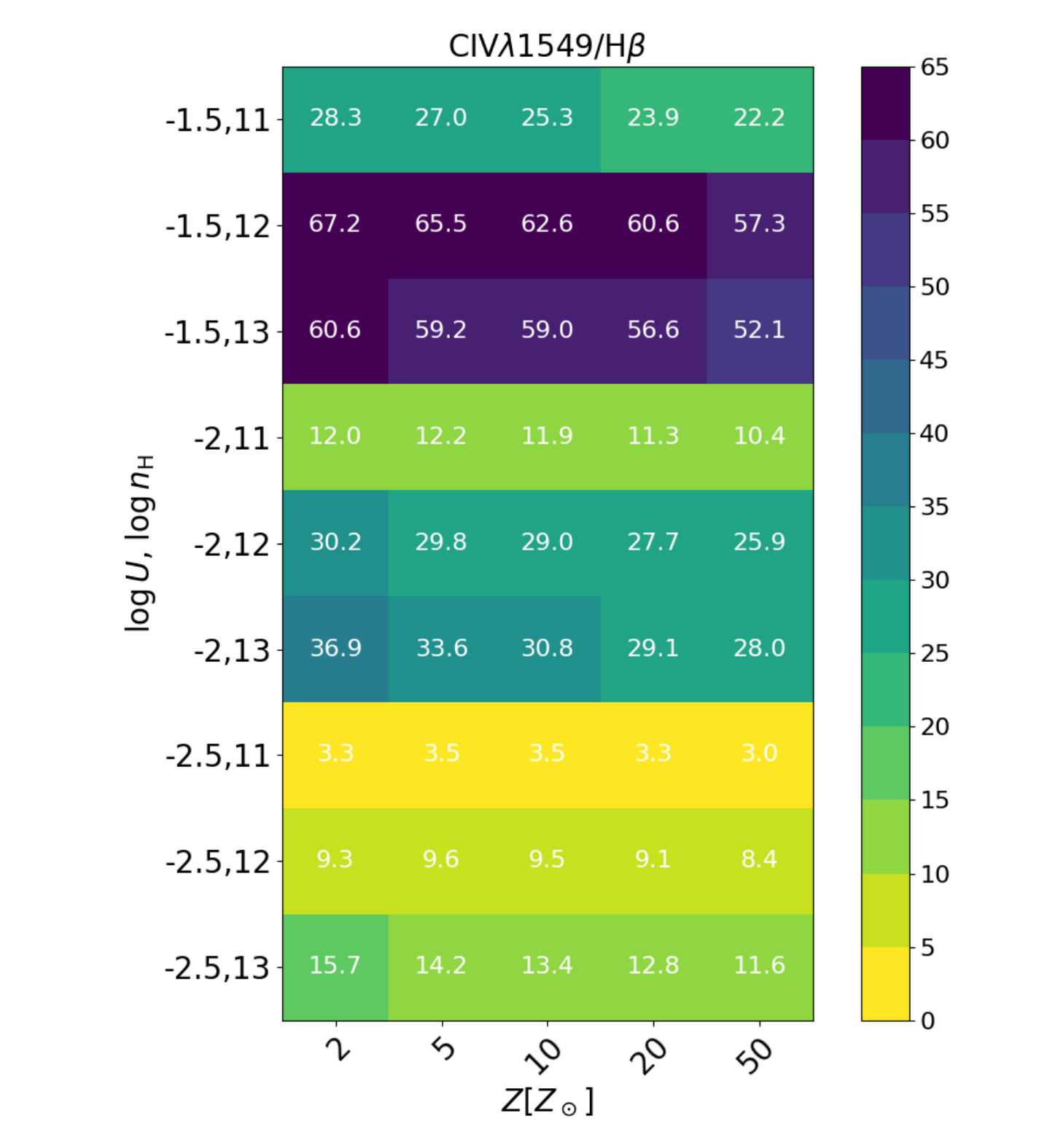}\\
\includegraphics[scale=0.25]{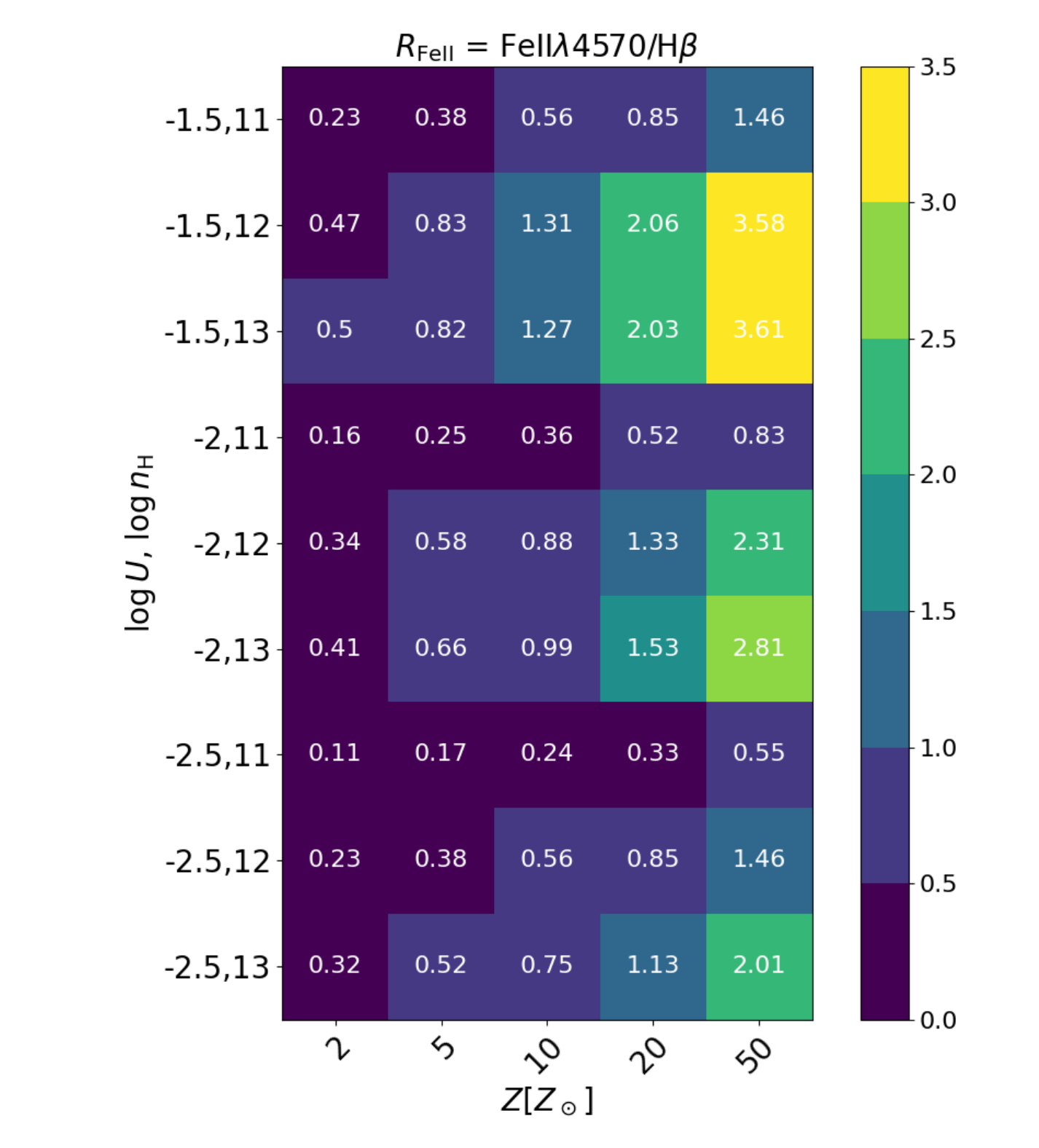}\hspace{-0.45cm}
\includegraphics[scale=0.25]{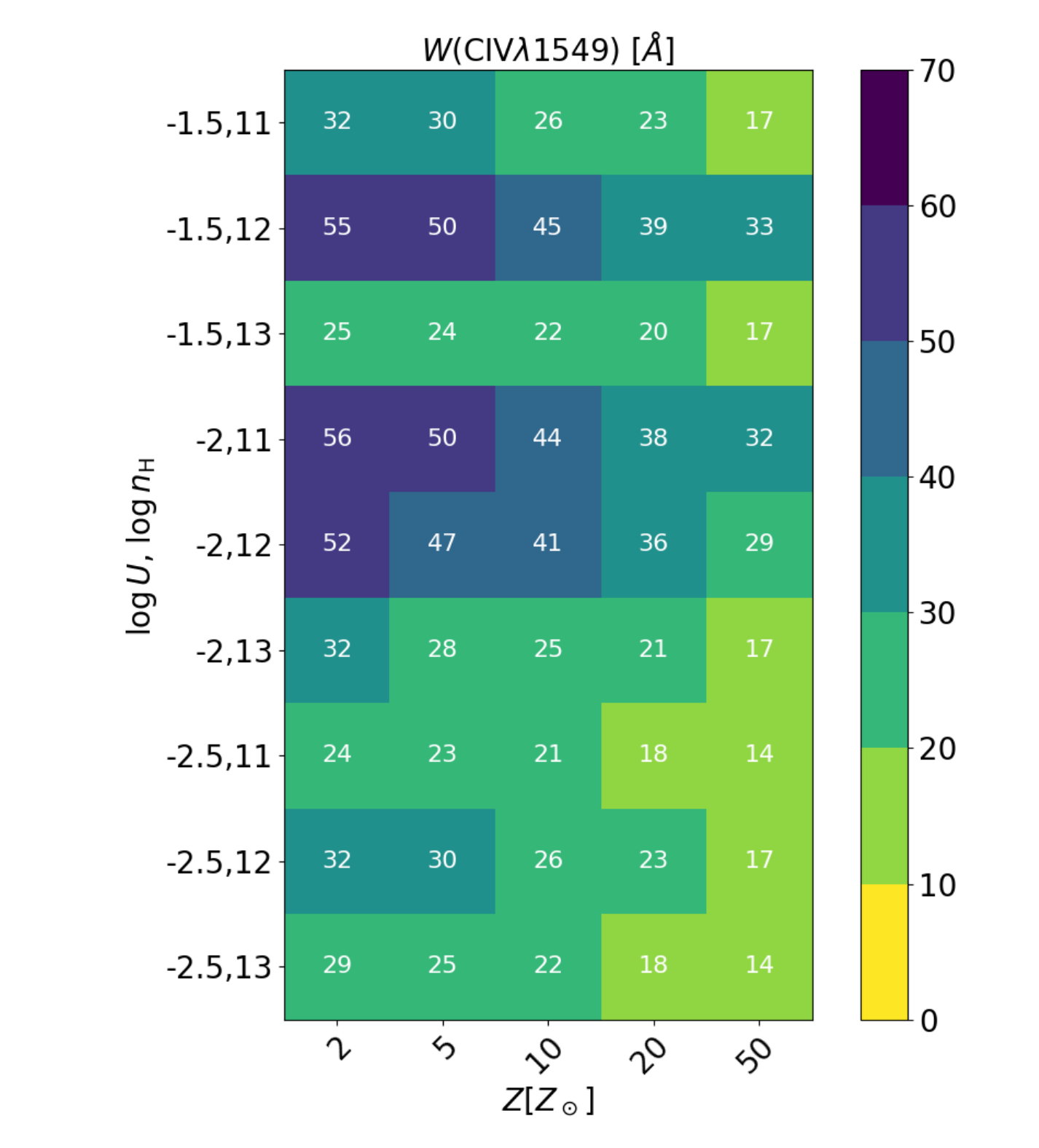}\hspace{-0.45cm}
\includegraphics[scale=0.25]{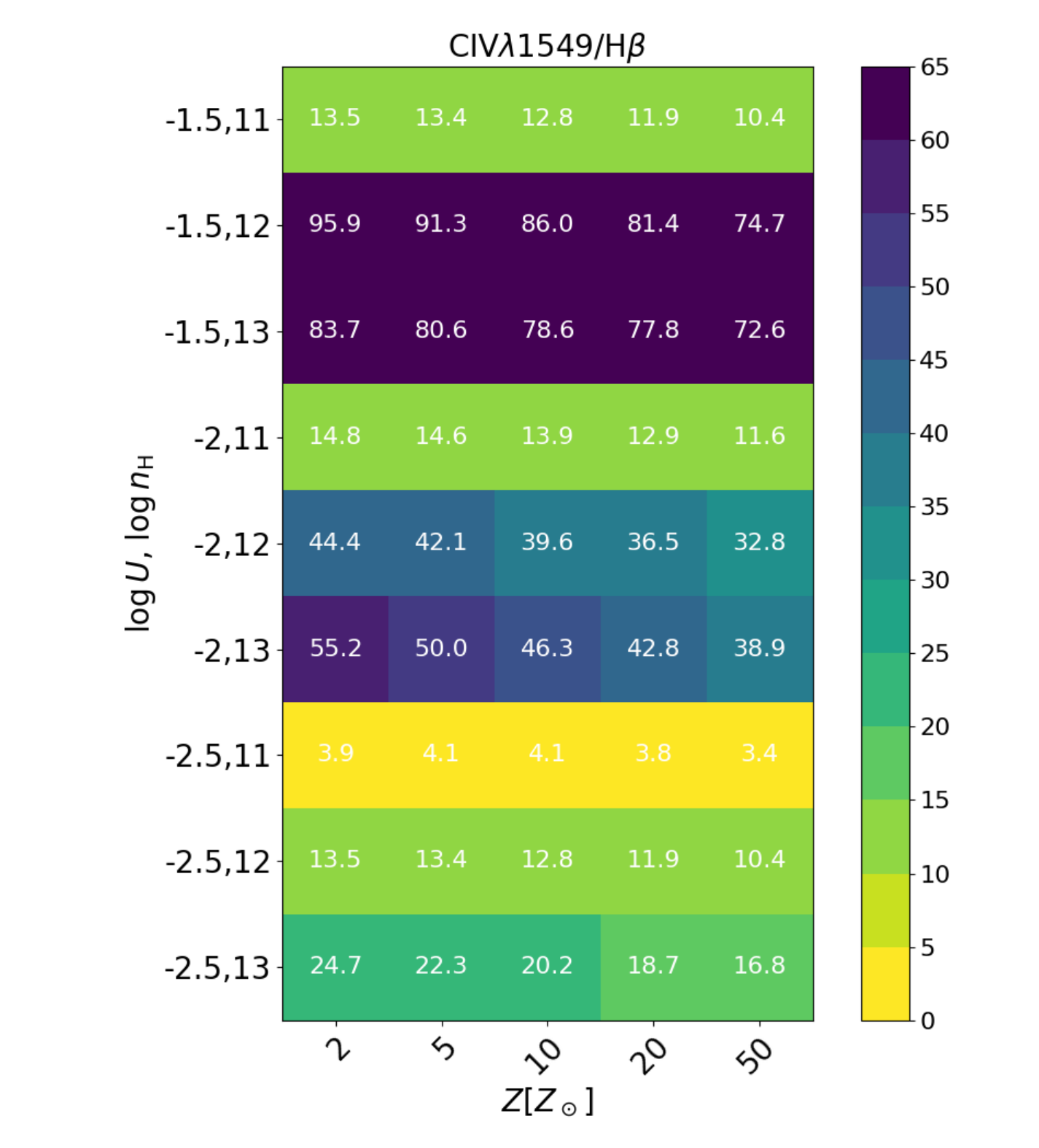}\\

\caption{{ Upper panels: }Predictions for \rfe\ (left), equivalent width of \civ\ (middle), intensity ratio \civ/\hb\, as a function of $\log U$ and $\log n_\mathrm{H}$ (ordinate) and metallicity $Z$\ in solar units. The $q2$ SED normalized to the flux at 5100 \AA\ has been used as an input to {\tt CLOUDY 23.01} photoionization simulations. { Bottom panels: Prediction for the three parameters as a function of $\log U$ and $\log n_\mathrm{H}$ (ordinate) and metallicity $Z$\ for the luminosity-scaled median SED, assuming an optical luminosity, and scaling the radii values to obtain the exact $\log U$ = -2.5,-2,-1.5, as in the previous cases. A radius $\log r \approx 17.013$ \ yields $\log U = -2$\ for $\log n_\mathrm{H} = 12$ [cm$^{-3}$].   }\label{fig:photoion} }
\end{figure*}

\subsection{The role of viewing angle in the SED appearance and in the xA classification}

A relevant question is: how are the accretion disks in the quasars of our sample oriented toward the observer? The present analysis is based on a very heterogeneous collection of measurements. There is no uniform calibration for the specific fluxes and no correction for the outflow component in the \hb\ profiles. However, we can count on the previous results of the virial equation in which the dependence on the viewing angle is considered explicitly \citep{2018negrete,dultzinetal20}. The differences between the luminosity or the distance moduli estimated from the virial luminosity equation and the standard $\Lambda$CDM\ cosmology can be entirely explained based on orientation effects. The sample of \citet{2018negrete} has been reanalyzed by \citet{donofrioetal24}, who show the distribution of viewing angles $\theta$\ (their Fig. 11). The $\theta$\ distribution peaks at $\theta \approx 17$ degrees, and is constrained within $\theta \lesssim 35$ degrees. The geometrically thick advection-dominated accretion flow (ADAF) models predict that the emission should have minimal dependence on the viewing angle up to $\theta \approx 30$ \citep[][c.f. \citealt{czerny19}]{wangzhou99}.       

However, the continuum observed by the distant observer is not necessarily the same as that seen by the emitting regions. This is particularly relevant in the case of an optically thick, geometrically thick disk \citep[e.g.,][]{2014wang, kubotadone19, panda2021, pandamarziani23}. Alternatively, the low-ionization emitting gas could be shielded by an inner outflow (\citealt{leighlymoore04, zappacostaetal20}; for a different context in terms of accretion power, see \citealt{mehdipouretal24}). These two scenarios are not mutually exclusive. The photoionization analysis in Sect. \ref{photoion} suggests that the degree of anisotropy should not be too large, as this could prevent us from reproducing the low-ionization emission spectrum (associated with \feii\ and Balmer lines) using the observed SED.

\subsection{Virial Luminosity and a tentative Hubble diagram}
\label{d}

Several authors \cite[e.g.,][]{2014marziani} explored the adaptation of the Faber-Jackson law for galaxies and quasars (see the extensive review by \citealt{donofrioetal24}).  This approach aims to establish a correlation between the luminosity and the line broadening of quasars, from a measure of line width, let it be the velocity dispersion $\sigma$ or the FWHM.  Here we focus on quasars with high Eddington ratios, which exhibit maximal radiative output relative to their black hole mass and are therefore expected to minimize their dispersion in Eddington ratio. The aforementioned  authors utilized spectroscopic data to analyze quasar emission lines, devising a method for using line width to measure cosmological distances independently from redshift. Fig. \ref{fig:dmodkg} shows the Hubble diagram for the { 155} quasars used in the computation of the SED. The agreement is remarkably good, as the averages over 8 bins are in agreement with the standard cosmology trend (the black line) within their rms. The dispersion of the average data are higher than in the case of careful data analyzed following a standardized procedure with consistent correction because of outflow component, \oiiiopt\ and \feii\ emission.   However, the agreement is even more remarkable right because the cosmological trend obtained by assuming a single value for the constant $\mathcal L_0 = 10^{44.897}$ erg  s$^{-1}$\ that was derived in \citet{2014marziani}, assuming $H_0 = 70$ km s$^{-1}$ Mpc$^{-1}$. 

The distribution indicates a majority of well-behaved points with a minority of $\approx 10$\ objects with larger $\mu$ than expected, by $\delta\mu \lesssim 5$. Upward-displaced sources might be due to an excess non-virial broadening: an outflow component that significantly increases the FWHM by a factor $\sim 1.5$, would imply a luminosity a factor $\approx 5$ larger. In turn, this implies a larger distance for the sources, if observed with the same optical flux: they would be placed a factor  2.25 farther out, implying  $\delta\mu \approx 1.76 $\ mag. However, the displacement seems to be too large for the strongest outliers,    $\delta\mu \approx 5 $\ mag, as the typical broadening of \hb\ is modest, $\approx 10$ \%\ \citep{2018negrete}.  We believe that these sources are due to poor data, as the FWHM data were retrieved from heterogeneous sources available in the literature. A check reveals that Mark 231 original data were not corrected for the outflow component strongly affecting the \hb\ profiles \citep[FWHM \hb\ passed from $\approx$6000 to $\approx$2000 \kms][]{sulenticetal06a}.  A few sources may show a very broad profile and are misclassified as xA while in reality, they are Seyfert 1.8 with some \feii\ and a weak, broad \hb. Median and average SEDs should be however not strongly affected by the possibility of a misclassification for a few sources.

\begin{figure}

\includegraphics[scale=0.4]{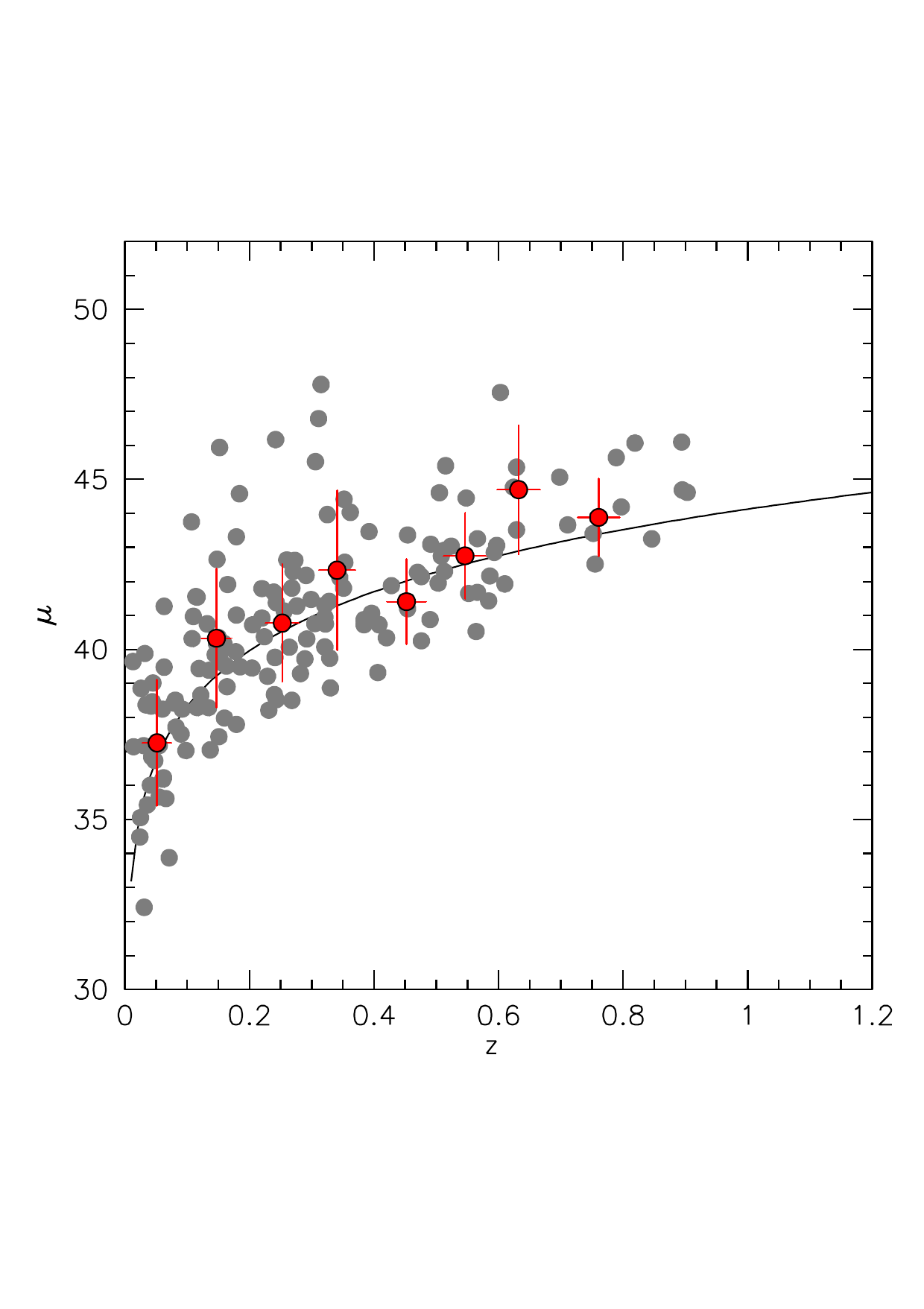} \\

\caption{A tentative Hubble diagram, distance modulus $\mu$ vs redshift $z$\ for the {155} quasars identified in this work and considered in the computation of the SED. Grey dots: individual data; red dots: averages over redshift bins of $\delta z = 0.1$. Error bars are sample standard deviations within each redshift bin. Note that the trend for $\Lambda$CDM cosmology is not fit to the data. The distance modulus was computed assuming $\log \mathcal L_0 \approx 44.897$  erg s$^{-1}$\ in the relation $L \approx{\mathcal L_0}  $FWHM$^4$.   \label{fig:dmodkg} }
\end{figure}  

\section{Conclusions}
\label{CONCLU}

For the first time, we modeled the Spectral Energy Distributions (SEDs) for extreme population A sources, also known as super-Eddington candidates \citep{duetal16a}. These sources were selected based on a criterion of \rfe $\ge 1$, indicating the highest prominence in singly-ionized iron emission along the quasar main sequence \citep{sulenticetal00a,2014marziani, pandaetal19}.

We mainly use observational data from radio (0.1GHz) through very hard X-rays (195 keV). The unobservable UV gap was interpolated by an accretion disc model including a Comptonized region associated with the X-ray emitting corona  \citep{kubotadone18, kubotadone19}. This SED will help us in the interpretation of the emission line spectra of xA sources at low redshift, and to discern their differences to the rest of population A   and the population B sources radiating at modest and low Eddington ratios respectively. The resulting SEDs are consistent with the ones derived in recent works {for high Eddington ratio sources} \citep{2020Feland}.

The xA SED { at energy $\gtrsim$ 1 Rydberg} is designed to serve as the ionizing continuum input parameter for {\tt CLOUDY} photoionization simulations, enabling more accurate predictions of line intensities and derived parameters in the context of xA sources. Accordingly, Appendix \ref{digitalSED} provides the digital version of the xA SEDs for any astrophysical application.

 {The 
 FIR and radio emission of the RQ of xA sources are consistently explained by dominance of star formation, consistently with past and ongoing work \citep[][Gendron-Marsolais et al., in preparation]{2019ganci,yue_novel_2024,yue_novel_2025}}. We also identify a minority of {  sources which shows an   IR excess   linked to a strongest \feii\ emission  and highe{st} \rfe\ parameter. } This connection highlights that the most extreme xA sources could show stronger star formation concomitant with a high accretion rate \citep{marzianietal21}. The source with the highest IR excess, Mark 231, is in this subsample and is rightly considered the nearest super-Eddington quasar \citep{vanderwerfetal10,rupkeVeilleux11,feruglioetal15} with extremely strong \feii\ and powerful multi-scale outflows, in part associated with its AGN and in part with enhanced star formation.

\section*{Acknowledgements}

{ We thank the anonymous reviewer for their helpful comments that helped improve the content of the manuscript.} DD acknowledges support from PAPIIT UNAM through grant IN111422. SP is supported by the International Gemini Observatory, a program of NSF NOIRLab, which is managed by the Association of Universities for Research in Astronomy (AURA) under a cooperative agreement with the U.S. National Science Foundation, on behalf of the Gemini partnership of Argentina, Brazil, Canada, Chile, the Republic of Korea, and the United States of America. SP acknowledges the financial support of the Conselho Nacional de Desenvolvimento Científico e Tecnol\'ogico (CNPq) Fellowship 301628/2024-6. 

{This research has made use of the NASA/IPAC Extragalactic Database (NED),
which is operated by the Jet Propulsion Laboratory, California Institute of Technology,
under contract with the National Aeronautics and Space Administration.}

\section*{Data Availability}

The sample used in this article is reported in table \ref{tab:150_basic}. The photometry was retrieved from public databases, NASA Extragalactic Database  (NED, \url{https://ned.ipac.caltech.edu/}) and VizieR (\url{https://vizier.cds.unistra.fr/}). The bibliographical references of other data are specified in the article on Section \ref{sec:sample_retrieve}. The resulting SEDs from this study are appended to this paper as online supplementary material.



\bibliographystyle{mnras}

\begin{thebibliography}{}
\makeatletter
\relax
\def\mn@urlcharsother{\let\do\@makeother \do\$\do\&\do\#\do\^\do\_\do\%\do\~}
\def\mn@doi{\begingroup\mn@urlcharsother \@ifnextchar [ {\mn@doi@}
  {\mn@doi@[]}}
\def\mn@doi@[#1]#2{\def\@tempa{#1}\ifx\@tempa\@empty \href
  {http://dx.doi.org/#2} {doi:#2}\else \href {http://dx.doi.org/#2} {#1}\fi
  \endgroup}
\def\mn@eprint#1#2{\mn@eprint@#1:#2::\@nil}
\def\mn@eprint@arXiv#1{\href {http://arxiv.org/abs/#1} {{\tt arXiv:#1}}}
\def\mn@eprint@dblp#1{\href {http://dblp.uni-trier.de/rec/bibtex/#1.xml}
  {dblp:#1}}
\def\mn@eprint@#1:#2:#3:#4\@nil{\def\@tempa {#1}\def\@tempb {#2}\def\@tempc
  {#3}\ifx \@tempc \@empty \let \@tempc \@tempb \let \@tempb \@tempa \fi \ifx
  \@tempb \@empty \def\@tempb {arXiv}\fi \@ifundefined
  {mn@eprint@\@tempb}{\@tempb:\@tempc}{\expandafter \expandafter \csname
  mn@eprint@\@tempb\endcsname \expandafter{\@tempc}}}

\bibitem[\protect\citeauthoryear{{Akylas} \& {Georgantopoulos}}{{Akylas} \&
  {Georgantopoulos}}{2021}]{2021Akylas}
{Akylas} A.,  {Georgantopoulos} I.,  2021, \mn@doi [\aap]
  {10.1051/0004-6361/202141186}, \href
  {https://ui.adsabs.harvard.edu/abs/2021A&A...655A..60A} {655, A60}

\bibitem[\protect\citeauthoryear{{Bensch}, {del Olmo}, {Sulentic}, {Perea}  \&
  {Marziani}}{{Bensch} et~al.}{2015}]{benschetal15}
{Bensch} K.,  {del Olmo} A.,  {Sulentic} J.,  {Perea} J.,   {Marziani} P.,
  2015, \mn@doi [Journal of Astrophysics and Astronomy]
  {10.1007/s12036-015-9355-8}, \href
  {http://adsabs.harvard.edu/abs/2015JApA...36..467B} {36, 467}

\bibitem[\protect\citeauthoryear{{Bischetti} et~al.,}{{Bischetti}
  et~al.}{2017}]{bischettietal17}
{Bischetti} M.,  et~al., 2017, \mn@doi [\aap] {10.1051/0004-6361/201629301},
  \href {http://adsabs.harvard.edu/abs/2017A%26A...598A.122B} {598, A122}

\bibitem[\protect\citeauthoryear{{Boksenberg}, {Carswell}, {Allen}, {Fosbury},
  {Penston}  \& {Sargent}}{{Boksenberg} et~al.}{1977}]{boksenbergetal77}
{Boksenberg} A.,  {Carswell} R.~F.,  {Allen} D.~A.,  {Fosbury} R.~A.~E.,
  {Penston} M.~V.,   {Sargent} W.~L.~W.,  1977, \mn@doi [\mnras]
  {10.1093/mnras/178.3.451}, \href
  {https://ui.adsabs.harvard.edu/abs/1977MNRAS.178..451B} {178, 451}

\bibitem[\protect\citeauthoryear{{Boller}, {Brandt}  \& {Fink}}{{Boller}
  et~al.}{1996}]{1996boller}
{Boller} T.,  {Brandt} W.~N.,   {Fink} H.,  1996, \mn@doi [\aap]
  {10.48550/arXiv.astro-ph/9504093}, \href
  {https://ui.adsabs.harvard.edu/abs/1996A&A...305...53B} {305, 53}

\bibitem[\protect\citeauthoryear{{Boller}, {Freyberg}, {Tr{\"u}mper}, {Haberl},
  {Voges}  \& {Nandra}}{{Boller} et~al.}{2016}]{2016Boller}
{Boller} T.,  {Freyberg} M.~J.,  {Tr{\"u}mper} J.,  {Haberl} F.,  {Voges} W.,
  {Nandra} K.,  2016, \mn@doi [\aap] {10.1051/0004-6361/201525648}, \href
  {https://ui.adsabs.harvard.edu/abs/2016A&A...588A.103B} {588, A103}

\bibitem[\protect\citeauthoryear{{Bonzini} et~al.,}{{Bonzini}
  et~al.}{2015}]{bonzinietal15}
{Bonzini} M.,  et~al., 2015, \mn@doi [\mnras] {10.1093/mnras/stv1675}, \href
  {http://adsabs.harvard.edu/abs/2015MNRAS.453.1079B} {453, 1079}

\bibitem[\protect\citeauthoryear{{Boroson} \& {Green}}{{Boroson} \&
  {Green}}{1992}]{borosongreen92}
{Boroson} T.~A.,  {Green} R.~F.,  1992, \mn@doi [\apjs] {10.1086/191661}, \href
  {https://ui.adsabs.harvard.edu/abs/1992ApJS...80..109B} {80, 109}

\bibitem[\protect\citeauthoryear{{Bressan}, {Silva}  \& {Granato}}{{Bressan}
  et~al.}{2002}]{bressanetal02}
{Bressan} A.,  {Silva} L.,   {Granato} G.~L.,  2002, \mn@doi [\aap]
  {10.1051/0004-6361:20020960}, \href
  {https://ui.adsabs.harvard.edu/abs/2002A&A...392..377B} {392, 377}

\bibitem[\protect\citeauthoryear{{Brightman} et~al.,}{{Brightman}
  et~al.}{2013}]{brightmanetal13}
{Brightman} M.,  et~al., 2013, \mn@doi [\mnras] {10.1093/mnras/stt920}, \href
  {https://ui.adsabs.harvard.edu/abs/2013MNRAS.433.2485B} {433, 2485}

\bibitem[\protect\citeauthoryear{{Brown}, {Duncan}, {Landt}, {Kirk}, {Ricci},
  {Kamraj}, {Salvato}  \& {Ananna}}{{Brown} et~al.}{2019}]{brownetal19}
{Brown} M.~J.~I.,  {Duncan} K.~J.,  {Landt} H.,  {Kirk} M.,  {Ricci} C.,
  {Kamraj} N.,  {Salvato} M.,   {Ananna} T.,  2019, \mn@doi [\mnras]
  {10.1093/mnras/stz2324}, \href
  {https://ui.adsabs.harvard.edu/abs/2019MNRAS.489.3351B} {489, 3351}

\bibitem[\protect\citeauthoryear{{Caccianiga} et~al.,}{{Caccianiga}
  et~al.}{2015}]{caccianigaetal15}
{Caccianiga} A.,  et~al., 2015, \mn@doi [\mnras] {10.1093/mnras/stv939}, \href
  {https://ui.adsabs.harvard.edu/abs/2015MNRAS.451.1795C} {451, 1795}

\bibitem[\protect\citeauthoryear{{Cao}, {Nampalliwar}, {Bambi}, {Dauser}  \&
  {Garc{\'\i}a}}{{Cao} et~al.}{2018}]{2018Cao}
{Cao} Z.,  {Nampalliwar} S.,  {Bambi} C.,  {Dauser} T.,   {Garc{\'\i}a} J.~A.,
  2018, \mn@doi [\prl] {10.1103/PhysRevLett.120.051101}, \href
  {https://ui.adsabs.harvard.edu/abs/2018PhRvL.120e1101C} {120, 051101}

\bibitem[\protect\citeauthoryear{{Cardelli}, {Clayton}  \& {Mathis}}{{Cardelli}
  et~al.}{1989}]{1989Cardelli}
{Cardelli} J.~A.,  {Clayton} G.~C.,   {Mathis} J.~S.,  1989, \mn@doi [\apj]
  {10.1086/167900}, \href
  {https://ui.adsabs.harvard.edu/abs/1989ApJ...345..245C} {345, 245}

\bibitem[\protect\citeauthoryear{{Chatzikos} et~al.,}{{Chatzikos}
  et~al.}{2023}]{2023Chatzikos}
{Chatzikos} M.,  et~al., 2023, \mn@doi [\rmxaa]
  {10.22201/ia.01851101p.2023.59.02.12}, \href
  {https://ui.adsabs.harvard.edu/abs/2023RMxAA..59..327C} {59, 327}

\bibitem[\protect\citeauthoryear{{Coatman}, {Hewett}, {Banerji}  \&
  {Richards}}{{Coatman} et~al.}{2016}]{coatmanetal16}
{Coatman} L.,  {Hewett} P.~C.,  {Banerji} M.,   {Richards} G.~T.,  2016,
  \mn@doi [\mnras] {10.1093/mnras/stw1360}, \href
  {http://adsabs.harvard.edu/abs/2016MNRAS.461..647C} {461, 647}

\bibitem[\protect\citeauthoryear{{Coatman}, {Hewett}, {Banerji}, {Richards},
  {Hennawi}  \& {Xavier Prochaska}}{{Coatman} et~al.}{2019}]{coatmanetal19}
{Coatman} L.,  {Hewett} P.~C.,  {Banerji} M.,  {Richards} G.~T.,  {Hennawi}
  J.~F.,   {Xavier Prochaska} J.,  2019, \mn@doi [\mnras]
  {10.1093/mnras/stz1167}, \href
  {https://ui.adsabs.harvard.edu/abs/2019MNRAS.tmp.1122C} {p.~1122}

\bibitem[\protect\citeauthoryear{{Condon}, {Kellermann}, {Kimball},
  {Ivezi{\'c}}  \& {Perley}}{{Condon} et~al.}{2013}]{condonetal13}
{Condon} J.~J.,  {Kellermann} K.~I.,  {Kimball} A.~E.,  {Ivezi{\'c}} {\v{Z}}.,
   {Perley} R.~A.,  2013, \mn@doi [\apj] {10.1088/0004-637X/768/1/37}, \href
  {https://ui.adsabs.harvard.edu/abs/2013ApJ...768...37C} {768, 37}

\bibitem[\protect\citeauthoryear{{Czerny}}{{Czerny}}{2019}]{czerny19}
{Czerny} B.,  2019, \mn@doi [Universe] {10.3390/universe5050131}, \href
  {https://ui.adsabs.harvard.edu/abs/2019Univ....5..131C} {5, 131}

\bibitem[\protect\citeauthoryear{{Czerny} \& {Elvis}}{{Czerny} \&
  {Elvis}}{1987}]{czerny_elvis_1987}
{Czerny} B.,  {Elvis} M.,  1987, \mn@doi [\apj] {10.1086/165630}, \href
  {https://ui.adsabs.harvard.edu/abs/1987ApJ...321..305C} {321, 305}

\bibitem[\protect\citeauthoryear{{D'Ammando}}{{D'Ammando}}{2019}]{dammando19}
{D'Ammando} F.,  2019, \mn@doi [Galaxies] {10.3390/galaxies7040087}, \href
  {https://ui.adsabs.harvard.edu/abs/2019Galax...7...87D} {7, 87}

\bibitem[\protect\citeauthoryear{{D'Ammando}}{{D'Ammando}}{2020}]{dammando20}
{D'Ammando} F.,  2020, \mn@doi [\mnras] {10.1093/mnras/staa1580}, \href
  {https://ui.adsabs.harvard.edu/abs/2020MNRAS.496.2213D} {496, 2213}

\bibitem[\protect\citeauthoryear{{D'Onofrio} \& {Marziani}}{{D'Onofrio} \&
  {Marziani}}{2018}]{2018Donofrio}
{D'Onofrio} M.,  {Marziani} P.,  2018, \mn@doi [Frontiers in Astronomy and
  Space Sciences] {10.3389/fspas.2018.00031}, \href
  {https://ui.adsabs.harvard.edu/abs/2018FrASS...5...31D} {5, 31}

\bibitem[\protect\citeauthoryear{D'Onofrio, Marziani, Chiosi  \&
  Negrete}{D'Onofrio et~al.}{2024}]{donofrioetal24}
D'Onofrio M.,  Marziani P.,  Chiosi C.,   Negrete C.~A.,  2024, \mn@doi
  [Universe] {10.3390/universe10060254}, 10

\bibitem[\protect\citeauthoryear{{Danehkar}}{{Danehkar}}{2024}]{danekhar24}
{Danehkar} A.,  2024, \mn@doi [Frontiers in Astronomy and Space Sciences]
  {10.3389/fspas.2024.1479301}, \href
  {https://ui.adsabs.harvard.edu/abs/2024FrASS..1179301D} {11, 1479301}

\bibitem[\protect\citeauthoryear{{Deconto-Machado}, {del Olmo Orozco},
  {Marziani}, {Perea}  \& {Stirpe}}{{Deconto-Machado}
  et~al.}{2023}]{deconto-machadoetal23}
{Deconto-Machado} A.,  {del Olmo Orozco} A.,  {Marziani} P.,  {Perea} J.,
  {Stirpe} G.~M.,  2023, \mn@doi [\aap] {10.1051/0004-6361/202243801}, \href
  {https://ui.adsabs.harvard.edu/abs/2023A&A...669A..83D} {669, A83}

\bibitem[\protect\citeauthoryear{{Deconto-Machado}, {del Olmo}  \&
  {Marziani}}{{Deconto-Machado} et~al.}{2024}]{deconto-machadoetal24}
{Deconto-Machado} A.,  {del Olmo} A.,   {Marziani} P.,  2024, \mn@doi [\aap]
  {10.1051/0004-6361/202449976}, \href
  {https://ui.adsabs.harvard.edu/abs/2024A&A...691A..15D} {691, A15}

\bibitem[\protect\citeauthoryear{{Done}, {Davis}, {Jin}, {Blaes}  \&
  {Ward}}{{Done} et~al.}{2012}]{2012Done}
{Done} C.,  {Davis} S.~W.,  {Jin} C.,  {Blaes} O.,   {Ward} M.,  2012, \mn@doi
  [\mnras] {10.1111/j.1365-2966.2011.19779.x}, \href
  {https://ui.adsabs.harvard.edu/abs/2012MNRAS.420.1848D} {420, 1848}

\bibitem[\protect\citeauthoryear{{Du} \& {Wang}}{{Du} \&
  {Wang}}{2019}]{2019Du&Wang}
{Du} P.,  {Wang} J.-M.,  2019, \mn@doi [\apj] {10.3847/1538-4357/ab4908}, \href
  {https://ui.adsabs.harvard.edu/abs/2019ApJ...886...42D} {886, 42}

\bibitem[\protect\citeauthoryear{{Du} et~al.,}{{Du} et~al.}{2014}]{duetal14}
{Du} P.,  et~al., 2014, \mn@doi [\apj] {10.1088/0004-637X/782/1/45}, \href
  {https://ui.adsabs.harvard.edu/abs/2014ApJ...782...45D} {782, 45}

\bibitem[\protect\citeauthoryear{{Du} et~al.,}{{Du} et~al.}{2015}]{duetal15}
{Du} P.,  et~al., 2015, \mn@doi [\apj] {10.1088/0004-637X/806/1/22}, \href
  {https://ui.adsabs.harvard.edu/abs/2015ApJ...806...22D} {806, 22}

\bibitem[\protect\citeauthoryear{{Du}, {Wang}, {Hu}, {Ho}, {Li}  \& {Bai}}{{Du}
  et~al.}{2016a}]{duetal16a}
{Du} P.,  {Wang} J.-M.,  {Hu} C.,  {Ho} L.~C.,  {Li} Y.-R.,   {Bai} J.-M.,
  2016a, \mn@doi [\apjl] {10.3847/2041-8205/818/1/L14}, \href
  {http://adsabs.harvard.edu/abs/2016ApJ...818L..14D} {818, L14}

\bibitem[\protect\citeauthoryear{{Du} et~al.,}{{Du} et~al.}{2016b}]{duetal16}
{Du} P.,  et~al., 2016b, \mn@doi [\apj] {10.3847/0004-637X/820/1/27}, \href
  {https://ui.adsabs.harvard.edu/abs/2016ApJ...820...27D} {820, 27}

\bibitem[\protect\citeauthoryear{{Du} et~al.,}{{Du} et~al.}{2018}]{duetal18}
{Du} P.,  et~al., 2018, \mn@doi [\apj] {10.3847/1538-4357/aaae6b}, \href
  {https://ui.adsabs.harvard.edu/abs/2018ApJ...856....6D} {856, 6}

\bibitem[\protect\citeauthoryear{{Dultzin} et~al.,}{{Dultzin}
  et~al.}{2020}]{dultzinetal20}
{Dultzin} D.,  et~al., 2020, \mn@doi [Frontiers in Astronomy and Space
  Sciences] {10.3389/fspas.2019.00080}, \href
  {https://ui.adsabs.harvard.edu/abs/2020FrASS...6...80M} {6, 80}

\bibitem[\protect\citeauthoryear{{Elvis} et~al.,}{{Elvis}
  et~al.}{1994}]{1994elvis}
{Elvis} M.,  et~al., 1994, \mn@doi [\apjs] {10.1086/192093}, \href
  {https://ui.adsabs.harvard.edu/abs/1994ApJS...95....1E} {95, 1}

\bibitem[\protect\citeauthoryear{{Esparza-Arredondo}, {Gonzalez-Mart{\'\i}n},
  {Dultzin}, {Masegosa}, {Ramos-Almeida}, {Garc{\'\i}a-Bernete}, {Fritz}  \&
  {Osorio-Clavijo}}{{Esparza-Arredondo}
  et~al.}{2021}]{esparza-arrendondoetal21}
{Esparza-Arredondo} D.,  {Gonzalez-Mart{\'\i}n} O.,  {Dultzin} D.,  {Masegosa}
  J.,  {Ramos-Almeida} C.,  {Garc{\'\i}a-Bernete} I.,  {Fritz} J.,
  {Osorio-Clavijo} N.,  2021, \mn@doi [\aap] {10.1051/0004-6361/202040043},
  \href {https://ui.adsabs.harvard.edu/abs/2021A&A...651A..91E} {651, A91}

\bibitem[\protect\citeauthoryear{{Evans} et~al.,}{{Evans}
  et~al.}{2020}]{2020Evans}
{Evans} P.~A.,  et~al., 2020, \mn@doi [\apjs] {10.3847/1538-4365/ab7db9}, \href
  {https://ui.adsabs.harvard.edu/abs/2020ApJS..247...54E} {247, 54}

\bibitem[\protect\citeauthoryear{{Fanali}, {Caccianiga}, {Severgnini}, {Della
  Ceca}, {Marchese}, {Carrera}, {Corral}  \& {Mateos}}{{Fanali}
  et~al.}{2013}]{fanalietal13}
{Fanali} R.,  {Caccianiga} A.,  {Severgnini} P.,  {Della Ceca} R.,  {Marchese}
  E.,  {Carrera} F.~J.,  {Corral} A.,   {Mateos} S.,  2013, \mn@doi [\mnras]
  {10.1093/mnras/stt757}, \href
  {https://ui.adsabs.harvard.edu/abs/2013MNRAS.433..648F} {433, 648}

\bibitem[\protect\citeauthoryear{{Ferland}, {Done}, {Jin}, {Landt}  \&
  {Ward}}{{Ferland} et~al.}{2020}]{2020Feland}
{Ferland} G.~J.,  {Done} C.,  {Jin} C.,  {Landt} H.,   {Ward} M.~J.,  2020,
  \mn@doi [\mnras] {10.1093/mnras/staa1207}, \href
  {https://ui.adsabs.harvard.edu/abs/2020MNRAS.494.5917F} {494, 5917}

\bibitem[\protect\citeauthoryear{{Feruglio} et~al.,}{{Feruglio}
  et~al.}{2015}]{feruglioetal15}
{Feruglio} C.,  et~al., 2015, \mn@doi [\aap] {10.1051/0004-6361/201526020},
  \href {https://ui.adsabs.harvard.edu/abs/2015A&A...583A..99F} {583, A99}

\bibitem[\protect\citeauthoryear{{Floris}, {Marziani}, {Panda}, {Sniegowska},
  {D'Onofrio}, {Deconto-Machado}, {Del Olmo}  \& {Czerny}}{{Floris}
  et~al.}{2024}]{florisetal24}
{Floris} A.,  {Marziani} P.,  {Panda} S.,  {Sniegowska} M.,  {D'Onofrio} M.,
  {Deconto-Machado} A.,  {Del Olmo} A.,   {Czerny} B.,  2024, \mn@doi [arXiv
  e-prints] {10.48550/arXiv.2405.04456}, \href
  {https://ui.adsabs.harvard.edu/abs/2024arXiv240504456F} {p. arXiv:2405.04456}

\bibitem[\protect\citeauthoryear{{Foschini} et~al.,}{{Foschini}
  et~al.}{2021}]{foschinietal21}
{Foschini} L.,  et~al., 2021, \mn@doi [Universe] {10.3390/universe7100372},
  \href {https://ui.adsabs.harvard.edu/abs/2021Univ....7..372F} {7, 372}

\bibitem[\protect\citeauthoryear{{Ganci}, {Marziani}, {D'Onofrio}, {del Olmo},
  {Bon}, {Bon}  \& {Negrete}}{{Ganci} et~al.}{2019}]{2019ganci}
{Ganci} V.,  {Marziani} P.,  {D'Onofrio} M.,  {del Olmo} A.,  {Bon} E.,  {Bon}
  N.,   {Negrete} C.~A.,  2019, \mn@doi [\aap] {10.1051/0004-6361/201936270},
  \href {https://ui.adsabs.harvard.edu/abs/2019A&A...630A.110G} {630, A110}

\bibitem[\protect\citeauthoryear{{Garnica}, {Negrete}, {Marziani}, {Dultzin},
  {{\'S}niegowska}  \& {Panda}}{{Garnica} et~al.}{2022}]{2022garnica}
{Garnica} K.,  {Negrete} C.~A.,  {Marziani} P.,  {Dultzin} D.,
  {{\'S}niegowska} M.,   {Panda} S.,  2022, \mn@doi [\aap]
  {10.1051/0004-6361/202142837}, \href
  {https://ui.adsabs.harvard.edu/abs/2022A&A...667A.105G} {667, A105}

\bibitem[\protect\citeauthoryear{{Grupe}}{{Grupe}}{2004}]{grupe04}
{Grupe} D.,  2004, \mn@doi [\aj] {10.1086/382516}, \href
  {http://adsabs.harvard.edu/abs/2004AJ....127.1799G} {127, 1799}

\bibitem[\protect\citeauthoryear{{Grupe}, {Beuermann}, {Thomas}, {Mannheim}  \&
  {Fink}}{{Grupe} et~al.}{1998}]{grupeetal98}
{Grupe} D.,  {Beuermann} K.,  {Thomas} H.~C.,  {Mannheim} K.,   {Fink} H.~H.,
  1998, \aap, \href {https://ui.adsabs.harvard.edu/abs/1998A&A...330...25G}
  {330, 25}

\bibitem[\protect\citeauthoryear{{Grupe}, {Wills}, {Leighly}  \&
  {Meusinger}}{{Grupe} et~al.}{2004}]{2004grupe}
{Grupe} D.,  {Wills} B.~J.,  {Leighly} K.~M.,   {Meusinger} H.,  2004, \mn@doi
  [\aj] {10.1086/380233}, \href
  {https://ui.adsabs.harvard.edu/abs/2004AJ....127..156G} {127, 156}

\bibitem[\protect\citeauthoryear{{Grupe}, {Komossa}, {Leighly}  \&
  {Page}}{{Grupe} et~al.}{2010}]{grupeetal10}
{Grupe} D.,  {Komossa} S.,  {Leighly} K.~M.,   {Page} K.~L.,  2010, \mn@doi
  [\apjs] {10.1088/0067-0049/187/1/64}, \href
  {http://adsabs.harvard.edu/abs/2010ApJS..187...64G} {187, 64}

\bibitem[\protect\citeauthoryear{{Gunasekera}, {van Hoof}, {Chatzikos}  \&
  {Ferland}}{{Gunasekera} et~al.}{2023}]{gunasekeraetal23}
{Gunasekera} C.~M.,  {van Hoof} P. A.~M.,  {Chatzikos} M.,   {Ferland} G.~J.,
  2023, \mn@doi [Research Notes of the American Astronomical Society]
  {10.3847/2515-5172/ad0e75}, \href
  {https://ui.adsabs.harvard.edu/abs/2023RNAAS...7..246G} {7, 246}

\bibitem[\protect\citeauthoryear{{Haardt} \& {Maraschi}}{{Haardt} \&
  {Maraschi}}{1991}]{haardtmaraschi91}
{Haardt} F.,  {Maraschi} L.,  1991, \mn@doi [\apjl] {10.1086/186171}, \href
  {https://ui.adsabs.harvard.edu/abs/1991ApJ...380L..51H} {380, L51}

\bibitem[\protect\citeauthoryear{{Haardt} \& {Maraschi}}{{Haardt} \&
  {Maraschi}}{1993}]{haardtmaraschi93}
{Haardt} F.,  {Maraschi} L.,  1993, \mn@doi [\apj] {10.1086/173020}, \href
  {https://ui.adsabs.harvard.edu/abs/1993ApJ...413..507H} {413, 507}

\bibitem[\protect\citeauthoryear{{Hamann} \& {Ferland}}{{Hamann} \&
  {Ferland}}{1993}]{hamannferland93}
{Hamann} F.,  {Ferland} G.,  1993, \mn@doi [\apj] {10.1086/173366}, \href
  {http://adsabs.harvard.edu/abs/1993ApJ...418...11H} {418, 11}

\bibitem[\protect\citeauthoryear{{Harrison} et~al.,}{{Harrison}
  et~al.}{2013}]{harrisonetal13}
{Harrison} F.~A.,  et~al., 2013, \mn@doi [\apj] {10.1088/0004-637X/770/2/103},
  \href {https://ui.adsabs.harvard.edu/abs/2013ApJ...770..103H} {770, 103}

\bibitem[\protect\citeauthoryear{{Huang}, {Luo}, {Du}, {Hu}, {Wang}  \&
  {Li}}{{Huang} et~al.}{2020}]{huangetal20}
{Huang} J.,  {Luo} B.,  {Du} P.,  {Hu} C.,  {Wang} J.-M.,   {Li} Y.-J.,  2020,
  \mn@doi [\apj] {10.3847/1538-4357/ab9019}, \href
  {https://ui.adsabs.harvard.edu/abs/2020ApJ...895..114H} {895, 114}

\bibitem[\protect\citeauthoryear{{Jin}, {Ward}, {Done}  \& {Gelbord}}{{Jin}
  et~al.}{2012}]{2012Jin}
{Jin} C.,  {Ward} M.,  {Done} C.,   {Gelbord} J.,  2012, \mn@doi [\mnras]
  {10.1111/j.1365-2966.2011.19805.x}, \href
  {https://ui.adsabs.harvard.edu/abs/2012MNRAS.420.1825J} {420, 1825}

\bibitem[\protect\citeauthoryear{{Kara}, {Miller}, {Reynolds}  \& {Dai}}{{Kara}
  et~al.}{2016}]{karaetal16}
{Kara} E.,  {Miller} J.~M.,  {Reynolds} C.,   {Dai} L.,  2016, \mn@doi [\nat]
  {10.1038/nature18007}, \href
  {https://ui.adsabs.harvard.edu/abs/2016Natur.535..388K} {535, 388}

\bibitem[\protect\citeauthoryear{{Kauffmann}, {Mill{\'a}n-Irigoyen}, {Crowther}
   \& {Maraston}}{{Kauffmann} et~al.}{2024}]{kauffmannetal24}
{Kauffmann} G.,  {Mill{\'a}n-Irigoyen} I.,  {Crowther} P.,   {Maraston} C.,
  2024, \mn@doi [\mnras] {10.1093/mnras/stad3096}, \href
  {https://ui.adsabs.harvard.edu/abs/2024MNRAS.527..150K} {527, 150}

\bibitem[\protect\citeauthoryear{{Kellermann}, {Sramek}, {Schmidt}, {Shaffer}
  \& {Green}}{{Kellermann} et~al.}{1989}]{kellermannetal89}
{Kellermann} K.~I.,  {Sramek} R.,  {Schmidt} M.,  {Shaffer} D.~B.,   {Green}
  R.,  1989, \mn@doi [\aj] {10.1086/115207}, \href
  {http://adsabs.harvard.edu/abs/1989AJ.....98.1195K} {98, 1195}

\bibitem[\protect\citeauthoryear{{King} \& {Pounds}}{{King} \&
  {Pounds}}{2015}]{2015KingPunds}
{King} A.,  {Pounds} K.,  2015, \mn@doi [\araa]
  {10.1146/annurev-astro-082214-122316}, \href
  {https://ui.adsabs.harvard.edu/abs/2015ARA&A..53..115K} {53, 115}

\bibitem[\protect\citeauthoryear{{Kubota} \& {Done}}{{Kubota} \&
  {Done}}{2018}]{kubotadone18}
{Kubota} A.,  {Done} C.,  2018, \mn@doi [\mnras] {10.1093/mnras/sty1890}, \href
  {https://ui.adsabs.harvard.edu/abs/2018MNRAS.480.1247K} {480, 1247}

\bibitem[\protect\citeauthoryear{{Kubota} \& {Done}}{{Kubota} \&
  {Done}}{2019}]{kubotadone19}
{Kubota} A.,  {Done} C.,  2019, \mn@doi [\mnras] {10.1093/mnras/stz2140}, \href
  {https://ui.adsabs.harvard.edu/abs/2019MNRAS.489..524K} {489, 524}

\bibitem[\protect\citeauthoryear{{Lanzuisi} et~al.,}{{Lanzuisi}
  et~al.}{2016}]{lanzuisietal16}
{Lanzuisi} G.,  et~al., 2016, \mn@doi [\aap] {10.1051/0004-6361/201628325},
  \href {https://ui.adsabs.harvard.edu/abs/2016A&A...590A..77L} {590, A77}

\bibitem[\protect\citeauthoryear{{Laor}, {Fiore}, {Elvis}, {Wilkes}  \&
  {McDowell}}{{Laor} et~al.}{1997}]{laoretal97b}
{Laor} A.,  {Fiore} F.,  {Elvis} M.,  {Wilkes} B.~J.,   {McDowell} J.~C.,
  1997, \mn@doi [ApJ] {10.1086/303696}, \href
  {http://adsabs.harvard.edu/abs/1997ApJ...477...93L} {477, 93}

\bibitem[\protect\citeauthoryear{{Larson} et~al.,}{{Larson}
  et~al.}{2023}]{larsonetal23}
{Larson} R.~L.,  et~al., 2023, \mn@doi [\apjl] {10.3847/2041-8213/ace619},
  \href {https://ui.adsabs.harvard.edu/abs/2023ApJ...953L..29L} {953, L29}

\bibitem[\protect\citeauthoryear{{Laurenti} et~al.,}{{Laurenti}
  et~al.}{2022}]{laurentietal22}
{Laurenti} M.,  et~al., 2022, \mn@doi [\aap] {10.1051/0004-6361/202141829},
  \href {https://ui.adsabs.harvard.edu/abs/2022A&A...657A..57L} {657, A57}

\bibitem[\protect\citeauthoryear{{Leighly}}{{Leighly}}{1999}]{1999leighly}
{Leighly} K.~M.,  1999, \mn@doi [\apjs] {10.1086/313287}, \href
  {https://ui.adsabs.harvard.edu/abs/1999ApJS..125..317L} {125, 317}

\bibitem[\protect\citeauthoryear{{Leighly} \& {Moore}}{{Leighly} \&
  {Moore}}{2004}]{leighlymoore04}
{Leighly} K.~M.,  {Moore} J.~R.,  2004, \mn@doi [\apj] {10.1086/422088}, \href
  {http://adsabs.harvard.edu/abs/2004ApJ...611..107L} {611, 107}

\bibitem[\protect\citeauthoryear{{Lewin} et~al.,}{{Lewin}
  et~al.}{2022}]{2022Lewin}
{Lewin} C.,  et~al., 2022, \mn@doi [\apj] {10.3847/1538-4357/ac978f}, \href
  {https://ui.adsabs.harvard.edu/abs/2022ApJ...939..109L} {939, 109}

\bibitem[\protect\citeauthoryear{{Li}, {Calzetti}, {Kennicutt}, {Hong},
  {Engelbracht}, {Dale}  \& {Moustakas}}{{Li} et~al.}{2010}]{lietal10}
{Li} Y.,  {Calzetti} D.,  {Kennicutt} R.~C.,  {Hong} S.,  {Engelbracht} C.~W.,
  {Dale} D.~A.,   {Moustakas} J.,  2010, \mn@doi [\apj]
  {10.1088/0004-637X/725/1/677}, \href
  {https://ui.adsabs.harvard.edu/abs/2010ApJ...725..677L} {725, 677}

\bibitem[\protect\citeauthoryear{{Li} et~al.,}{{Li} et~al.}{2025}]{2025Li}
{Li} Y.-R.,  et~al., 2025, \mn@doi [arXiv e-prints]
  {10.48550/arXiv.2502.18856}, \href
  {https://ui.adsabs.harvard.edu/abs/2025arXiv250218856L} {p. arXiv:2502.18856}

\bibitem[\protect\citeauthoryear{{Lipari}}{{Lipari}}{1994}]{liparietal94}
{Lipari} S.,  1994, \mn@doi [\apj] {10.1086/174884}, \href
  {https://ui.adsabs.harvard.edu/abs/1994ApJ...436..102L} {436, 102}

\bibitem[\protect\citeauthoryear{{Lipari}, {Terlevich}  \&
  {Macchetto}}{{Lipari} et~al.}{1993}]{1993Lipari}
{Lipari} S.,  {Terlevich} R.,   {Macchetto} F.,  1993, \mn@doi [\apj]
  {10.1086/172456}, \href
  {https://ui.adsabs.harvard.edu/abs/1993ApJ...406..451L} {406, 451}

\bibitem[\protect\citeauthoryear{{L{\'\i}pari}, {Terlevich}, {D{\'\i}az},
  {Taniguchi}, {Zheng}, {Tsvetanov}, {Carranza}  \& {Dottori}}{{L{\'\i}pari}
  et~al.}{2003}]{liparietal03}
{L{\'\i}pari} S.,  {Terlevich} R.,  {D{\'\i}az} R.~J.,  {Taniguchi} Y.,
  {Zheng} W.,  {Tsvetanov} Z.,  {Carranza} G.,   {Dottori} H.,  2003, \mn@doi
  [\mnras] {10.1046/j.1365-8711.2003.06309.x}, \href
  {https://ui.adsabs.harvard.edu/abs/2003MNRAS.340..289L} {340, 289}

\bibitem[\protect\citeauthoryear{{Liu}, {Luo}, {Brandt}, {Brotherton},
  {Gallagher}, {Ni}, {Shemmer}  \& {Timlin}}{{Liu} et~al.}{2021}]{liuetal21}
{Liu} H.,  {Luo} B.,  {Brandt} W.~N.,  {Brotherton} M.~S.,  {Gallagher} S.~C.,
  {Ni} Q.,  {Shemmer} O.,   {Timlin} J.~D. I.,  2021, \mn@doi [\apj]
  {10.3847/1538-4357/abe37f}, \href
  {https://ui.adsabs.harvard.edu/abs/2021ApJ...910..103L} {910, 103}

\bibitem[\protect\citeauthoryear{{MacAlpine}, {Davidson}, {Gull}  \&
  {Wu}}{{MacAlpine} et~al.}{1985}]{macalpineetal85}
{MacAlpine} G.~M.,  {Davidson} K.,  {Gull} T.~R.,   {Wu} C.~C.,  1985, \mn@doi
  [\apj] {10.1086/163282}, \href
  {https://ui.adsabs.harvard.edu/abs/1985ApJ...294..147M} {294, 147}

\bibitem[\protect\citeauthoryear{{Malkan}}{{Malkan}}{1983}]{malkan83}
{Malkan} M.~A.,  1983, \mn@doi [\apj] {10.1086/160981}, \href
  {https://ui.adsabs.harvard.edu/abs/1983ApJ...268..582M} {268, 582}

\bibitem[\protect\citeauthoryear{{Malkan} \& {Sargent}}{{Malkan} \&
  {Sargent}}{1982}]{malkansargent82}
{Malkan} M.~A.,  {Sargent} W.~L.~W.,  1982, \mn@doi [\apj] {10.1086/159701},
  \href {http://adsabs.harvard.edu/abs/1982ApJ...254...22M} {254, 22}

\bibitem[\protect\citeauthoryear{{Marinello}, {Rodr{\'\i}guez-Ardila},
  {Marziani}, {Sigut}  \& {Pradhan}}{{Marinello} et~al.}{2020}]{2020marinello}
{Marinello} M.,  {Rodr{\'\i}guez-Ardila} A.,  {Marziani} P.,  {Sigut} A.,
  {Pradhan} A.,  2020, \mn@doi [\mnras] {10.1093/mnras/staa934}, \href
  {https://ui.adsabs.harvard.edu/abs/2020MNRAS.494.4187M} {494, 4187}

\bibitem[\protect\citeauthoryear{{Mart{\'{\i}}nez-Aldama}, {Dultzin},
  {Marziani}, {Sulentic}, {Bressan}, {Chen}  \&
  {Stirpe}}{{Mart{\'{\i}}nez-Aldama} et~al.}{2015}]{martinez-aldamaetal15}
{Mart{\'{\i}}nez-Aldama} M.~L.,  {Dultzin} D.,  {Marziani} P.,  {Sulentic}
  J.~W.,  {Bressan} A.,  {Chen} Y.,   {Stirpe} G.~M.,  2015, \mn@doi [ApJS]
  {10.1088/0067-0049/217/1/3}, \href
  {http://adsabs.harvard.edu/abs/2015ApJS..217....3L} {217, 3}

\bibitem[\protect\citeauthoryear{Mart{\'\i}nez-Aldama, Del~Olmo, Marziani,
  Sulentic, Negrete, Dultzin, Perea  \& D'Onofrio}{Mart{\'\i}nez-Aldama
  et~al.}{2018}]{martinez-aldamaetal18}
Mart{\'\i}nez-Aldama M.~L.,  Del~Olmo A.,  Marziani P.,  Sulentic J.~W.,
  Negrete C.~A.,  Dultzin D.,  Perea J.,   D'Onofrio M.,  2018, \mn@doi
  [Frontiers in Astronomy and Space Sciences] {10.3389/fspas.2017.00065}, 4, 65

\bibitem[\protect\citeauthoryear{{Mart{\'\i}nez-Aldama}, {Panda}, {Czerny},
  {Marinello}, {Marziani}  \& {Dultzin}}{{Mart{\'\i}nez-Aldama}
  et~al.}{2021}]{martinez-aldama_etal_2021}
{Mart{\'\i}nez-Aldama} M.~L.,  {Panda} S.,  {Czerny} B.,  {Marinello} M.,
  {Marziani} P.,   {Dultzin} D.,  2021, \mn@doi [\apj]
  {10.3847/1538-4357/ac03b6}, \href
  {https://ui.adsabs.harvard.edu/abs/2021ApJ...918...29M} {918, 29}

\bibitem[\protect\citeauthoryear{{Marziani} \& {Sulentic}}{{Marziani} \&
  {Sulentic}}{2014}]{2014marziani}
{Marziani} P.,  {Sulentic} J.~W.,  2014, \mn@doi [\mnras]
  {10.1093/mnras/stu951}, \href
  {https://ui.adsabs.harvard.edu/abs/2014MNRAS.442.1211M} {442, 1211}

\bibitem[\protect\citeauthoryear{{Marziani}, {Sulentic}, {Dultzin-Hacyan},
  {Calvani}  \& {Moles}}{{Marziani} et~al.}{1996}]{marzianietal96}
{Marziani} P.,  {Sulentic} J.~W.,  {Dultzin-Hacyan} D.,  {Calvani} M.,
  {Moles} M.,  1996, \mn@doi [ApJS] {10.1086/192291}, \href
  {http://adsabs.harvard.edu/abs/1996ApJS..104...37M} {104, 37}

\bibitem[\protect\citeauthoryear{{Marziani}, {Sulentic}, {Zwitter},
  {Dultzin-Hacyan}  \& {Calvani}}{{Marziani} et~al.}{2001}]{marzianietal01}
{Marziani} P.,  {Sulentic} J.~W.,  {Zwitter} T.,  {Dultzin-Hacyan} D.,
  {Calvani} M.,  2001, \mn@doi [ApJ] {10.1086/322286}, \href
  {http://adsabs.harvard.edu/abs/2001ApJ...558..553M} {558, 553}

\bibitem[\protect\citeauthoryear{{Marziani}, {Sulentic}, {Zamanov}, {Calvani},
  {Dultzin-Hacyan}, {Bachev}  \& {Zwitter}}{{Marziani}
  et~al.}{2003a}]{marzianietal03a}
{Marziani} P.,  {Sulentic} J.~W.,  {Zamanov} R.,  {Calvani} M.,
  {Dultzin-Hacyan} D.,  {Bachev} R.,   {Zwitter} T.,  2003a, \mn@doi [ApJS]
  {10.1086/346025}, \href {http://adsabs.harvard.edu/abs/2003ApJS..145..199M}
  {145, 199}

\bibitem[\protect\citeauthoryear{{Marziani}, {Sulentic}, {Zamanov}, {Calvani},
  {Dultzin-Hacyan}, {Bachev}  \& {Zwitter}}{{Marziani}
  et~al.}{2003b}]{2003marziani}
{Marziani} P.,  {Sulentic} J.~W.,  {Zamanov} R.,  {Calvani} M.,
  {Dultzin-Hacyan} D.,  {Bachev} R.,   {Zwitter} T.,  2003b, \mn@doi [\apjs]
  {10.1086/346025}, \href
  {https://ui.adsabs.harvard.edu/abs/2003ApJS..145..199M} {145, 199}

\bibitem[\protect\citeauthoryear{{Marziani}, {Zamanov}, {Sulentic}  \&
  {Calvani}}{{Marziani} et~al.}{2003c}]{marzianietal03b}
{Marziani} P.,  {Zamanov} R.~K.,  {Sulentic} J.~W.,   {Calvani} M.,  2003c,
  \mn@doi [MNRAS] {10.1046/j.1365-2966.2003.07033.x}, \href
  {http://adsabs.harvard.edu/abs/2003MNRAS.345.1133M} {345, 1133}

\bibitem[\protect\citeauthoryear{{Marziani}, {Dultzin-Hacyan}  \&
  {Sulentic}}{{Marziani} et~al.}{2006}]{marzianietal06}
{Marziani} P.,  {Dultzin-Hacyan} D.,   {Sulentic} J.~W.,  2006, in {Kreitler}
  P.~V.,  ed., , New Developments in Black Hole Research.
Nova Press, New York, p.~123

\bibitem[\protect\citeauthoryear{{Marziani}, {Sulentic}, {Negrete}, {Dultzin},
  {Zamfir}  \& {Bachev}}{{Marziani} et~al.}{2010}]{marzianietal10}
{Marziani} P.,  {Sulentic} J.~W.,  {Negrete} C.~A.,  {Dultzin} D.,  {Zamfir}
  S.,   {Bachev} R.,  2010, \mn@doi [\mnras]
  {10.1111/j.1365-2966.2010.17357.x}, \href
  {http://adsabs.harvard.edu/abs/2010MNRAS.409.1033M} {409, 1033}

\bibitem[\protect\citeauthoryear{{Marziani}, {Sulentic}, {Plauchu-Frayn}  \&
  {del Olmo}}{{Marziani} et~al.}{2013a}]{2013Marziani_composite}
{Marziani} P.,  {Sulentic} J.~W.,  {Plauchu-Frayn} I.,   {del Olmo} A.,  2013a,
  \mn@doi [\aap] {10.1051/0004-6361/201321374}, \href
  {https://ui.adsabs.harvard.edu/abs/2013A&A...555A..89M} {555, A89}

\bibitem[\protect\citeauthoryear{{Marziani}, {Sulentic}, {Plauchu-Frayn}  \&
  {del Olmo}}{{Marziani} et~al.}{2013b}]{marzianietal13a}
{Marziani} P.,  {Sulentic} J.~W.,  {Plauchu-Frayn} I.,   {del Olmo} A.,  2013b,
  AAp, \href {http://adsabs.harvard.edu/abs/2013arXiv1305.1096M} {555, 89,
  16pp}

\bibitem[\protect\citeauthoryear{{Marziani}, {Mart{\'{\i}}nez Carballo},
  {Sulentic}, {Del Olmo}, {Stirpe}  \& {Dultzin}}{{Marziani}
  et~al.}{2016}]{marzianietal16a}
{Marziani} P.,  {Mart{\'{\i}}nez Carballo} M.~A.,  {Sulentic} J.~W.,  {Del
  Olmo} A.,  {Stirpe} G.~M.,   {Dultzin} D.,  2016, \mn@doi [\apss]
  {10.1007/s10509-015-2611-1}, \href
  {http://adsabs.harvard.edu/abs/2016Ap%26SS.361...29M} {361, 29}

\bibitem[\protect\citeauthoryear{{Marziani}, {Negrete}, {Dultzin},
  {Mart{\'\i}nez-Aldama}, {Del Olmo}, {D'Onofrio}  \& {Stirpe}}{{Marziani}
  et~al.}{2017}]{marzianietal17}
{Marziani} P.,  {Negrete} C.~A.,  {Dultzin} D.,  {Mart{\'\i}nez-Aldama} M.~L.,
  {Del Olmo} A.,  {D'Onofrio} M.,   {Stirpe} G.~M.,  2017, \mn@doi [Frontiers
  in Astronomy and Space Sciences] {10.3389/fspas.2017.00016}, \href
  {https://ui.adsabs.harvard.edu/abs/2017FrASS...4...16M} {4, 16}

\bibitem[\protect\citeauthoryear{{Marziani} et~al.,}{{Marziani}
  et~al.}{2021a}]{marzianietal21}
{Marziani} P.,  et~al., 2021a, \mn@doi [Research Notes of the American
  Astronomical Society] {10.3847/2515-5172/abe46a}, \href
  {https://ui.adsabs.harvard.edu/abs/2021RNAAS...5...25M} {5, 25}

\bibitem[\protect\citeauthoryear{{Marziani}, {Berton}, {Panda}  \&
  {Bon}}{{Marziani} et~al.}{2021b}]{marzianietal21a}
{Marziani} P.,  {Berton} M.,  {Panda} S.,   {Bon} E.,  2021b, \mn@doi
  [Universe] {10.3390/universe7120484}, \href
  {https://ui.adsabs.harvard.edu/abs/2021Univ....7..484M} {7, 484}

\bibitem[\protect\citeauthoryear{{Marziani}, {Panda}, {Deconto Machado}  \&
  {Del Olmo}}{{Marziani} et~al.}{2023}]{marzianietal23}
{Marziani} P.,  {Panda} S.,  {Deconto Machado} A.,   {Del Olmo} A.,  2023,
  \mn@doi [Galaxies] {10.3390/galaxies11020052}, \href
  {https://ui.adsabs.harvard.edu/abs/2023Galax..11...52M} {11, 52}

\bibitem[\protect\citeauthoryear{Marziani, Garnica~Luna, Floris, del Olmo,
  Deconto-Machado, Buendia-Rios, Negrete  \& Dultzin}{Marziani
  et~al.}{2025}]{marzianietal25}
Marziani P.,  Garnica~Luna K.,  Floris A.,  del Olmo A.,  Deconto-Machado A.,
  Buendia-Rios T.~M.,  Negrete C.~A.,   Dultzin D.,  2025, \mn@doi [Universe]
  {10.3390/universe11020069}, 11

\bibitem[\protect\citeauthoryear{{Mathews} \& {Ferland}}{{Mathews} \&
  {Ferland}}{1987}]{1987mathewsferland}
{Mathews} W.~G.,  {Ferland} G.~J.,  1987, \mn@doi [ApJ] {10.1086/165843}, \href
  {http://adsabs.harvard.edu/abs/1987ApJ...323..456M} {323, 456}

\bibitem[\protect\citeauthoryear{{Mehdipour}, {Kriss}, {Kaastra}, {Costantini},
  {Gu}, {Landt}, {Mao}  \& {Rogantini}}{{Mehdipour}
  et~al.}{2024}]{mehdipouretal24}
{Mehdipour} M.,  {Kriss} G.~A.,  {Kaastra} J.~S.,  {Costantini} E.,  {Gu} L.,
  {Landt} H.,  {Mao} J.,   {Rogantini} D.,  2024, \mn@doi [\apj]
  {10.3847/1538-4357/ad1bcb}, \href
  {https://ui.adsabs.harvard.edu/abs/2024ApJ...962..155M} {962, 155}

\bibitem[\protect\citeauthoryear{{Mej{\'\i}a-Restrepo}, {Lira}, {Netzer},
  {Trakhtenbrot}  \& {Capellupo}}{{Mej{\'\i}a-Restrepo}
  et~al.}{2018}]{2018mejia-restrepo}
{Mej{\'\i}a-Restrepo} J.~E.,  {Lira} P.,  {Netzer} H.,  {Trakhtenbrot} B.,
  {Capellupo} D.~M.,  2018, \mn@doi [Nature Astronomy]
  {10.1038/s41550-017-0305-z}, \href
  {https://ui.adsabs.harvard.edu/abs/2018NatAs...2...63M} {2, 63}

\bibitem[\protect\citeauthoryear{{Miniutti}, {Fabian}, {Brandt}, {Gallo}  \&
  {Boller}}{{Miniutti} et~al.}{2009}]{2009miniutti}
{Miniutti} G.,  {Fabian} A.~C.,  {Brandt} W.~N.,  {Gallo} L.~C.,   {Boller} T.,
   2009, \mn@doi [\mnras] {10.1111/j.1745-3933.2009.00669.x}, \href
  {https://ui.adsabs.harvard.edu/abs/2009MNRAS.396L..85M} {396, L85}

\bibitem[\protect\citeauthoryear{{Mochizuki}, {Mizumoto}  \&
  {Ebisawa}}{{Mochizuki} et~al.}{2023}]{2023Mochizuki_mrk766}
{Mochizuki} Y.,  {Mizumoto} M.,   {Ebisawa} K.,  2023, \mn@doi [\mnras]
  {10.1093/mnras/stad2329}, \href
  {https://ui.adsabs.harvard.edu/abs/2023MNRAS.525..922M} {525, 922}

\bibitem[\protect\citeauthoryear{{Nandra} \& {Pounds}}{{Nandra} \&
  {Pounds}}{1994}]{nandrapounds94}
{Nandra} K.,  {Pounds} K.~A.,  1994, \mn@doi [\mnras]
  {10.1093/mnras/268.2.405}, \href
  {https://ui.adsabs.harvard.edu/abs/1994MNRAS.268..405N} {268, 405}

\bibitem[\protect\citeauthoryear{{Nardini} et~al.,}{{Nardini}
  et~al.}{2015}]{nardinietal15}
{Nardini} E.,  et~al., 2015, \mn@doi [Science] {10.1126/science.1259202}, \href
  {https://ui.adsabs.harvard.edu/abs/2015Sci...347..860N} {347, 860}

\bibitem[\protect\citeauthoryear{{Negrete}, {Dultzin}, {Marziani}  \&
  {Sulentic}}{{Negrete} et~al.}{2012}]{negreteetal12}
{Negrete} A.,  {Dultzin} D.,  {Marziani} P.,   {Sulentic} J.,  2012, ApJ, \href
  {http://adsabs.harvard.edu/abs/2011arXiv1107.3188N} {757, 62}

\bibitem[\protect\citeauthoryear{{Negrete} et~al.,}{{Negrete}
  et~al.}{2018}]{2018negrete}
{Negrete} C.~A.,  et~al., 2018, \mn@doi [\aap] {10.1051/0004-6361/201833285},
  \href {https://ui.adsabs.harvard.edu/abs/2018A&A...620A.118N} {620, A118}

\bibitem[\protect\citeauthoryear{{Netzer}, {Lani}, {Nordon}, {Trakhtenbrot},
  {Lira}  \& {Shemmer}}{{Netzer} et~al.}{2016}]{netzeretal16}
{Netzer} H.,  {Lani} C.,  {Nordon} R.,  {Trakhtenbrot} B.,  {Lira} P.,
  {Shemmer} O.,  2016, \mn@doi [\apj] {10.3847/0004-637X/819/2/123}, \href
  {https://ui.adsabs.harvard.edu/abs/2016ApJ...819..123N} {819, 123}

\bibitem[\protect\citeauthoryear{{Onoue} et~al.,}{{Onoue}
  et~al.}{2023}]{onoueetal23}
{Onoue} M.,  et~al., 2023, \mn@doi [\apjl] {10.3847/2041-8213/aca9d3}, \href
  {https://ui.adsabs.harvard.edu/abs/2023ApJ...942L..17O} {942, L17}

\bibitem[\protect\citeauthoryear{{Padovani}}{{Padovani}}{2017a}]{padovani17a}
{Padovani} P.,  2017a, \mn@doi [Nature Astronomy] {10.1038/s41550-017-0194},
  \href {https://ui.adsabs.harvard.edu/abs/2017NatAs...1E.194P} {1, 0194}

\bibitem[\protect\citeauthoryear{{Padovani}}{{Padovani}}{2017b}]{padovani17}
{Padovani} P.,  2017b, \mn@doi [Frontiers in Astronomy and Space Sciences]
  {10.3389/fspas.2017.00035}, \href
  {https://ui.adsabs.harvard.edu/abs/2017FrASS...4...35P} {4, 35}

\bibitem[\protect\citeauthoryear{{Paliya}}{{Paliya}}{2019}]{paliya19}
{Paliya} V.~S.,  2019, \mn@doi [Journal of Astrophysics and Astronomy]
  {10.1007/s12036-019-9604-3}, \href
  {https://ui.adsabs.harvard.edu/abs/2019JApA...40...39P} {40, 39}

\bibitem[\protect\citeauthoryear{{Panagiotou} \& {Walter}}{{Panagiotou} \&
  {Walter}}{2020}]{2020Panagiotou}
{Panagiotou} C.,  {Walter} R.,  2020, \mn@doi [\aap]
  {10.1051/0004-6361/201937390}, \href
  {https://ui.adsabs.harvard.edu/abs/2020A&A...640A..31P} {640, A31}

\bibitem[\protect\citeauthoryear{{Panda}}{{Panda}}{2021}]{panda2021}
{Panda} S.,  2021, \mn@doi [\aap] {10.1051/0004-6361/202140393}, \href
  {https://ui.adsabs.harvard.edu/abs/2021A&A...650A.154P} {650, A154}

\bibitem[\protect\citeauthoryear{{Panda} \& {Marziani}}{{Panda} \&
  {Marziani}}{2023}]{pandamarziani23}
{Panda} S.,  {Marziani} P.,  2023, \mn@doi [Frontiers in Astronomy and Space
  Sciences] {10.3389/fspas.2023.1130103}, \href
  {https://ui.adsabs.harvard.edu/abs/2023FrASS..1030103P} {10, 1130103}

\bibitem[\protect\citeauthoryear{{Panda}, {Czerny}, {Adhikari}, {Hryniewicz},
  {Wildy}, {Kuraszkiewicz}  \& {{\'S}niegowska}}{{Panda}
  et~al.}{2018}]{panda2018}
{Panda} S.,  {Czerny} B.,  {Adhikari} T.~P.,  {Hryniewicz} K.,  {Wildy} C.,
  {Kuraszkiewicz} J.,   {{\'S}niegowska} M.,  2018, \mn@doi [\apj]
  {10.3847/1538-4357/aae209}, \href
  {https://ui.adsabs.harvard.edu/abs/2018ApJ...866..115P} {866, 115}

\bibitem[\protect\citeauthoryear{{Panda}, {Czerny}, {Done}  \&
  {Kubota}}{{Panda} et~al.}{2019a}]{panda_etal_2019_wc}
{Panda} S.,  {Czerny} B.,  {Done} C.,   {Kubota} A.,  2019a, \mn@doi [\apj]
  {10.3847/1538-4357/ab11cb}, \href
  {https://ui.adsabs.harvard.edu/abs/2019ApJ...875..133P} {875, 133}

\bibitem[\protect\citeauthoryear{{Panda}, {Marziani}  \& {Czerny}}{{Panda}
  et~al.}{2019b}]{pandaetal19}
{Panda} S.,  {Marziani} P.,   {Czerny} B.,  2019b, \mn@doi [\apj]
  {10.3847/1538-4357/ab3292}, \href
  {https://ui.adsabs.harvard.edu/abs/2019ApJ...882...79P} {882, 79}

\bibitem[\protect\citeauthoryear{{Panda}, {Marziani}  \& {Czerny}}{{Panda}
  et~al.}{2020}]{pandaetal20}
{Panda} S.,  {Marziani} P.,   {Czerny} B.,  2020, \mn@doi [Contributions of the
  Astronomical Observatory Skalnate Pleso] {10.31577/caosp.2020.50.1.293},
  \href {https://ui.adsabs.harvard.edu/abs/2020CoSka..50..293P} {50, 293}

\bibitem[\protect\citeauthoryear{{Panda} et~al.,}{{Panda}
  et~al.}{2024}]{panda_etal_2024}
{Panda} S.,  et~al., 2024, \mn@doi [\apjs] {10.3847/1538-4365/ad3549}, \href
  {https://ui.adsabs.harvard.edu/abs/2024ApJS..272...11P} {272, 11}

\bibitem[\protect\citeauthoryear{{Panessa} et~al.,}{{Panessa}
  et~al.}{2011}]{2011panessa}
{Panessa} F.,  et~al., 2011, \mn@doi [\mnras]
  {10.1111/j.1365-2966.2011.19268.x}, \href
  {https://ui.adsabs.harvard.edu/abs/2011MNRAS.417.2426P} {417, 2426}

\bibitem[\protect\citeauthoryear{{Perry} \& {Dyson}}{{Perry} \&
  {Dyson}}{1992}]{perrydyson92}
{Perry} J.,  {Dyson} J.,  1992, in {Holt} S.~S.,  {Neff} S.~G.,   {Urry} C.~M.,
   eds,  American Institute of Physics Conference Series Vol. 254, Testing the
  AGN paradigm. AIP, pp 553--555, \mn@doi{10.1063/1.42236}

\bibitem[\protect\citeauthoryear{{Puchnarewicz} et~al.,}{{Puchnarewicz}
  et~al.}{1992}]{1992Puchnarewicz}
{Puchnarewicz} E.~M.,  et~al., 1992, \mn@doi [\mnras]
  {10.1093/mnras/256.3.589}, \href
  {https://ui.adsabs.harvard.edu/abs/1992MNRAS.256..589P} {256, 589}

\bibitem[\protect\citeauthoryear{{Rafanelli}, {La Mura}, {Bindoni}, {Ciroi},
  {Cracco}, {Di Mille}  \& {Vaona}}{{Rafanelli} et~al.}{2011}]{rafanellietal11}
{Rafanelli} P.,  {La Mura} G.,  {Bindoni} D.,  {Ciroi} S.,  {Cracco} V.,  {Di
  Mille} F.,   {Vaona} L.,  2011, \mn@doi [Baltic Astronomy]
  {10.1515/astro-2017-0313}, \href
  {https://ui.adsabs.harvard.edu/abs/2011BaltA..20..419R} {20, 419}

\bibitem[\protect\citeauthoryear{{Reeves}, {O'Brien}, {Vaughan}, {Law-Green},
  {Ward}, {Simpson}, {Pounds}  \& {Edelson}}{{Reeves}
  et~al.}{2000}]{reevesetal00}
{Reeves} J.~N.,  {O'Brien} P.~T.,  {Vaughan} S.,  {Law-Green} D.,  {Ward} M.,
  {Simpson} C.,  {Pounds} K.~A.,   {Edelson} R.,  2000, \mn@doi [\mnras]
  {10.1046/j.1365-8711.2000.03282.x}, \href
  {https://ui.adsabs.harvard.edu/abs/2000MNRAS.312L..17R} {312, L17}

\bibitem[\protect\citeauthoryear{{Reeves}, {O'Brien}  \& {Ward}}{{Reeves}
  et~al.}{2003}]{reevesetal03}
{Reeves} J.~N.,  {O'Brien} P.~T.,   {Ward} M.~J.,  2003, \mn@doi [\apjl]
  {10.1086/378218}, \href
  {https://ui.adsabs.harvard.edu/abs/2003ApJ...593L..65R} {593, L65}

\bibitem[\protect\citeauthoryear{{Ricci} et~al.,}{{Ricci}
  et~al.}{2017}]{2017ricci}
{Ricci} C.,  et~al., 2017, \mn@doi [\apjs] {10.3847/1538-4365/aa96ad}, \href
  {https://ui.adsabs.harvard.edu/abs/2017ApJS..233...17R} {233, 17}

\bibitem[\protect\citeauthoryear{{Richards} et~al.,}{{Richards}
  et~al.}{2006}]{2006richards}
{Richards} G.~T.,  et~al., 2006, \mn@doi [\apjs] {10.1086/506525}, \href
  {https://ui.adsabs.harvard.edu/abs/2006ApJS..166..470R} {166, 470}

\bibitem[\protect\citeauthoryear{{Risaliti}, {Young}  \& {Elvis}}{{Risaliti}
  et~al.}{2009}]{risalitietal09}
{Risaliti} G.,  {Young} M.,   {Elvis} M.,  2009, \mn@doi [\apjl]
  {10.1088/0004-637X/700/1/L6}, \href
  {https://ui.adsabs.harvard.edu/abs/2009ApJ...700L...6R} {700, L6}

\bibitem[\protect\citeauthoryear{{Rodr{\'\i}guez-Ardila}, {Fonseca-Faria},
  {Dias dos Santos}, {Panda}  \& {Marinello}}{{Rodr{\'\i}guez-Ardila}
  et~al.}{2024}]{rodriguez-ardila_etal_2024}
{Rodr{\'\i}guez-Ardila} A.,  {Fonseca-Faria} M.~A.,  {Dias dos Santos} D.,
  {Panda} S.,   {Marinello} M.,  2024, \mn@doi [\aj]
  {10.3847/1538-3881/ad36bf}, \href
  {https://ui.adsabs.harvard.edu/abs/2024AJ....167..244R} {167, 244}

\bibitem[\protect\citeauthoryear{{Runnoe}, {Shang}  \& {Brotherton}}{{Runnoe}
  et~al.}{2013}]{runnoeetal13}
{Runnoe} J.~C.,  {Shang} Z.,   {Brotherton} M.~S.,  2013, \mn@doi [\mnras]
  {10.1093/mnras/stt1528}, \href
  {http://adsabs.harvard.edu/abs/2013MNRAS.435.3251R} {435, 3251}

\bibitem[\protect\citeauthoryear{{Rupke} \& {Veilleux}}{{Rupke} \&
  {Veilleux}}{2011}]{rupkeVeilleux11}
{Rupke} D. S.~N.,  {Veilleux} S.,  2011, \mn@doi [\apjl]
  {10.1088/2041-8205/729/2/L27}, \href
  {https://ui.adsabs.harvard.edu/abs/2011ApJ...729L..27R} {729, L27}

\bibitem[\protect\citeauthoryear{{Saccheo} et~al.,}{{Saccheo}
  et~al.}{2023}]{saccheoetal23}
{Saccheo} I.,  et~al., 2023, \mn@doi [\aap] {10.1051/0004-6361/202244296},
  \href {https://ui.adsabs.harvard.edu/abs/2023A&A...671A..34S} {671, A34}

\bibitem[\protect\citeauthoryear{{Sanders} \& {Mirabel}}{{Sanders} \&
  {Mirabel}}{1996}]{sandersmirabel96}
{Sanders} D.~B.,  {Mirabel} I.~F.,  1996, \mn@doi [\araa]
  {10.1146/annurev.astro.34.1.749}, \href
  {https://ui.adsabs.harvard.edu/abs/1996ARA&A..34..749S} {34, 749}

\bibitem[\protect\citeauthoryear{{Sanders}, {Soifer}, {Elias}, {Madore},
  {Matthews}, {Neugebauer}  \& {Scoville}}{{Sanders}
  et~al.}{1988}]{sandersetal88}
{Sanders} D.~B.,  {Soifer} B.~T.,  {Elias} J.~H.,  {Madore} B.~F.,  {Matthews}
  K.,  {Neugebauer} G.,   {Scoville} N.~Z.,  1988, \mn@doi [\apj]
  {10.1086/165983}, \href
  {https://ui.adsabs.harvard.edu/abs/1988ApJ...325...74S} {325, 74}

\bibitem[\protect\citeauthoryear{{Sani}, {Lutz}, {Risaliti}, {Netzer}, {Gallo},
  {Trakhtenbrot}, {Sturm}  \& {Boller}}{{Sani} et~al.}{2010}]{sanietal10}
{Sani} E.,  {Lutz} D.,  {Risaliti} G.,  {Netzer} H.,  {Gallo} L.~C.,
  {Trakhtenbrot} B.,  {Sturm} E.,   {Boller} T.,  2010, \mn@doi [MNRAS]
  {10.1111/j.1365-2966.2009.16217.x}, \href
  {http://adsabs.harvard.edu/abs/2010MNRAS.403.1246S} {403, 1246}

\bibitem[\protect\citeauthoryear{{Shang} et~al.,}{{Shang}
  et~al.}{2011}]{shangetal11}
{Shang} Z.,  et~al., 2011, \mn@doi [\apjs] {10.1088/0067-0049/196/1/2}, \href
  {http://adsabs.harvard.edu/abs/2011ApJS..196....2S} {196, 2}

\bibitem[\protect\citeauthoryear{{Shemmer}, {Brandt}, {Netzer}, {Maiolino}  \&
  {Kaspi}}{{Shemmer} et~al.}{2008}]{shemmeretal08}
{Shemmer} O.,  {Brandt} W.~N.,  {Netzer} H.,  {Maiolino} R.,   {Kaspi} S.,
  2008, \mn@doi [\apj] {10.1086/588776}, \href
  {https://ui.adsabs.harvard.edu/abs/2008ApJ...682...81S} {682, 81}

\bibitem[\protect\citeauthoryear{{Shen} \& {Ho}}{{Shen} \&
  {Ho}}{2014}]{shenho14}
{Shen} Y.,  {Ho} L.~C.,  2014, \mn@doi [\nat] {10.1038/nature13712}, \href
  {http://adsabs.harvard.edu/abs/2014Natur.513..210S} {513, 210}

\bibitem[\protect\citeauthoryear{{{\'S}niegowska}, {Marziani}, {Czerny},
  {Panda}, {Mart{\'\i}nez-Aldama}, {del Olmo}  \& {D'Onofrio}}{{{\'S}niegowska}
  et~al.}{2021}]{2021sniegowska}
{{\'S}niegowska} M.,  {Marziani} P.,  {Czerny} B.,  {Panda} S.,
  {Mart{\'\i}nez-Aldama} M.~L.,  {del Olmo} A.,   {D'Onofrio} M.,  2021,
  \mn@doi [\apj] {10.3847/1538-4357/abe1c8}, \href
  {https://ui.adsabs.harvard.edu/abs/2021ApJ...910..115S} {910, 115}

\bibitem[\protect\citeauthoryear{{Spinoglio}, {Fern{\'a}ndez-Ontiveros}  \&
  {Malkan}}{{Spinoglio} et~al.}{2024}]{spinolioetal24}
{Spinoglio} L.,  {Fern{\'a}ndez-Ontiveros} J.~A.,   {Malkan} M.~A.,  2024,
  \mn@doi [\apj] {10.3847/1538-4357/ad23e4}, \href
  {https://ui.adsabs.harvard.edu/abs/2024ApJ...964..117S} {964, 117}

\bibitem[\protect\citeauthoryear{{Storchi-Bergmann}, {Schmitt}, {Calzetti}  \&
  {Kinney}}{{Storchi-Bergmann} et~al.}{1998}]{storchi-Bergmannetal98}
{Storchi-Bergmann} T.,  {Schmitt} H.~R.,  {Calzetti} D.,   {Kinney} A.~L.,
  1998, \mn@doi [\aj] {10.1086/300242}, \href
  {https://ui.adsabs.harvard.edu/abs/1998AJ....115..909S} {115, 909}

\bibitem[\protect\citeauthoryear{{Sulentic}, {Marziani}  \&
  {Dultzin-Hacyan}}{{Sulentic} et~al.}{2000a}]{sulenticetal00a}
{Sulentic} J.~W.,  {Marziani} P.,   {Dultzin-Hacyan} D.,  2000a, \mn@doi
  [ARA\&A] {10.1146/annurev.astro.38.1.521}, \href
  {http://adsabs.harvard.edu/abs/2000ARA\% 26A..38..521S} {38, 521}

\bibitem[\protect\citeauthoryear{{Sulentic}, {Zwitter}, {Marziani}  \&
  {Dultzin-Hacyan}}{{Sulentic} et~al.}{2000b}]{sulenticetal00c}
{Sulentic} J.~W.,  {Zwitter} T.,  {Marziani} P.,   {Dultzin-Hacyan} D.,  2000b,
  \mn@doi [ApJL] {10.1086/312717}, \href
  {http://adsabs.harvard.edu/abs/2000ApJ...536L...5S} {536, L5}

\bibitem[\protect\citeauthoryear{{Sulentic}, {Marziani}, {Zamanov}, {Bachev},
  {Calvani}  \& {Dultzin-Hacyan}}{{Sulentic} et~al.}{2002}]{sulenticetal02}
{Sulentic} J.~W.,  {Marziani} P.,  {Zamanov} R.,  {Bachev} R.,  {Calvani} M.,
  {Dultzin-Hacyan} D.,  2002, \mn@doi [ApJL] {10.1086/339594}, \href
  {http://adsabs.harvard.edu/abs/2002ApJ...566L..71S} {566, L71}

\bibitem[\protect\citeauthoryear{{Sulentic}, {Dultzin-Hacyan}, {Marziani},
  {Bongardo}, {Braito}, {Calvani}  \& {Zamanov}}{{Sulentic}
  et~al.}{2006a}]{sulenticetal06a}
{Sulentic} J.~W.,  {Dultzin-Hacyan} D.,  {Marziani} P.,  {Bongardo} C.,
  {Braito} V.,  {Calvani} M.,   {Zamanov} R.,  2006a, Revista Mexicana de
  Astronomia y Astrofisica, \href
  {http://adsabs.harvard.edu/abs/2006RMxAA..42...23S} {42, 23}

\bibitem[\protect\citeauthoryear{{Sulentic}, {Repetto}, {Stirpe}, {Marziani},
  {Dultzin-Hacyan}  \& {Calvani}}{{Sulentic} et~al.}{2006b}]{2006sulentic}
{Sulentic} J.~W.,  {Repetto} P.,  {Stirpe} G.~M.,  {Marziani} P.,
  {Dultzin-Hacyan} D.,   {Calvani} M.,  2006b, \mn@doi [A\&Ap]
  {10.1051/0004-6361:20054153}, \href
  {http://adsabs.harvard.edu/abs/2006A26A...456..929S} {456, 929}

\bibitem[\protect\citeauthoryear{{Sulentic}, {Bachev}, {Marziani}, {Negrete}
  \& {Dultzin}}{{Sulentic} et~al.}{2007}]{sulenticetal07}
{Sulentic} J.~W.,  {Bachev} R.,  {Marziani} P.,  {Negrete} C.~A.,   {Dultzin}
  D.,  2007, \mn@doi [\apj] {10.1086/519916}, \href
  {https://ui.adsabs.harvard.edu/abs/2007ApJ...666..757S} {666, 757}

\bibitem[\protect\citeauthoryear{{Sulentic} et~al.,}{{Sulentic}
  et~al.}{2017}]{sulenticetal17}
{Sulentic} J.~W.,  et~al., 2017, \mn@doi [\aap] {10.1051/0004-6361/201630309},
  \href {http://adsabs.harvard.edu/abs/2017A%26A...608A.122S} {608, A122}

\bibitem[\protect\citeauthoryear{{Sun} \& {Shen}}{{Sun} \&
  {Shen}}{2015}]{sunshen15}
{Sun} J.,  {Shen} Y.,  2015, \mn@doi [\apjl] {10.1088/2041-8205/804/1/L15},
  \href {http://adsabs.harvard.edu/abs/2015ApJ...804L..15S} {804, L15}

\bibitem[\protect\citeauthoryear{{Tortosa}, {Bianchi}, {Marinucci}, {Matt}  \&
  {Petrucci}}{{Tortosa} et~al.}{2018}]{2018Tortosa}
{Tortosa} A.,  {Bianchi} S.,  {Marinucci} A.,  {Matt} G.,   {Petrucci} P.~O.,
  2018, \mn@doi [\aap] {10.1051/0004-6361/201732382}, \href
  {https://ui.adsabs.harvard.edu/abs/2018A&A...614A..37T} {614, A37}

\bibitem[\protect\citeauthoryear{{Tortosa} et~al.,}{{Tortosa}
  et~al.}{2022}]{tortosaetal22}
{Tortosa} A.,  et~al., 2022, \mn@doi [\mnras] {10.1093/mnras/stab3152}, \href
  {https://ui.adsabs.harvard.edu/abs/2022MNRAS.509.3599T} {509, 3599}

\bibitem[\protect\citeauthoryear{{Tortosa} et~al.,}{{Tortosa}
  et~al.}{2023}]{2023Tortosa}
{Tortosa} A.,  et~al., 2023, \mn@doi [\mnras] {10.1093/mnras/stac3590}, \href
  {https://ui.adsabs.harvard.edu/abs/2023MNRAS.519.6267T} {519, 6267}

\bibitem[\protect\citeauthoryear{{Trakhtenbrot} et~al.,}{{Trakhtenbrot}
  et~al.}{2017}]{trakhtenbrotetal17}
{Trakhtenbrot} B.,  et~al., 2017, \mn@doi [\mnras] {10.1093/mnras/stx1117},
  \href {https://ui.adsabs.harvard.edu/abs/2017MNRAS.470..800T} {470, 800}

\bibitem[\protect\citeauthoryear{{Vasudevan} \& {Fabian}}{{Vasudevan} \&
  {Fabian}}{2007}]{2007Vasudeban&Fabian}
{Vasudevan} R.~V.,  {Fabian} A.~C.,  2007, \mn@doi [\mnras]
  {10.1111/j.1365-2966.2007.12328.x}, \href
  {https://ui.adsabs.harvard.edu/abs/2007MNRAS.381.1235V} {381, 1235}

\bibitem[\protect\citeauthoryear{{V{\'e}ron-Cetty} \&
  {V{\'e}ron}}{{V{\'e}ron-Cetty} \& {V{\'e}ron}}{2001}]{2001veron-cetty}
{V{\'e}ron-Cetty} M.~P.,  {V{\'e}ron} P.,  2001, \mn@doi [\aap]
  {10.1051/0004-6361:20010718}, \href
  {https://ui.adsabs.harvard.edu/abs/2001A&A...374...92V} {374, 92}

\bibitem[\protect\citeauthoryear{{Vestergaard} \& {Peterson}}{{Vestergaard} \&
  {Peterson}}{2006}]{vestergaardpeterson06}
{Vestergaard} M.,  {Peterson} B.~M.,  2006, \mn@doi [ApJ] {10.1086/500572},
  \href {http://adsabs.harvard.edu/abs/2006ApJ...641..689V} {641, 689}

\bibitem[\protect\citeauthoryear{{Vietri} et~al.,}{{Vietri}
  et~al.}{2018}]{vietrietal18}
{Vietri} G.,  et~al., 2018, \mn@doi [\aap] {10.1051/0004-6361/201732335}, \href
  {https://ui.adsabs.harvard.edu/abs/2018A&A...617A..81V} {617, A81}

\bibitem[\protect\citeauthoryear{{Vietri} et~al.,}{{Vietri}
  et~al.}{2020}]{vietrietal20}
{Vietri} G.,  et~al., 2020, \mn@doi [\aap] {10.1051/0004-6361/202039136}, \href
  {https://ui.adsabs.harvard.edu/abs/2020A&A...644A.175V} {644, A175}

\bibitem[\protect\citeauthoryear{{Walter} \& {Fink}}{{Walter} \&
  {Fink}}{1993}]{1993Walter&Fink}
{Walter} R.,  {Fink} H.~H.,  1993, \aap, \href
  {https://ui.adsabs.harvard.edu/abs/1993A&A...274..105W} {274, 105}

\bibitem[\protect\citeauthoryear{{Wang} \& {Netzer}}{{Wang} \&
  {Netzer}}{2003}]{wangnetzer03}
{Wang} J.-M.,  {Netzer} H.,  2003, \mn@doi [\aap] {10.1051/0004-6361:20021511},
  \href {http://adsabs.harvard.edu/abs/2003A%26A...398..927W} {398, 927}

\bibitem[\protect\citeauthoryear{{Wang} \& {Zhou}}{{Wang} \&
  {Zhou}}{1999}]{wangzhou99}
{Wang} J.-M.,  {Zhou} Y.-Y.,  1999, \mn@doi [\apj] {10.1086/307080}, \href
  {https://ui.adsabs.harvard.edu/abs/1999ApJ...516..420W} {516, 420}

\bibitem[\protect\citeauthoryear{{Wang}, {Du}, {Valls-Gabaud}, {Hu}  \&
  {Netzer}}{{Wang} et~al.}{2013}]{wangetal13}
{Wang} J.-M.,  {Du} P.,  {Valls-Gabaud} D.,  {Hu} C.,   {Netzer} H.,  2013,
  \mn@doi [Physical Review Letters] {10.1103/PhysRevLett.110.081301}, \href
  {http://adsabs.harvard.edu/abs/2013PhRvL.110h1301W} {110, 081301}

\bibitem[\protect\citeauthoryear{{Wang}, {Qiu}, {Du}  \& {Ho}}{{Wang}
  et~al.}{2014}]{2014wang}
{Wang} J.-M.,  {Qiu} J.,  {Du} P.,   {Ho} L.~C.,  2014, \mn@doi [\apj]
  {10.1088/0004-637X/797/1/65}, \href
  {https://ui.adsabs.harvard.edu/abs/2014ApJ...797...65W} {797, 65}

\bibitem[\protect\citeauthoryear{{Wang} et~al.,}{{Wang}
  et~al.}{2023}]{Wang2023}
{Wang} J.-M.,  et~al., 2023, \mn@doi [\apj] {10.3847/1538-4357/acdf48}, \href
  {https://ui.adsabs.harvard.edu/abs/2023ApJ...954...84W} {954, 84}

\bibitem[\protect\citeauthoryear{{Wildy}, {Czerny}  \& {Panda}}{{Wildy}
  et~al.}{2019}]{wildyetal19}
{Wildy} C.,  {Czerny} B.,   {Panda} S.,  2019, \mn@doi [\aap]
  {10.1051/0004-6361/201935620}, \href
  {https://ui.adsabs.harvard.edu/abs/2019A&A...632A..41W} {632, A41}

\bibitem[\protect\citeauthoryear{{Wilkes}, {Kuraszkiewicz}, {Green}, {Mathur}
  \& {McDowell}}{{Wilkes} et~al.}{1999}]{wilkesetal99}
{Wilkes} B.~J.,  {Kuraszkiewicz} J.,  {Green} P.~J.,  {Mathur} S.,   {McDowell}
  J.~C.,  1999, \mn@doi [\apj] {10.1086/306828}, \href
  {https://ui.adsabs.harvard.edu/abs/1999ApJ...513...76W} {513, 76}

\bibitem[\protect\citeauthoryear{{Wilkins}, {Gallo}, {Costantini}, {Brandt}  \&
  {Blandford}}{{Wilkins} et~al.}{2021}]{wilkinsetal21}
{Wilkins} D.~R.,  {Gallo} L.~C.,  {Costantini} E.,  {Brandt} W.~N.,
  {Blandford} R.~D.,  2021, \mn@doi [\nat] {10.1038/s41586-021-03667-0}, \href
  {https://ui.adsabs.harvard.edu/abs/2021Natur.595..657W} {595, 657}

\bibitem[\protect\citeauthoryear{{Wolf} et~al.,}{{Wolf}
  et~al.}{2020}]{wolfetal20}
{Wolf} J.,  et~al., 2020, \mn@doi [\mnras] {10.1093/mnras/staa018}, \href
  {https://ui.adsabs.harvard.edu/abs/2020MNRAS.492.3580W} {492, 3580}

\bibitem[\protect\citeauthoryear{{Xu}, {Bian}, {Shen}, {Zuo}, {Fan}  \&
  {Zhu}}{{Xu} et~al.}{2018}]{xuetal18}
{Xu} F.,  {Bian} F.,  {Shen} Y.,  {Zuo} W.,  {Fan} X.,   {Zhu} Z.,  2018,
  \mn@doi [\mnras] {10.1093/mnras/sty1763}, \href
  {https://ui.adsabs.harvard.edu/abs/2018MNRAS.480..345X} {480, 345}

\bibitem[\protect\citeauthoryear{Yue et~al.,}{Yue
  et~al.}{2024}]{yue_novel_2024}
Yue B.~H.,  et~al., 2024, \mn@doi [Monthly Notices of the Royal Astronomical
  Society] {10.1093/mnras/stae725}, 529, 3939

\bibitem[\protect\citeauthoryear{Yue et~al.,}{Yue
  et~al.}{2025}]{yue_novel_2025}
Yue B.~H.,  et~al., 2025, \mn@doi [Monthly Notices of the Royal Astronomical
  Society] {10.1093/mnras/staf077}, 537, 858

\bibitem[\protect\citeauthoryear{{Yun}, {Reddy}  \& {Condon}}{{Yun}
  et~al.}{2001}]{yunetal01}
{Yun} M.~S.,  {Reddy} N.~A.,   {Condon} J.~J.,  2001, \mn@doi [\apj]
  {10.1086/323145}, \href
  {https://ui.adsabs.harvard.edu/abs/2001ApJ...554..803Y} {554, 803}

\bibitem[\protect\citeauthoryear{{Zamfir}, {Sulentic}  \& {Marziani}}{{Zamfir}
  et~al.}{2008}]{zamfiretal08}
{Zamfir} S.,  {Sulentic} J.~W.,   {Marziani} P.,  2008, \mn@doi [MNRAS]
  {10.1111/j.1365-2966.2008.13290.x}, \href
  {http://adsabs.harvard.edu/abs/2008MNRAS.387..856Z} {387, 856}

\bibitem[\protect\citeauthoryear{{Zappacosta} et~al.,}{{Zappacosta}
  et~al.}{2020}]{zappacostaetal20}
{Zappacosta} L.,  et~al., 2020, \mn@doi [\aap] {10.1051/0004-6361/201937292},
  \href {https://ui.adsabs.harvard.edu/abs/2020A&A...635L...5Z} {635, L5}

\bibitem[\protect\citeauthoryear{{Zhang} et~al.,}{{Zhang}
  et~al.}{2024}]{zhangetal24}
{Zhang} Z.,  et~al., 2024, \mn@doi [arXiv e-prints]
  {10.48550/arXiv.2407.08596}, \href
  {https://ui.adsabs.harvard.edu/abs/2024arXiv240708596Z} {p. arXiv:2407.08596}

\bibitem[\protect\citeauthoryear{{Zhou}, {Wang}, {Yuan}, {Lu}, {Dong}, {Wang}
  \& {Lu}}{{Zhou} et~al.}{2006}]{2006zhou}
{Zhou} H.,  {Wang} T.,  {Yuan} W.,  {Lu} H.,  {Dong} X.,  {Wang} J.,   {Lu} Y.,
   2006, \mn@doi [\apjs] {10.1086/504869}, \href
  {https://ui.adsabs.harvard.edu/abs/2006ApJS..166..128Z} {166, 128}

\bibitem[\protect\citeauthoryear{{van der Werf} et~al.,}{{van der Werf}
  et~al.}{2010}]{vanderwerfetal10}
{van der Werf} P.~P.,  et~al., 2010, \mn@doi [\aap]
  {10.1051/0004-6361/201014682}, \href
  {https://ui.adsabs.harvard.edu/abs/2010A&A...518L..42V} {518, L42}

\makeatother
\end{thebibliography}

\appendix
\section{The SED sample photometric data}
\label{App:digitalSAMPLE}

{Our sample consists of 155 quasars with z $\lesssim$ 1 and covers a redshift that allows for the detection and observation of the \hb+\feii\ region from optical spectrometers. As mentioned in Sec. \ref{sec:sample_retrieve}, we retrieved the data from multiple catalogs.  The radio to near-UV photometric data were downloaded from the NASA Extragalactic Database (NED). We provide the data as downloaded from NED, at the beginning of this work, for the 155 sources used to model our SEDs, in a zip file{ } available online as supplementary material.}} 

{The zip archive contains 155 files with comma-separated values (CSV) , one file for each source. Each CSV file contains up to   {26} columns, including: observed passband, frequency, flux density, and refcode (reference code), for each photometric measurement listed on each file. A complete listing of the columns is provided in table \ref{tab:columnsNED}}

 \begin{table}
    \centering
    \tabcolsep=2pt
    \begin{tabular}{rlcccccc}\hline\hline
Column header & Description \\\hline 
No. &  row number\\ 
Observed Passband &  Observed passband reported by the authors\\
& who published the data originally \\
Photometry Measurement & Photometry measurement reported by the authors \\
Uncertainty & Uncertainty reported by the authors  \\
Units &  Units of the photometry measurement\\
Frequency & NED frequency in units of Hz\\
Flux Density & NED flux density in units of Jy\\
Upper limit of uncertainty & Upper limit of uncertainty\\
Lower limit of uncertainty & Lower limit of uncertainty\\
Upper limit of Flux Density & Upper limit of Flux Density\\
NED Uncertainty &  NED Uncertainty in units of Jy\\
NED Units & Uniform units  \\
Refcode &  ADS bibliographic code (bibcode) of reference\\
Significance & Significance of the measurement\\
Published frequency & Frequency published by the authors\\
Frequency Mode & Frequency mode of the measurement\\
Coordinates Targeted & Coordinates targeted\\
Spatial Mode & Spatial Mode\\
Qualifiers & Qualifiers of the measurement \\
Comments &  Comments about data processing \\
Unnamed 21-26 & Additional comments about data processing \\
\hline
    \end{tabular}
    \caption{Column header: name of each individual column in the CSV table; description: brief explanation of the corresponding  column content. Authors means the authors of the paper refenced in Column Refcode.  }
    \label{tab:columnsNED}
\end{table}

{A synopsis of the instrumental resolutions and apertures for the observations across the electromagnetic spectrum is provided in Table \ref{tab:apertures} more details are listed on each CSV file.  }

 \begin{table}
    \centering
    \tabcolsep=2pt
    \begin{tabular}{cccccccc}\hline\hline
 Range & Observatory/  & Channels & PSF & Aperture    \\ 
 & Survey \\\hline 
 X-ray & XMM & EPIC & 6"& 15" \\   
       & ASCA & GIS \& SIS & 2.9' & 30" \\
       & SWIFT & XRT & 18" & coded \\
       & EINSTEIN & IPC & 1' & 6-30"\\
       & ROSAT & PSPC & 5-25" & 30" \\ 
\hline
UV &  FUSE &SiC \& LiF &1.5-5" & 20-30" \\
   & SWIFT & UVOT & 2.5" & 2.5 - 12" \\
   & XMM-OM & OM & 1.5-3" &17.5"  \\
   & HST & FOS & 1.5" & 0.1-4.3" \\
\hline
 Optical & XMM-OM & OM & 1.5-3" & 6" \\ 
         & SDSS &   & 1.1-2.0" & PSF, model,  \\ 
     & & & &   Petrosian \\
         & KPNO &  & 1-1.5" & 1-2" \\
\hline
IR &  WISE & W1 & 6.1" & \\
   &  WISE & W2 & 6.4" & \\
   &  WISE & W3 & 6.5" &  \\
   &  WISE & W4 & 12.0" & \\
   & ISO & ISOCAM & 3" \\
   & ISO & ISOPHOT & 90"\\
   & 2MASS &  J (1.25 $\mu$m) & $\sim$2.9" &  7.0" \\
   & 2MASS &  H (1.65 $\mu$m) & $\sim$2.8" &  7.0" \\
   & 2MASS &  Ks (2.17 $\mu$m) & $\sim$2.5" &  7.0" \\
   & SPITZER & IRAC & 1.66-1.98"   \\
   & IRAS  &  12   $\mu m$ & \ldots &1' by 5'\\
   & IRAS  & 25 $\mu$m & \ldots & 1' by 5'\\
   & IRAS & 60 $\mu$m & \ldots\ & 2' by 5' \\ 
   & IRAS & 100 $\mu$m & \ldots & 4' by 5' \\
   & AKARI & 9$\mu$m & 5.5" & 9.36 × 9.36  \\
   & AKARI & 18$\mu$m  & 5.7" & 10.4 × 9.36\\
\hline
 Radio & VLA FIRST & 20 cm & 5" & \ldots  \\
       & NVSS   & 21 cm & 45" & \ldots \\
       & OVRO   & 100 GHz & 8" & \ldots \\
       
\hline
    \end{tabular}
    \caption{Columns list facility name, wavelength of frequency of the observation, resolution defined by instrumental PSF, and aperture, whenever appropriate.}
    \label{tab:apertures}
\end{table}

\appendix
\setcounter{section}{1}
\section{The SED data in machine-readable form}
\label{digitalSED}

Template median SED ($q2$)  for the RQ extreme population A (13{9} sources), along with the first and third quartile ($q1$ and $q3$), in units of normalized arbitrary intensity (scaled to the same value at 5100 \AA) $\nu f_\nu$ vs frequency $\nu$\ [Hz], on linear scale.  The following files can be converted to FITS:

\begin{itemize}
    \item {\tt\ sed$\_$q2$\_$cloudy$\_$f5100.{sed}} The file contains a working SED ready to be input to {\tt CLOUDY} containing the median SED. 
    \item {\tt\ sed$\_$q1$\_$cloudy$\_$f5100.{sed}} The first quartile can be tentatively interpreted as representative of the sub-Eddington elements in our sample. Given the rather poor agreement between the X-ray range as deduced from the observations and the disk models, it should not be used in photoionization computation.
    \item {\tt\ sed$\_$q3$\_$cloudy$\_$f5100.{sed}}\ The third quartile is probably representative of the most super-Eddington sources in our sample. 
       \end{itemize}

In addition, we provide slightly different SEDs with $\log \dot{m}$\ = 1.0 following \citet{kubotadone19}, 

\begin{itemize}
   \item {\tt\ sed$\_$q2$\_$se$\_$cloudy.{sed}} 
    \item {\tt\ sed$\_$q3$\_$se$\_$cloudy.{sed}} 
       \end{itemize}

In this case no $q1$\ SED is provided since the observed SED is definitely sub-Eddington. 

{ For the luminosity-scaled SED we provide only the median, in the same format of the previous datafiles: }
\begin{itemize}
    \item {\tt\ sed$\_$q2$\_$lum$\_$cloudy.{sed}} The file contains a working SED ready to be input to {\tt CLOUDY}. 
\end{itemize}

{ The luminosity at 51000 \AA\ of this SED  is slightly higher ($\lambda L_\lambda$(5100 \AA)    $\approx 10^{44.6}$ erg s$^{-1}$ than the median luminosity value of the RQ  sample at $\lambda L_\lambda$(5100 \AA),  $\approx 10^{44.4}$ erg s$^{-1}$ . }

{The SEDs are provided in a zip file as supplementary material available onlline.  
}



\appendix


\bsp	
\label{lastpage}
\end{document}